# Nonadiabatic Field:
# A Conceptually Novel Approach for Nonadiabatic Quantum Molecular Dynamics

*Baihua Wu[1,†], Bingqi Li[1,†], Xin He[1,†], Xiangsong Cheng[1], Jiajun Ren[2], Jian Liu[1,*]*

1. Beijing National Laboratory for Molecular Sciences, Institute of Theoretical and Computational Chemistry, College of Chemistry and Molecular Engineering, Peking University, Beijing 100871, China

2. Key Laboratory of Theoretical and Computational Photochemistry, Ministry of Education, College of Chemistry, Beijing Normal University, Beijing 100875, China

AUTHOR INFORMATION

**Corresponding Author**

* Electronic mail: jianliupku@pku.edu.cn

**Author Contributions**

† B. W., B. L. and X. H. contributed equally.

**ABSTRACT.**

Reliable trajectory-based nonadiabatic quantum dynamics methods at the atomic/molecular level are critical for the practical understanding and rational design of many important processes in real large/complex systems, where the quantum dynamical behavior of electrons and that of nuclei are coupled. The paper reports latest progress of nonadiabatic field (NaF), a conceptually novel approach for nonadiabatic quantum dynamics with independent trajectories. Substantially different from the mainstreams of Ehrenfest-like dynamics and surface hopping methods, the nuclear force in NaF involves the nonadiabatic force arising from the nonadiabatic coupling between different electronic states, in addition to the adiabatic force contributed by a single adiabatic electronic state. NaF is capable of faithfully describing the interplay between electronic and nuclear motion in a broad regime, which covers where the relevant electronic states keep coupled in a wide range or all the time and where the bifurcation characteristic of nuclear motion is essential. NaF is derived from the exact generalized phase space formulation with coordinate-momentum variables, where constraint phase space (CPS) is employed for discrete electronic-state degrees of freedom (DOFs) and infinite Wigner phase space is used for continuous nuclear DOFs. We propose efficient integrators for the equations of motion of NaF in both adiabatic and diabatic representations. Since the formalism in the CPS formulation is not unique, NaF can in principle be implemented with various phase space representations of the time correlation function (TCF) for the time-dependent property. They are applied to a suite of representative gas-phase and condensed-phase benchmark models where numerically exact results are available for comparison. It is shown that NaF is relatively insensitive to the phase space representation of the electronic TCF and will be a potential tool for practical and reliable simulations of the quantum mechanical behavior of both electronic and nuclear dynamics of nonadiabatic transition processes in real systems.

# TOC GRAPHICS

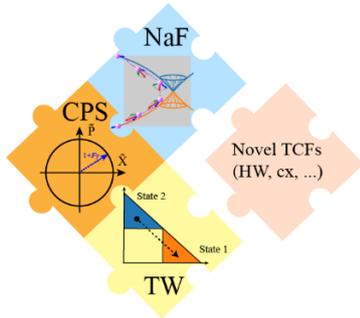

# 1. Introduction

A great deal of theoretical effort has been focused on developing trajectory-based approaches for including quantum mechanical effects in molecular dynamics simulations. These trajectory-based approaches consistently recover classical mechanical regions in chemical and biological processes, where nuclear quantum effects are negligible and classical molecular dynamics on a single-adiabatic-electronic-state potential energy surface (PES) is capable of describing the main features[1] when the Born-Oppenheimer (BO) approximation[2] is valid. Such classical mechanical regions are often too difficult to approach numerically by wave function methods, because a lot of destructive interference in the wave function interpretation (or in the quantum basis set expansion) is necessary to yield classical dynamics. Reasonable trajectory-based dynamics approaches are competent in practically illustrating the transition from the classical mechanical region to the region where nuclear quantum effects become non-negligible and then important. The phase space formulation with coordinate-momentum variables offers not only an exact representation of classical mechanics[3, 4], but also a rigorous interpretation of quantum mechanics[5-12]. It naturally bridges quantum and classical concepts, which offers insight into developing practical trajectory-based quantum dynamics approaches[11-18] for real molecular systems in the age of postmodern quantum mechanics[19].

When the BO approximation holds, the conventional infinite coordinate-momentum phase space[6, 9, 10, 15] for nuclear degrees of freedom (DOFs) is often used in the linearized semiclassical initial value representation (LSC-IVR)/classical Wigner[14, 20-27], forward-backward semiclassical dynamics[28-35], path integral Liouville dynamics[36-39], equilibrium continuity dynamics[17, 40], and other practical phase space quantum dynamics approaches[18, 41-48]. Some of these methods have been successfully applied to study electronically adiabatic processes where nuclear quantum effects are important[20, 21, 23, 24, 42, 49-65], such as transport properties[21, 50, 52, 55], dynamic structure factors[51, 65], vibrational spectra[38-40, 53, 54, 64, 66], chemical reaction rates[20, 21, 23, 24, 44], vibrational relaxation rates[56-61], and so forth.  It is important to note that the LSC-IVR/classical Wigner and some other approximate trajectory-based phase space approaches can in principle systematically be improved with more numerical effort by more advanced SC-IVRs of Miller and co-workers[67-70], the SC-IVR series of Pollak and co-workers[71-73], or by higher order corrections of the exact series expansion of the phase space propagator of Shao and Pollak[74] and those of ref [40] by us.

The BO approximation, however, fails in many important fundamental processes in electron/hole/charge transfer, photoactivated, energy conversion, strong electromagnetic field/ vacuum field manipulated processes in physical, chemical, biological, materials, geological, astronomic, quantum computation and quantum information systems[75-86], which involve the quantum mechanical behavior of both electrons and nuclei in the context of nonadiabatic transition dynamics[11, 12, 87-107].  Comprehensive topics on the nonadiabatic transition are presented in the seminal reviews by Domcke, Yarkony, Koppel, Cederbaum, and co-workers[87-93]. In nonadiabatic transition processes, electrons are often depicted by $F$ coupled discrete electronic states, while nuclei are described in continuous coordinate space.  The state-state nonadiabatic coupling can be either inherent in molecular systems or induced by the external field.  The composite/nonadiabatic system then includes both discrete electronic-state DOFs and continuous nuclear DOFs.  Since the pioneering work of Makri and co-workers for providing benchmark results for the spin-boson model (a condensed-phase two-state nonadiabatic model) by quasi-adiabatic propagator path integral (QuAPI)[108-111], a few numerically exact methods, including more advanced real time path integral methods[112-121], hierarchy equations of motion (HEOM)[122-127], dissipaton equation of motion[128-133], (multilayer) multiconfiguration time-dependent Hartree [(ML-)MCTDH][134-150], time-dependent density matrix renormalization group (TD-DMRG)[151-160], and other tensor network methods[161-166], have been developed for benchmark model systems for nonadiabatic dynamics.

When real (large) molecular systems are studied, most practical nonadiabatic dynamics methods with independent trajectories fall into two mainstreams. The first mainstream employs mean-field trajectories. The pioneering work of the Meyer-Miller (MM) mapping Hamiltonian model originally proposed from the "classical electron-analog" by Meyer and Miller in 1979[167] treats both electronic and nuclear DOFs on the same footing. In 1997 Stock and Thoss used the Schwinger oscillator theory of angular momentum to show the mapping Hamiltonian is exact in quantum mechanics[168]. Although the MM mapping Hamiltonian model approach[104, 168-229] also employs mean field trajectories, in many regards it consistently outperforms traditional Ehrenfest dynamics[230-233] in spirit of the Ehrenfest theorem[234]. Nuclear DOFs of the mean field trajectory evolve on an averaged PES, of which the nuclear force can be decomposed into two components in the adiabatic representation, one is the nonadiabatic nuclear force, and the other is the mean of the adiabatic (BO) forces contributed by all electronic states. Ehrenfest dynamics fails to capture the bifurcation characteristic of nuclear motion in the asymptotic region where the nonadiabatic coupling disappears (e. g., for the gas-phase photodissociation event). When the MM Hamiltonian is used in the forward-backward or fully SC-IVR framework[29, 67, 69], the interference between different mean field trajectories naturally leads to the nuclear bifurcation characteristic in the asymptotic region as shown in refs [186, 187]. When the infinite Wigner phase space is applied to both electronic and nuclear DOFs of the MM Hamiltonian and implemented in the full LSC-IVR framework for electronically nonadiabatic processes[183], the performance is much less satisfying even for electronic dynamics[185-187]. This is mainly because only a physical subspace of the harmonic oscillator is involved in the Schwinger mapping scheme, where the bosonic commutation relation does not necessarily hold, as pointed out in Appendix A of ref [235]. A few methods[190, 191, 193, 195, 197, 200-203, 207, 209] have been proposed on the (practical) quasi-classical level to improve the numerical performance over the full LSC-IVR for the MM mapping Hamiltonian model, of which the most prevailing one is the symmetrical quasi-classical (SQC) approach with triangle window functions (TWFs) developed by Cotton and Miller[195, 200] and widely used in refs [209, 210, 236-251]. This mainstream of trajectory-based methods is often competent in describing dynamics in the nonadiabatic coupling region, but difficult to produce the bifurcation characteristic of nuclear motion in the asymptotic region unless more numerically-demanding strategies are applied. Another mainstream employs various hopping mechanisms for connecting two independent BO trajectories generated on two different adiabatic PESs in the nonadiabatic coupling region[252-255]. The most popular approach is the fewest-switches surface hopping (SH) originally developed by Tully in 1990[253]. It has been successfully implemented for studying both gas phase models and realistic molecular systems[99, 256-269]. Quite a few other SH algorithms[270-295], which introduce either stochastic or deterministic hopping events, have been further developed. This mainstream naturally satisfies the Born-Oppenheimer limit and captures the bifurcation characteristic of nuclear motion in the asymptotic limit where no nonadiabatic coupling exists, but encounters the challenge in nonadiabatic processes where the states keep coupled in a wide region or all the time. In addition to the two mainstreams, there exist some other methods employing independent trajectories[296-305]. Nonadiabatic dynamics approaches with coupled trajectories include multiple spawning by Martinez and co-workers[306-309], exact factorization by Gross and co-workers[310-312], multiconfiguration Ehrenfest by Shalashilin and co-workers[313-316], *etc*.

The generalized coordinate-momentum phase space formulation[11, 12, 235, 317-326] rigorously maps the composite system onto phase space with continuous coordinate-momentum variables, where nuclear DOFs are still depicted by the conventional infinite coordinate-momentum phase space but $F$ discrete electronic-states are represented by the *constraint coordinate-momentum phase space* (CPS), which is diffeomorphic to the complex Stiefel manifolds U($F$)/U($F$-$r$) (with $1 \leq r < F$ )[325-328]. After ref [235] first presented the key idea of a complete space with coordinate-momentum variables for constructing mapping Hamiltonian models for finite-state quantum systems, ref [318] further established the CPS formulation related to the quotient space $\mathrm{U}(F)/\mathrm{U}(F-1)$ for general $F$-state systems by the sphere representation with coordinate-momentum variables in its main text and by the simplex representation with action-angle variables in its Appendix A. Reference [320] then presented CPS with commutator variables that is related to the complex Stiefel manifolds $\mathrm{U}(F)/\mathrm{U}(F-r)$. Interestingly, the quotient space $\mathrm{U}(F)/\mathrm{U}(F-2)$ is related to the equations of motion (EOMs) of the first model of ref [235] or ref [318] as well as the method used in ref [329].

The exact EOMs of mapping coordinate-momentum variables of CPS for the pure $F$-state quantum system are linear[11, 12, 235, 317-326]. As pointed out in Appendix 3 of ref [12], the CPS formulation with coordinate-momentum variables is superior to the conventional Stratonovich phase space approaches with angle variables[330-333] used for studying composite/nonadiabatic systems[334-339], because the EOMs of the latter are highly nonlinear and tedious, where inevitable singularities need to be excluded in dynamics especially when $F$ is large. More importantly, the CPS formulation is versatile for yielding more new phase space representations of the finite-state quantum system[323, 326]. Some of these phase space representations will be discussed in Sub-Section 2.4. The generalized coordinate-momentum phase space formulation of quantum mechanics exactly represents the three key elements[12] for nonadiabatic dynamics: the phase space integral expression for the expectation or ensemble average of the physical property of interest, the initial condition on phase space, and the EOMs on phase space.

Even when we introduce the independent trajectory approximation in the generalized coordinate-momentum phase space formulation, the first two key elements are still exactly represented, and only the third key element is approximated. That is, the general Wigner-Moyal equation on quantum phase space, which is a partial differential equation, is replaced by a set of ordinary-differential equations to produce the independent trajectory. This involves the same strategy discussed in ref [40]. When the phase space function corresponding to the quantum Hamiltonian operator is used to generate the EOMs for the independent trajectory, it leads to the classical mapping model (CMM)[318, 319] when the quotient space $\mathrm{U}(F)/\mathrm{U}(F-1)$ is employed, or the CMM with commutator variables (CMMcv)[320] when the quotient space $\mathrm{U}(F)/\mathrm{U}(F-r)$ with $1 \leq r < F$ is used. More recently, we have proposed nonadiabatic field (NaF)[322, 324], a conceptually novel nonadiabatic dynamics approach with independent trajectories on quantum phase space. NaF is competent in faithfully describing both electronic and nuclear motion in the nonadiabatic coupling region and in the asymptotic region where the state-state coupling vanishes. The nuclear force in NaF includes two terms, one is the nonadiabatic nuclear force contributed by the product of the nonadiabatic coupling vector (NACV) and the electronic coherence between different electronic states, and the other is the adiabatic nuclear force of a single adiabatic electronic state (either stochastically with electronic weights or deterministically with the dominant electronic weight). NaF is then fundamentally different from the two prevailing conventional mainstreams with independent trajectories, because of two key elements: the exact expressions of the initial condition and of the time-dependent properties on generalized quantum phase space with coordinate-momentum variables for electronic and nuclear DOFs, and the nuclear EOMs on quantum phase space with the aforementioned adiabatic and nonadiabatic force. As shown in the main text of and Section S7 of the Supporting Information[340] of ref [322], the comparison among NaF, Ehrenfest dynamics, SH methods, and the brute-force implementation of the EOMs of NaF to either Ehrenfest dynamics or SH methods demonstrates the importance of the two key elements.

Recent progress on the CPS formulation has revealed new classes[323-326]. For instance, we have proposed a novel class of CPS for two-state systems[323], which satisfies a relation derived from the *Abel integral equation* leading to the exact population dynamics in the frozen nuclei limit. In any case of this class, each trajectory on CPS makes non-negative contribution to the electronic population dynamics. Interestingly, the TWF approach of Cotton and Miller[195] used for population dynamics in the SQC/MM method is proved as a special case of this class. More classes are proposed in ref [326]. In this paper, we apply NaF with a few new formalisms of CPS and test their performance in a suite of typical benchmark model systems, including linear vibronic coupling models, one-dimensional scattering models in gas-phase, system-bath models and atom-in-cavity models, where numerically exact results are available for comparison.

The paper is organized as follows: Section 2 presents the theory of NaF methods, including phase space mapping formalisms, equations of motion, and a review of various time correlation functions (TCFs) in the CPS formulation. Section 3 demonstrates the numerical performance of NaF methods for the suite of gas-phase and condensed-phase model systems. Conclusion remarks are presented in Section 4.

## 2. Theory

### 2.1 Background.

In atomic units, the full Hamiltonian for $N_{atom}$ nuclei and $N_{ele}$ electrons reads

$$\hat{H} = \sum_{J=1}^{N_{atom}} \frac{\hat{\mathbf{P}}_J \cdot \hat{\mathbf{P}}_J}{2\tilde{M}_J} + \sum_{j=1}^{N_{ele}} \frac{\hat{\underline{\mathbf{p}}}_j \cdot \hat{\underline{\mathbf{p}}}_j}{2} - \sum_{j=1}^{N_{ele}} \sum_{J=1}^{N_{atom}} \frac{Z_J}{\left|\hat{\mathbf{r}}_j - \hat{\mathbf{R}}_J\right|} + \sum_{j=1}^{N_{ele}} \sum_{i>j}^{N_{ele}} \frac{1}{\left|\hat{\mathbf{r}}_j - \hat{\mathbf{r}}_i\right|} + \sum_{J=1}^{N_{atom}} \sum_{I>J}^{N_{atom}} \frac{Z_I Z_J}{\left|\hat{\mathbf{R}}_I - \hat{\mathbf{R}}_J\right|} \quad . \tag{1}$$

Here, $\{\mathbf{R}_J, \mathbf{P}_J\}$ are the coordinates and momenta of the $J$-th nucleus, $\tilde{M}_J$ is the ratio of the mass of the $J$-th nucleus to the mass of an electron, $Z_J$ is the atomic (charge) number of the $J$-th nucleus, and $\{\mathbf{r}_j, \underline{\mathbf{p}}_j\}$ are the coordinates and momenta of the $j$-th electron. The first term of the right-hand side (RHS) of eq (1) represents the kinetic energy of the nuclei; the second term is the kinetic energy of the electrons; the third term stands for the Coulomb attraction potential between electrons and nuclei; the fourth and fifth terms are the repulsion potential between electrons and that between nuclei, respectively. The total number of nuclear DOFs is $N_{nuc} = 3N_{atom}$. The reduced Planck constant, $\hbar$, is set to 1 and then omitted for electronic DOFs throughout the paper. In Sub-Sections 2.1 and 2.2, to illustrate nuclear quantum effects, $\hbar$ for nuclear DOFs is explicitly expressed in formulas. But in other parts of the paper, $\hbar$ for nuclear DOFs is also set to 1 and then omitted, because atomic units are used in all the benchmark tests.

Define the electronic Hamiltonian $\hat{H}_{el}(\hat{\mathbf{R}})$ as the sum of the last four terms in the RHS of eq (1). The full Hamiltonian becomes

$$\hat{H} = \frac{1}{2}\hat{\mathbf{P}} \cdot \mathbf{M}^{-1} \hat{\mathbf{P}} + \hat{H}_{el}(\hat{\mathbf{R}}) \quad, \tag{2}$$

where $\mathbf{M} = diag\{\tilde{M}_1, \tilde{M}_1, \tilde{M}_1, \tilde{M}_2, \tilde{M}_2, \tilde{M}_2, \cdots, \tilde{M}_{N_{atom}}, \tilde{M}_{N_{atom}}, \tilde{M}_{N_{atom}}\}$ is the diagonal nuclear mass matrix, $\{\mathbf{R}, \mathbf{P}\}$ are the coordinate and momentum vectors of nuclear DOFs, and the physical nuclear kinetic energy $\sum_{J=1}^{N_{atom}} \frac{\hat{\mathbf{P}}_J \cdot \hat{\mathbf{P}}_J}{2\tilde{M}_J} = \frac{1}{2}\hat{\mathbf{P}} \cdot \mathbf{M}^{-1}\hat{\mathbf{P}} = \sum_{I=1}^{N_{nuc}} \frac{\hat{P}_I^2}{2M_I}$ is the same as the first term of the RHS of eq (1) for the full Hamiltonian. Here, $M_I$ is the mass of the $I$-th nuclear DOF, $P_I$ is the $I$-th component of the momentum vector, and $R_I$ is the $I$-th component of the coordinate vector.

Assume that $\{|\phi_k(\mathbf{R})\rangle\}$ is the complete set of orthonormal adiabatic electronic states for a given nuclear configuration, $\mathbf{R}$. The representation of $\hat{H}_{el}(\hat{\mathbf{R}})$ in the adiabatic basis reads

$$\hat{H}_{el}(\mathbf{R}) = \sum_k E_k(\mathbf{R})|\phi_k(\mathbf{R})\rangle\langle\phi_k(\mathbf{R})| \quad , \tag{3}$$

where $E_k(\mathbf{R})$ denotes the adiabatic potential energy surface of the *k*-th adiabatic electronic state $|\phi_k(\mathbf{R})\rangle$. The rigorous expression of eq (3) in general involves a complete set of infinite adiabatic electronic states. It was in refs [12, 341] where eq (2) with the expression of eq (3) for the electronic Hamiltonian, $\hat{H}_{el}(\mathbf{R})$, was first employed for phase space mapping methods for nonadiabatic dynamics. E.g., see eqs (55) and (57) of ref [12].

The physical nuclear kinetic energy operator in either eq (1) or eq (2) should intrinsically be expressed in the full coordinate space of nuclear DOFs, which is independent of the electronic space. When the complete adiabatic electronic basis set is available, the expression of the physical nuclear kinetic energy operator reads

$$\frac{1}{2}\hat{\mathbf{P}}\cdot\mathbf{M}^{-1}\hat{\mathbf{P}} = \sum_{k,l}\left(\frac{1}{2}\hat{\mathbf{P}}_{can}\cdot\mathbf{M}^{-1}\hat{\mathbf{P}}_{can}\delta_{kl} - i\hbar\mathbf{d}_{kl}(\mathbf{R})\cdot\mathbf{M}^{-1}\hat{\mathbf{P}}_{can} - \hbar^2\sum_{J=1}^{N_{nuc}}\frac{1}{2M_J}D_{kl}^{(J)}(\mathbf{R})\right)|\phi_k(\mathbf{R})\rangle\langle\phi_l(\mathbf{R})| \tag{4}$$

where $\hat{\mathbf{P}}_{can}$ is the canonical nuclear momentum operator that does not *explicitly* operate on any adiabatic basis state $|\phi_k(\mathbf{R})\rangle$,

$$\mathbf{d}_{kl}(\mathbf{R}) = \left\langle \phi_k(\mathbf{R}) \middle| \frac{\partial \phi_l(\mathbf{R})}{\partial \mathbf{R}} \right\rangle \quad (5)$$

is the first-order nonadiabatic coupling vector between the *k*-th and *l*-th adiabatic electronic states, of which the $J$-th component is $d_{kl}^{(J)}(\mathbf{R})$, and

$$D_{kl}^{(J)}(\mathbf{R}) = \left\langle \phi_k(\mathbf{R}) \middle| \frac{\partial^2 \phi_l(\mathbf{R})}{\partial R_J^2} \right\rangle \quad (6)$$

is the second-order derivative term with respect to the $J$-th nuclear DOF in the nonadiabatic coupling between the *k*-th and *l*-th adiabatic electronic states. It is easy to show

$$\mathbf{d}_{lk}(\mathbf{R}) = -\mathbf{d}_{kl}^{*}(\mathbf{R}) \quad (7)$$

and

$$D_{kl}^{(J)}(\mathbf{R}) = \sum_m d_{km}^{(J)}(\mathbf{R}) d_{ml}^{(J)}(\mathbf{R}) + \frac{\partial}{\partial R_J} d_{kl}^{(J)}(\mathbf{R}) \ . \quad (8)$$

Equations (4), (7), and (8) lead to an equivalent expression of the physical nuclear kinetic energy operator

$$\frac{1}{2}\hat{\mathbf{P}} \cdot \mathbf{M}^{-1}\hat{\mathbf{P}} = \sum_{k,l,n} \left( \frac{1}{2} \left( \hat{\mathbf{P}}_{\text{can}} \delta_{kn} - i\hbar \mathbf{d}_{kn}(\mathbf{R}) \right) \cdot \mathbf{M}^{-1} \left( \hat{\mathbf{P}}_{\text{can}} \delta_{nl} - i\hbar \mathbf{d}_{nl}(\mathbf{R}) \right) \right) |\phi_k(\mathbf{R})\rangle \langle \phi_l(\mathbf{R})| \ , \quad (9)$$

or in a more compact form,

$$\frac{1}{2}\hat{\mathbf{P}} \cdot \mathbf{M}^{-1}\hat{\mathbf{P}} = \frac{1}{2} \left( \hat{\mathbf{P}}_{\text{can}} \hat{\mathbf{1}}_{\text{ele}}^{(\text{adia})} - i\hbar \hat{\mathbf{d}}(\mathbf{R}) \right) \cdot \mathbf{M}^{-1} \left( \hat{\mathbf{P}}_{\text{can}} \hat{\mathbf{1}}_{\text{ele}}^{(\text{adia})} - i\hbar \hat{\mathbf{d}}(\mathbf{R}) \right) \ , \quad (10)$$

where $\hat{\mathbf{1}}_{\text{ele}}^{(\text{adia})} = \sum_k |\phi_k(\mathbf{R})\rangle \langle \phi_k(\mathbf{R})|$ is the identity operator in electronic space represented by the adiabatic electronic basis set and $\hat{\mathbf{d}}(\mathbf{R}) = \sum_{k,l} (\mathbf{d}_{kl}(\mathbf{R})) |\phi_k(\mathbf{R})\rangle \langle \phi_l(\mathbf{R})|$ is an electronic operator, of which each element is a vector in the nuclear space as defined by eq (5). The RHS of eq (9) or that of eq (10) was already used in the literature, e.g., refs [198, 276, 342, 343]. It is important to note that eqs (4), (8), (9), or (10) only hold when the adiabatic electronic basis set is complete. That is, the summation in eqs (4), (8), and (9) intrinsically include infinite adiabatic electronic states. We note that eq (10) inherently hints the relation between the physical nuclear momentum operator, $\hat{\mathbf{P}}$, and the canonical nuclear momentum operator in the adiabatic representation, $\hat{\mathbf{P}}_{can}$.

Most processes in chemistry, materials, biology, and so forth involve finite energy where only a finite number of electronic states are effectively included. Assume that only $F$ lowest adiabatic electronic basis states are relevant. The truncation of the electronic basis set in principle assumes that the third term of the equation below,

$$D_{kl}^{(J)}(\mathbf{R}) = \sum_{m=1}^{F} d_{km}^{(J)}(\mathbf{R}) d_{ml}^{(J)}(\mathbf{R}) + \frac{\partial}{\partial R_J} d_{kl}^{(J)}(\mathbf{R}) + \sum_{m=F+1}^{\infty} d_{km}^{(J)}(\mathbf{R}) d_{ml}^{(J)}(\mathbf{R}) \quad, \tag{11}$$

vanishes. When $\delta D_{kl}^{(J)}(\mathbf{R}) = \sum_{m=F+1}^{\infty} d_{km}^{(J)}(\mathbf{R}) d_{ml}^{(J)}(\mathbf{R})$ is not negligible for the system, it implies a nonabelian/Yang-Mills gauge field $-i\mathbf{d}(\mathbf{R})$ [12, 344] with the gauge field tensor

$$\mathcal{F}_{IJ} = \frac{\partial(-i\mathbf{d}^{(J)})}{\partial R_I} - \frac{\partial(-i\mathbf{d}^{(I)})}{\partial R_J} + i\left[-i\mathbf{d}^{(I)}, -i\mathbf{d}^{(J)}\right]_{ele} \quad. \tag{12}$$

When the truncation of the adiabatic electronic basis set is reasonable, $\delta D_{kl}^{(J)}(\mathbf{R})$ or $\mathcal{F}_{IJ}$ is often close to zero but should be ignored with caution.

When the *full* space of electrons is involved, which means infinite adiabatic electronic states are included, the expression of eq (3) is indeed complete for the electronic Hamiltonian. Under such a circumstance, an orthonormal diabatic electronic basis set, $\{|n\rangle\}$, that is independent of the nuclear coordinate vector, $\mathbf{R}$, can rigorously be defined,

$$|n\rangle = \sum_k T_{nk}^*(\mathbf{R}) |\phi_k(\mathbf{R})\rangle \quad, \tag{13}$$

where $T_{nk}^*(\mathbf{R}) = \langle \phi_k(\mathbf{R})|n\rangle$ or $T_{nk}(\mathbf{R}) = \langle n|\phi_k(\mathbf{R})\rangle$ is the element of the diabatic-to-adiabatic transformation matrix, $\mathbf{T}(\mathbf{R})$. Then the representation of $\hat{H}_{el}(\mathbf{R})$ in the diabatic basis set becomes

$$\hat{H}_{el}(\mathbf{R}) = \sum_{n,m} V_{nm}(\mathbf{R})|n\rangle\langle m| \equiv \mathbf{V}(\mathbf{R}) \quad , \tag{14}$$

where $V_{nm}(\mathbf{R}) = \langle n|\hat{H}_{el}(\mathbf{R})|m\rangle$ is the matrix element for the electronic Hamiltonian, $\hat{H}_{el}(\mathbf{R})$. When the complete diabatic electronic set is available, the physical nuclear kinetic energy operator reads

$$\frac{1}{2}\hat{\mathbf{P}}\cdot\mathbf{M}^{-1}\hat{\mathbf{P}} = \sum_{n,m}\left(\frac{1}{2}\hat{\mathbf{P}}\cdot\mathbf{M}^{-1}\hat{\mathbf{P}}\delta_{mn}\right)|n\rangle\langle m| \quad . \tag{15}$$

When only $F$ diabatic electronic states are effectively involved, the truncation in the diabatic basis set and eqs (2), (14), (15) leads to the expression of the full Hamiltonian operator

$$\hat{H}(\hat{\mathbf{R}},\hat{\mathbf{P}}) = \sum_{n,m=1}^{F}\left(\frac{1}{2}\hat{\mathbf{P}}^{\mathrm{T}}\mathbf{M}^{-1}\hat{\mathbf{P}}\delta_{nm} + V_{nm}(\hat{\mathbf{R}})\right)|n\rangle\langle m| \quad . \tag{16}$$

Equation (16) in principle assumes that $\{|n\rangle\}, n \in \{1,\cdots,F\}$ is the "complete" set of diabatic electronic states and $\{|\phi_k(\mathbf{R})\rangle\}, k \in \{1,\cdots,F\}$ is the "complete" set of orthonormal adiabatic electronic states.

We will first review the EOMs of NaF (on quantum phase space) in Sub-Section 2.2 and Sub-Section 2.3, then discuss the phase space integral expression of the TCF for evaluating the time-dependent property in Sub-Section 2.4.

## 2.2 Phase Space Mapping Hamiltonian for Nonadiabatic Systems

The one-to-one correspondence mapping function of the full Hamiltonian of eq (16) in the generalized coordinate-momentum phase space representation[11, 12, 235, 318-320] reads

$$H(\mathbf{R},\mathbf{P},\mathbf{x},\mathbf{p},\mathbf{\Gamma}) = \text{Tr}_{n,e}\left[\hat{H}\hat{K}_{\text{nuc}}(\mathbf{R},\mathbf{P}) \otimes \hat{K}_{\text{ele}}(\mathbf{x},\mathbf{p},\mathbf{\Gamma})\right]$$
$$= \frac{1}{2}\mathbf{P}^{\text{T}}\mathbf{M}^{-1}\mathbf{P} + \sum_{n,m=1}^{F} V_{mn}(\mathbf{R})\left(\frac{1}{2}(x^{(n)} + ip^{(n)})(x^{(m)} - ip^{(m)}) - \Gamma_{nm}\right), \quad (17)$$

where

$$\hat{K}_{\text{nuc}}(\mathbf{R},\mathbf{P}) = \left(\frac{\hbar}{2\pi}\right)^{N_{\text{nuc}}} \int d\boldsymbol{\xi}d\boldsymbol{\eta}\, e^{i\boldsymbol{\xi}\cdot(\hat{\mathbf{R}}-\mathbf{R})+i\boldsymbol{\eta}\cdot(\hat{\mathbf{P}}-\mathbf{P})} \quad (18)$$

denotes the mapping kernel of the Wigner phase space[6, 10, 319] for nuclear DOFs, and

$$\hat{K}_{\text{ele}}(\mathbf{x},\mathbf{p},\mathbf{\Gamma}) = \sum_{n,m=1}^{F}\left[\frac{1}{2}(x^{(n)} + ip^{(n)})(x^{(m)} - ip^{(m)}) - \Gamma_{nm}\right]|n\rangle\langle m| \quad (19)$$

denotes the mapping kernel of CPS[319, 320, 322, 326] for electronic DOFs. Here, the notation $\text{Tr}_n[\cdot]$ and $\text{Tr}_e[\cdot]$ represent the trace over nuclear DOFs and that over electronic DOFs, respectively. The commutator matrix $\mathbf{\Gamma}$ can be represented by a series of extended phase space variables (namely the commutator variables) through its spectral decomposition[320]. The mapping CPS[235, 318-320, 322], characterized by constraint $\mathcal{S}(\mathbf{x},\mathbf{p},\mathbf{\Gamma};\boldsymbol{\gamma})$, is related to the normalization of eq (19) (i.e., $\text{Tr}_e\left[\hat{K}_{\text{ele}}(\mathbf{x},\mathbf{p},\mathbf{\Gamma})\right]=1$), where $\boldsymbol{\gamma}$ denotes the parameter vector involved in the mapping CPS. The covariant form of eq (17) in the adiabatic representation reads[12]

$$H(\mathbf{R},\mathbf{P},\tilde{\mathbf{x}},\tilde{\mathbf{p}},\tilde{\mathbf{\Gamma}}) = \frac{1}{2}\mathbf{P}^{\text{T}}\mathbf{M}^{-1}\mathbf{P} + \sum_{n=1}^{F} E_n(\mathbf{R})\left(\frac{(\tilde{x}^{(n)})^2 + (\tilde{p}^{(n)})^2}{2} - \tilde{\Gamma}_{nn}\right), \quad (20)$$

where $\tilde{\mathbf{x}}(\mathbf{R}) + i\tilde{\mathbf{p}}(\mathbf{R}) = \mathbf{T}^{\dagger}(\mathbf{R})(\mathbf{x}+i\mathbf{p})$, $\tilde{\mathbf{\Gamma}}(\mathbf{R}) = \mathbf{T}^{\dagger}(\mathbf{R})\mathbf{\Gamma}\mathbf{T}(\mathbf{R})$ are covariant phase space variables of electronic DOFs in the adiabatic representation. As shown in refs [12, 345], the nuclear canonical momentum in the diabatic representation, $\mathbf{P}$, corresponds to the nuclear kinematic momentum in the adiabatic representation. The relation between the nuclear kinematic momentum in the adiabatic representation, $\mathbf{P}$, and the nuclear canonical momentum in the adiabatic representation, $\tilde{\mathbf{P}}$, reads[12, 198]

$$\tilde{\mathbf{P}} = \mathbf{P} + i \sum_{n,m=1}^{F} \left[ \frac{1}{2}\left(\tilde{x}^{(n)} + i\tilde{p}^{(n)}\right)\left(\tilde{x}^{(m)} - i\tilde{p}^{(m)}\right) - \tilde{\Gamma}_{nm} \right] \mathbf{d}_{mn}(\mathbf{R}) \quad . \tag{21}$$

When the electronic basis set is complete, the physical nuclear momentum, the nuclear canonical momentum in the diabatic representation, and the nuclear kinematic momentum in the adiabatic representation are in principle equivalent[12, 345]. This is inherently implied in eq (10) and eq (15).

Cotton, Liang and Miller directly started from the adiabatic representation and first showed the use of the nuclear kinematic momentum, $\mathbf{P}$, in Ehrenfest-like dynamics of the Meyer-Miller mapping model to avoid second-order nonadiabatic coupling terms[198]. The derivation of the nuclear kinematic momentum, $\mathbf{P}$, in the generalized phase space formulation has been demonstrated in Section S1 of the Supporting Information[346] of ref [320], Section 4 of ref [12] and Appendix 2 of its Supporting Information[345], where we have explicitly pointed out that the first-order nonadiabatic coupling corresponds to a nonabelian/Yang-Mills gauge field $-i\mathbf{d}^{(1)}(\mathbf{R})$ [12, 344] with the gauge field tensor defined by eq (12) and that the EOM of nuclear kinematic momentum $\mathbf{P}$ in the adiabatic representation does not include second-order nonadiabatic coupling terms *only* when such a gauge field tensor is zero, or equivalently, when the diabatic representation is rigorously defined[12, 87] or the adiabatic basis set is complete. NaF on quantum phase space employs nuclear kinematic momentum $\mathbf{P}$ in the adiabatic representation [322, 324]. In addition, as first explicitly pointed out in ref [12], and already used in refs [11, 320, 321, 347], nuclear kinematic momentum $\mathbf{P}$ in the adiabatic representation should be inherently utilized in SH methods. When the gauge field tensor (defined by eq (12)) is zero, in SH methods the use of nuclear kinematic momentum $\mathbf{P}$ in the adiabatic representation can also avoid second-order nonadiabatic coupling terms.

When the (electronic) adiabatic basis is incomplete, $\mathcal{F}_{IJ}$ in eq (12) deviates from zero. A correction term, $\mathbf{F}_{\text{residue}}$, arises from the nonabelian/Yang-Mills gauge field tensor in the force for the update of nuclear kinematic momentum vector $\mathbf{P}$ in the adiabatic representation[12, 167, 341, 348, 349], of which the $I$-th component reads

$$F_{\text{residue}}^{(I)} = \sum_{J=1}^{N_{\text{nuc}}} \frac{P_J}{M_J} \text{Tr}_e\left[\tilde{\boldsymbol{\rho}} \mathcal{F}_{IJ}\right]. \tag{22}$$

Here $\tilde{\boldsymbol{\rho}}$ denotes the corresponding effective electronic density matrix in the adiabatic representation, which depends on the EOMs of the nuclear DOFs. (An example is defined in eq (34).) Such a term was shown in eq (S30) of Appendix 2 of the Supporting Information[345] of ref [12] for the EOMs in the CPS formulation. Only when this term vanishes, the EOM of nuclear kinematic momentum $\mathbf{P}$ involves but the first-order nonadiabatic coupling term[12]. When nuclear canonical momentum $\tilde{\mathbf{P}}$ in the adiabatic representation was used instead, a term equivalent to the RHS of eq (22) appeared in Appendix B for the EOMs of the Meyer-Miller mapping model in ref [167] by Meyer and Miller and in eq (17) for Ehrenfest dynamics in ref [349] by Amano and Takatsuka. The relativistic analogue of the EOMs with the canonical momentum was proposed in ref [348] by Wong, where the nonadiabatic coupling plays the role of the nonabelian gauge field interacting with the isotopic-spin-carrying particle.

In all the benchmark tests presented in this paper where numerically exact results in the diabatic representation are available, the gauge field tensor defined by eq (12) is zero. When on-the-fly *ab initio* NaF simulations are performed for real systems where only a finite number of adiabatic electronic states are involved, the gauge field tensor of eq (12) should be disregarded with caution for the update of nuclear kinematic momentum $\mathbf{P}$.

## 2.3 Nonadiabatic Field

As derived in refs [12, 322, 324], the EOMs of the electronic phase space variables read

$$\dot{x}^{(n)} + i\dot{p}^{(n)} = -i\sum_{m=1}^{F} V_{nm}(\mathbf{R})\left(x^{(m)} + ip^{(m)}\right), \tag{23}$$

$$\dot{\Gamma}_{nm}(\mathbf{R}) = i\sum_{k=1}^{F}\left[\Gamma_{nk}V_{km}(\mathbf{R}) - V_{nk}(\mathbf{R})\Gamma_{km}\right] \tag{24}$$

in the diabatic representation[11, 12, 320], and

$$\dot{\tilde{x}}^{(n)}(\mathbf{R}) + i\dot{\tilde{p}}^{(n)}(\mathbf{R}) = -i\sum_{m=1}^{F} V_{nm}^{(\text{eff})}(\mathbf{R},\mathbf{P})\left[\tilde{x}^{(m)}(\mathbf{R}) + i\tilde{p}^{(m)}(\mathbf{R})\right], \tag{25}$$

$$\dot{\tilde{\Gamma}}_{nm}(\mathbf{R}) = i\sum_{k=1}^{F}\left[\tilde{\Gamma}_{nk}(\mathbf{R})V_{km}^{(\text{eff})}(\mathbf{R},\mathbf{P}) - V_{nk}^{(\text{eff})}(\mathbf{R},\mathbf{P})\tilde{\Gamma}_{km}(\mathbf{R})\right] \tag{26}$$

in the adiabatic representation[11, 12, 320]. Here, the elements of the effective potential matrix in the adiabatic representation are

$$V_{nm}^{(\text{eff})}(\mathbf{R},\mathbf{P}) = E_n(\mathbf{R})\delta_{nm} - i\mathbf{M}^{-1}\mathbf{P}\cdot\mathbf{d}_{nm}(\mathbf{R}) . \tag{27}$$

Define $\mathbf{g} = \mathbf{x} + i\mathbf{p}$ and $\tilde{\mathbf{g}}(\mathbf{R}) = \tilde{\mathbf{x}}(\mathbf{R}) + i\tilde{\mathbf{p}}(\mathbf{R})$. Integrating eqs (23)-(24) over $t$ with fixed $\mathbf{R}$ leads to the propagation of electronic phase space variables in the diabatic representation within a finite time-step $\Delta t$ [320, 350]

$$\mathbf{g}_{t+\Delta t} = \mathbf{U}(\mathbf{R};\Delta t)\mathbf{g}_t , \tag{28}$$

$$\mathbf{\Gamma}_{t+\Delta t} = \mathbf{U}(\mathbf{R};\Delta t)\mathbf{\Gamma}_t\mathbf{U}^{\dagger}(\mathbf{R};\Delta t) , \tag{29}$$

where the propagator in the diabatic representation reads

$$\mathbf{U}(\mathbf{R};\Delta t) = \exp(-i\mathbf{V}(\mathbf{R})\Delta t) . \tag{30}$$

Similarly, integrating eqs (25)-(26) over $t$ with fixed $\mathbf{R}$ and $\mathbf{P}$ yields[320]

$$\tilde{\mathbf{g}}_{t+\Delta t}(\mathbf{R}) = \tilde{\mathbf{U}}(\mathbf{R},\mathbf{P};\Delta t)\tilde{\mathbf{g}}_t(\mathbf{R}) , \tag{31}$$

$$\tilde{\Gamma}_{t+\Delta t}(\mathbf{R}) = \tilde{\mathbf{U}}(\mathbf{R},\mathbf{P};\Delta t)\tilde{\Gamma}_t(\mathbf{R})\tilde{\mathbf{U}}^\dagger(\mathbf{R},\mathbf{P};\Delta t) \tag{32}$$

with the propagator in the adiabatic representation

$$\tilde{\mathbf{U}}(\mathbf{R},\mathbf{P};\Delta t) = \exp\left(-i\mathbf{V}^{(\text{eff})}(\mathbf{R},\mathbf{P})\Delta t\right) . \tag{33}$$

For a given effective electronic density matrix

$$\tilde{\boldsymbol{\rho}}(\mathbf{R}) \equiv \tilde{\boldsymbol{\rho}}\left(\tilde{\mathbf{x}}(\mathbf{R}),\tilde{\mathbf{p}}(\mathbf{R}),\tilde{\Gamma}(\mathbf{R})\right) = \frac{1+\text{Tr}_e[\tilde{\Gamma}]}{\left(\tilde{\mathbf{x}}^T\tilde{\mathbf{x}}+\tilde{\mathbf{p}}^T\tilde{\mathbf{p}}\right)}(\tilde{\mathbf{x}}+i\tilde{\mathbf{p}})(\tilde{\mathbf{x}}-i\tilde{\mathbf{p}})^T - \tilde{\Gamma} \tag{34}$$

in the adiabatic representation, the EOMs of NaF for nuclear DOFs read[322, 324]

$$\begin{aligned} \dot{\mathbf{R}} &= \mathbf{M}^{-1}\mathbf{P} \\ \dot{\mathbf{P}} &= \mathbf{F}_{\text{adia.}} + \mathbf{F}_{\text{nonadia.}} \end{aligned}, \tag{35}$$

where

$$\mathbf{F}_{\text{adia.}} = -\nabla_\mathbf{R} E_{j_{\text{occ}}}(\mathbf{R}) \tag{36}$$

denotes the adiabatic force provided by the adiabatic electronic state, $\left|\phi_{j_{\text{occ}}}(\mathbf{R})\right\rangle$, and

$$\mathbf{F}_{\text{nonadia.}} = -\sum_{n\neq m}^F (E_n(\mathbf{R}) - E_m(\mathbf{R}))\mathbf{d}_{mn}(\mathbf{R})\tilde{\rho}_{nm}(\mathbf{R}) \tag{37}$$

represents the nonadiabatic force. Here, $\tilde{\rho}_{nm}(\mathbf{R})$ denotes the matrix element of $\tilde{\boldsymbol{\rho}}(\tilde{\mathbf{x}}(\mathbf{R}),\tilde{\mathbf{p}}(\mathbf{R}),\tilde{\Gamma}(\mathbf{R}))$. Equation (37) has an equivalent form to the nonadiabatic force in Ehrenfest-like dynamics of CMMcv[11, 320]. When the time-reversal symmetry holds, $\mathbf{F}_{\text{residue}}$ of eq (22) is a correction term to the nonadiabatic nuclear force in eq (37). This correction term involves the derivative with respective to nuclear coordinate $\mathbf{R}$ of nonadiabatic coupling vector $\mathbf{d}(\mathbf{R})$, which is often costly to obtain. Nevertheless, the contribution from eq (22) to the nonadiabatic nuclear force is typically small in most cases and should be neglected with caution. When the nonadiabatic nuclear force of eq (37) is neglected, the EOMs in eq (35) lead to the BO trajectory employed in SH methods (where nuclear kinematic momentum $\mathbf{P}$ should be used). Since eq (37) intrinsically provides feedback of electronic coherence to nuclear motion, it should never be disregarded in the state-state (nonadiabatic) coupling regions where the term, $(E_n(\mathbf{R}) - E_m(\mathbf{R}))\mathbf{d}_{mn}(\mathbf{R}) = \langle \phi_m(\mathbf{R}) | \nabla_\mathbf{R} \hat{H}_{el}(\mathbf{R}) | \phi_n(\mathbf{R}) \rangle$ $(n \neq m)$, plays a role. The adiabatic nuclear force in eq (36) can be determined either by the weight stochastically or by the dominant component deterministically[322, 324]. In the latter case, eq (36) reads

$$\mathbf{F}_{\text{adia.}} = -\sum_{k=1}^{F} \nabla_\mathbf{R} E_k(\mathbf{R}) \prod_{j \neq k}^{F} h\left(\tilde{\rho}_{kk}\left(\tilde{\mathbf{x}}(\mathbf{R}), \tilde{\mathbf{p}}(\mathbf{R}), \tilde{\boldsymbol{\Gamma}}(\mathbf{R})\right) - \tilde{\rho}_{jj}\left(\tilde{\mathbf{x}}(\mathbf{R}), \tilde{\mathbf{p}}(\mathbf{R}), \tilde{\boldsymbol{\Gamma}}(\mathbf{R})\right)\right) \quad (38)$$

with $h(x)$ denotes the Heaviside step function. The relation between the EOMs of NaF and the exact EOMs in the generalized coordinate-momentum phase space formulation of quantum mechanics is discussed in Section S9 of the Supporting Information.

In NaF the mapping energy on quantum phase space

$$H_{\text{NaF}}(\mathbf{R}, \mathbf{P}, \tilde{\boldsymbol{\rho}}) = \frac{1}{2}\mathbf{P}^T \mathbf{M}^{-1} \mathbf{P} + E_{j_{\text{occ}}}(\mathbf{R}) \quad (39)$$

in the adiabatic representation for each trajectory should be conserved. Here, $E_{j_{\text{occ}}}(\mathbf{R})$ is the adiabatic PES of the single electronic state which contributes to the nuclear adiabatic force. As discussed in ref [12] and already implemented in refs [11, 12, 320, 321, 347], the RHS of eq (39) is the same mapping energy (with nuclear kinematic momentum $\mathbf{P}$) on quantum phase space that the SH trajectory should conserve, even when nuclear DOFs are treated quantum mechanically and the nuclear initial condition is sampled on Wigner phase space. Since the nuclear kinematic momentum rather than the nuclear canonical momentum is involved in the mapping energy, in NaF the corresponding mapping energy (eq (39)) should not be used to generate the EOMs for $(\mathbf{R}, \mathbf{P}; \tilde{\mathbf{x}}, \tilde{\mathbf{p}}, \tilde{\boldsymbol{\Gamma}})$ on quantum phase space. In our previous works[322, 324], we employed a momentum rescaling approach to ensure that the trajectory generated by eqs (35)-(37) conserves the corresponding mapping energy (eq (39)) for NaF. That is, the mapping energy of eq (39) on quantum phase space is conserved by rescaling **P** along its direction after each time-step. In Section S1 of the Supporting Information, we demonstrate that this approach corresponds to the *effective* nonadiabatic force perpendicular to nuclear velocity $\mathbf{M}^{-1}\mathbf{P}$ in the infinitesimal time-step limit. Similar treatments have been discussed in Section S2 of the Supporting Information[351] of ref [324]. When the state with the dominant weight is switched, the nuclear kinematic momentum is also rescaled along its direction to achieve energy conservation if possible. In cases of frustrated switching, the occupied state, $j_{occ}$, is not changed, even though this state no longer holds the dominant weight. The corresponding complete integrator is described in refs [322, 324]. Below we introduce a more efficient integrator for NaF.

Here we propose two important elements for efficiently integrating the EOMs of NaF within a finite time-step $\Delta t$. The first element is the numerical integrator for the effective nonadiabatic force as derived in Section S1 of the Supporting Information, and the second one is the more efficient numerical integrator scheme "P-e-R-e-P". Combining these two elements, the complete integrator for the EOMs of NaF for a finite time-step $\Delta t$ then reads:

1. Update the nuclear kinematic momentum within a half time-step $\Delta t/2$ using the adiabatic force

$$\mathbf{P}_{t+\Delta t/2} \leftarrow \mathbf{P}_t - \nabla_\mathbf{R} E_{j_{\text{old}}}(\mathbf{R}_t)\frac{\Delta t}{2} \ . \qquad (40)$$

2. Update the nuclear kinematic momentum within a half time-step $\Delta t/2$ using the numerical integrator for the effective nonadiabatic force for the $N_{\text{nuc}} \geq 2$ case,

$$\begin{aligned}\mathbf{P}_{t+\Delta t/2} \leftarrow\ & c_1\left(\mathbf{R}_t, \mathbf{P}_{t+\Delta t/2}, \tilde{\boldsymbol{\rho}}_t, \Delta t/2\right)\mathbf{M}^{1/2}\mathbf{e}_{\|}\left(\mathbf{R}_t, \tilde{\boldsymbol{\rho}}_t\right) \\ & + c_2\left(\mathbf{R}_t, \mathbf{P}_{t+\Delta t/2}, \tilde{\boldsymbol{\rho}}_t, \Delta t/2\right)\mathbf{M}^{1/2}\mathbf{\Pi}_{\perp}\left(\mathbf{R}_t, \mathbf{P}_{t+\Delta t/2}, \tilde{\boldsymbol{\rho}}_t\right)\end{aligned} , \qquad (41)$$

where

$$c_1(\mathbf{R},\mathbf{P},\tilde{\boldsymbol{\rho}},\Delta t) = \sqrt{2E_{\text{kin}}} \frac{\left(\alpha_\parallel - \sqrt{2E_{\text{kin}}}\right) + \left(\alpha_\parallel + \sqrt{2E_{\text{kin}}}\right)\exp\left[-\frac{2B\Delta t}{\sqrt{2E_{\text{kin}}}}\right]}{\left(\sqrt{2E_{\text{kin}}} - \alpha_\parallel\right) + \left(\alpha_\parallel + \sqrt{2E_{\text{kin}}}\right)\exp\left[-\frac{2B\Delta t}{\sqrt{2E_{\text{kin}}}}\right]} \quad . \tag{42}$$

$$c_2(\mathbf{R},\mathbf{P},\tilde{\boldsymbol{\rho}},\Delta t) = \frac{2\sqrt{2E_{\text{kin}}}\exp\left[-\frac{B\Delta t}{\sqrt{2E_{\text{kin}}}}\right]}{\left(\sqrt{2E_{\text{kin}}} - \alpha_\parallel\right) + \left(\alpha_\parallel + \sqrt{2E_{\text{kin}}}\right)\exp\left[-\frac{2B\Delta t}{\sqrt{2E_{\text{kin}}}}\right]} \quad . \tag{43}$$

Here, $E_{\text{kin}} = \mathbf{P}^T\mathbf{M}^{-1}\mathbf{P}/2$ is the total kinetic energy, $\mathbf{e}_\parallel(\mathbf{R},\tilde{\boldsymbol{\rho}})$ denotes the unit vector for the direction of vector $\mathbf{B} = \mathbf{M}^{-1/2}\sum_{n\neq m}\tilde{\rho}_{nm}\left(E_n(\mathbf{R}) - E_m(\mathbf{R})\right)\mathbf{d}_{mn}(\mathbf{R})$

$= \mathbf{M}^{-1/2}\sum_{n\neq m}\tilde{\rho}_{nm}\langle\phi_m(\mathbf{R})|\nabla_\mathbf{R}\hat{H}_{\text{el}}(\mathbf{R})|\phi_n(\mathbf{R})\rangle$, $B$ represents the scalar length of vector $\mathbf{B}$,

$\boldsymbol{\Pi}_\parallel = \alpha_\parallel \mathbf{e}_\parallel = \left(\mathbf{M}^{-1/2}\mathbf{P}\cdot\mathbf{e}_\parallel\right)\mathbf{e}_\parallel$ and $\boldsymbol{\Pi}_\perp(\mathbf{R},\mathbf{P},\tilde{\boldsymbol{\rho}}) = \mathbf{M}^{-1/2}\mathbf{P} - \alpha_\parallel\mathbf{e}_\parallel$ are the components of

$\boldsymbol{\Pi} \equiv \mathbf{M}^{-1/2}\mathbf{P}$ parallel and perpendicular to $\mathbf{B}$, respectively. The integrator eq (41) conserves the total kinetic energy $E_{\text{kin}} = \mathbf{P}^T\mathbf{M}^{-1}\mathbf{P}/2$. For the $N_{\text{nuc}} = 1$ case, this step is skipped. When $\frac{B_t\Delta t/2}{\sqrt{2E_{\text{kin}}(t)}}$ is very small or very large, please refer to arguments below and Section S1 of the Supporting Information of this paper for details of additional treatments.

3. Update phase space variables of electronic DOFs within a half time-step $\Delta t/2$ according to

$$\tilde{\mathbf{g}}_{t+\Delta t/2} \leftarrow \tilde{\mathbf{U}}(\mathbf{R}_t,\mathbf{P}_{t+\Delta t/2};\Delta t/2)\tilde{\mathbf{g}}_t \quad . \tag{44}$$

$$\tilde{\boldsymbol{\Gamma}}_{t+\Delta t/2} \leftarrow \tilde{\mathbf{U}}(\mathbf{R}_t,\mathbf{P}_{t+\Delta t/2};\Delta t/2)\tilde{\boldsymbol{\Gamma}}_t\tilde{\mathbf{U}}^\dagger(\mathbf{R}_t,\mathbf{P}_{t+\Delta t/2};\Delta t/2) \quad . \tag{45}$$

4. Update the nuclear coordinate within a full time-step $\Delta t$

$$\mathbf{R}_{t+\Delta t} \leftarrow \mathbf{R}_t + \mathbf{M}^{-1}\mathbf{P}_{t+\Delta t/2}\Delta t \ . \tag{46}$$

5. Update phase space variables of electronic DOFs within the other half time-step $\Delta t/2$ according to

$$\tilde{\mathbf{g}}_{t+\Delta t} \leftarrow \tilde{\mathbf{U}}(\mathbf{R}_{t+\Delta t},\mathbf{P}_{t+\Delta t/2};\Delta t/2)\tilde{\mathbf{g}}_{t+\Delta t/2} \ . \tag{47}$$

$$\tilde{\mathbf{\Gamma}}_{t+\Delta t} \leftarrow \tilde{\mathbf{U}}(\mathbf{R}_{t+\Delta t},\mathbf{P}_{t+\Delta t/2};\Delta t/2)\tilde{\mathbf{\Gamma}}_{t+\Delta t/2}\tilde{\mathbf{U}}^{\dagger}(\mathbf{R}_{t+\Delta t},\mathbf{P}_{t+\Delta t/2};\Delta t/2) \ . \tag{48}$$

Calculate the effective electronic density matrix $\tilde{\boldsymbol{\rho}}$ according to eq (34).

6. Determine a new occupied state $j_{\text{new}}$ based on $\tilde{\boldsymbol{\rho}}$ and rescale $\mathbf{P}$ if $j_{\text{new}} \neq j_{\text{old}}$,

$$\mathbf{P}_{t+\Delta t/2} \leftarrow \mathbf{P}_{t+\Delta t/2}\sqrt{\left(H_{\text{NaF}}\left(\mathbf{R}_{t+\Delta t},\mathbf{P}_{t+\Delta t/2},\tilde{\boldsymbol{\rho}}_{t+\Delta t}\right) - E_{j_{\text{new}}}\left(\mathbf{R}_{t+\Delta t}\right)\right)/\left(\mathbf{P}_{t+\Delta t/2}^{\text{T}}\mathbf{M}^{-1}\mathbf{P}_{t+\Delta t/2}/2\right)} \ . \tag{49}$$

If $H_{\text{NaF}}\left(\mathbf{R}_{t+\Delta t},\mathbf{P}_{t+\Delta t/2},\tilde{\boldsymbol{\rho}}_{t+\Delta t}\right) < E_{j_{\text{new}}}\left(\mathbf{R}_{t+\Delta t}\right)$, the switching of the adiabatic nuclear force component is frustrated. In such a case we keep $j_{\text{new}} = j_{\text{old}}$ and the rescaling step (for the nuclear kinematic momentum) eq (49) is skipped.

7. Similar to Step 2, update the nuclear kinematic momentum within the other half time-step $\Delta t/2$ using the numerical integrator for the effective nonadiabatic force for the $N_{\text{nuc}} \geq 2$ case

$$\begin{aligned}\mathbf{P}_{t+\Delta t} \leftarrow &\ c_1\left(\mathbf{R}_{t+\Delta t},\mathbf{P}_{t+\Delta t/2},\tilde{\boldsymbol{\rho}}_{t+\Delta t},\Delta t/2\right)\mathbf{M}^{1/2}\mathbf{e}_{\parallel}\left(\mathbf{R}_{t+\Delta t},\tilde{\boldsymbol{\rho}}_{t+\Delta t}\right) \\ &\ +c_2\left(\mathbf{R}_{t+\Delta t},\mathbf{P}_{t+\Delta t/2},\tilde{\boldsymbol{\rho}}_{t+\Delta t},\Delta t/2\right)\mathbf{M}^{1/2}\mathbf{\Pi}_{\perp}\left(\mathbf{R}_{t+\Delta t},\mathbf{P}_{t+\Delta t/2},\tilde{\boldsymbol{\rho}}_{t+\Delta t}\right)\end{aligned} \ . \tag{50}$$

When $\dfrac{B_{t+\Delta t}\Delta t/2}{\sqrt{2E_{\text{kin}}(t+\Delta t/2)}}$ is very small or very large, please refer to arguments below and Section S1 of the Supporting Information for additional details.

8. Update the nuclear kinematic momentum within the other half time-step $\Delta t/2$ using the adiabatic force

$$\mathbf{P}_{t+\Delta t} \leftarrow \mathbf{P}_{t+\Delta t} - \nabla_{\mathbf{R}} E_{j_{\text{new}}}(\mathbf{R}_{t+\Delta t}) \frac{\Delta t}{2} . \tag{51}$$

In Steps 2 and 7 of the integrator above for the $N_{\text{nuc}} \geq 2$ case, two additional cases should be taken care of for achieving numerical stability. When $B\Delta t/\left(2\sqrt{2E_{\text{kin}}}\right)$ in eqs (42)-(43) is very small (e.g. $B\Delta t/\left(2\sqrt{2E_{\text{kin}}}\right) \leq 10^{-20}$) in the region where the nonadiabatic coupling almost vanishes, we use the following integrator to replace the one in Steps 2 and 7 (for a half time-step, $\Delta \tau = \Delta t/2$):

$$\mathbf{P} \leftarrow \left(1 + \Delta\tau \frac{\mathbf{B} \cdot \mathbf{M}^{-1/2}\mathbf{P}}{2E_{\text{kin}}}\right)\mathbf{P} - \Delta\tau \mathbf{M}^{1/2}\mathbf{B} . \tag{52}$$

Conversely, $B\Delta t/\left(2\sqrt{2E_{\text{kin}}}\right)$ is very large in the region where the NaF trajectory approaches the "classical" forbidden region ($E_{\text{kin}} \to 0$). In this region, if $\mathbf{M}^{-1/2}\mathbf{P}$ and $\mathbf{B}$ are nearly parallel with each other in the same direction ($\alpha_{\parallel}/\sqrt{2E_{\text{kin}}} \to 1$), a self-adaptive time-step strategy should be employed to avoid numerical instability in the integrator eq (41). (Please see Section S1 of the Supporting Information for more details.) This complete integrator for a finite time-step $\Delta t$ for NaF is applicable to general systems, regardless of whether the diabatic representation is available or not.

As mentioned in refs [322,324], several models defined in the diabatic representation (such as the FMO model and the singlet-fission model tested in the two references) require a considerably shorter time-step for numerical convergence when employing the propagator in eq (33) to propagate the electronic phase space variables in the adiabatic representation. When the diabatic representation is rigorously defined, we present two approaches to enhance the efficiency of NaF by taking advantage of that the diabatic basis set is available. The first approach is to use the covariant electronic propagator in the adiabatic representation according to its counterpart in the diabatic representation (eqs (28)-(29)), rather than using eqs (31)-(32) for propagating electronic DOFs. This approach is equivalent to evolving electronic DOFs in the diabatic representation while nuclear DOFs are propagated in the adiabatic representation. (See more details in Section S2 of the Supporting Information.) The second approach is to express the nuclear force of the second equation of eq (35) of NaF (the sum of the adiabatic force and the nonadiabatic force) in the diabatic representation and to evolve both nuclear and electronic DOFs in the diabatic representation, as described in Section S3 of the Supporting Information.

**2.4 Time Correlation Functions**

Consider the TCF of electronic DOFs

$$D_{nm,kl}(t) = \text{Tr}_e\left[|n\rangle\langle m|e^{i\hat{H}t}|k\rangle\langle l|e^{-i\hat{H}t}\right] , \qquad (53)$$

where the $(n=m, k=l)$, $(n=m, k \neq l)$, $(n \neq m, k=l)$, and $(n \neq m, k \neq l)$ cases of eq (53) stand for the (electronic) population-population, population-coherence, coherence-population, and coherence-coherence TCFs, respectively. The unified framework of ref [324] maps $D_{nm,kl}(t)$ to the phase space counterpart[11, 12, 319, 322-324]

$$D_{nm,kl}(t) \mapsto \frac{1}{\overline{C}_{nm,kl}(t)}\int d\gamma w(\gamma)\int_{\mathcal{S}(\mathbf{x},\mathbf{p},\Gamma;\gamma)} d\mu(\mathbf{x},\mathbf{p},\Gamma)\overline{\mathcal{Q}}_{nm,kl}(\mathbf{x},\mathbf{p},\Gamma;\gamma;t) , \qquad (54)$$

where $d\mu(\mathbf{x},\mathbf{p},\Gamma)$ denotes the integral measure over the mapping CPS, $w(\gamma)$ represents the normalized weight (i.e., quasi-probability distribution) function (of $\gamma$) as first proposed in ref [12], and $\bar{C}_{nm,kl}(t)$ denotes an element of the time-dependent normalization factor tensor[323, 324]. The integrand function of electronic phase space variables in eq (54) can be determined in various ways. One typical choice for developing trajectory-based dynamics methods is[12, 319, 322, 324]

$$\bar{Q}_{nm,kl}(\mathbf{x}_0,\mathbf{p}_0,\mathbf{\Gamma}_0;\mathbf{x}_t,\mathbf{p}_t,\mathbf{\Gamma}_t) = \text{Tr}_e\left[|n\rangle\langle m|\hat{K}_{\text{ele}}(\mathbf{x}_0,\mathbf{p}_0,\mathbf{\Gamma}_0)\right]\text{Tr}_e\left[|k\rangle\langle l|\hat{K}_{\text{ele}}^{-1}(\mathbf{x}_t,\mathbf{p}_t,\mathbf{\Gamma}_t)\right] , \quad (55)$$

where $\hat{K}_{\text{ele}}^{-1}(\mathbf{x},\mathbf{p},\mathbf{\Gamma})$ denotes the inverse mapping kernel of CPS, and $\{\mathbf{x}_t,\mathbf{p}_t,\mathbf{\Gamma}_t\}$ represents the phase variables of the trajectory at time $t$ with the initial condition $\{\mathbf{x}_0,\mathbf{p}_0,\mathbf{\Gamma}_0\}$. The simplest yet useful case where $\mathbf{\Gamma} = \gamma\mathbf{1}$ in the electronic mapping kernel and its inverse is often utilized[11, 12, 318-320, 322]. The constraint of the corresponding CPS reads

$$\mathcal{S}(\mathbf{x},\mathbf{p};\gamma) = \delta\left(\sum_{n=1}^{F}\frac{(x^{(n)})^2 + (p^{(n)})^2}{2} - (1+F\gamma)\right) \quad \text{with} \quad \gamma \in (-1/F,+\infty) \quad \text{[318, 319]}, \text{ yielding the}$$

U($F$)/U($F$-1) complex Stiefel manifold[12]. Our previous works have presented several different forms of the electronic mapping kernel and pointed out that even for the same electronic mapping kernel, inverse mapping kernels, are not unique[12, 324-326].

Below we list several types of TCFs in the CPS representation, and construct the corresponding NaF methods.

**Covariant-Covariant:** The mapping kernel and the inverse mapping kernel are given by[319]:

$$\hat{K}_{\text{ele}}(\mathbf{x},\mathbf{p}) = \sum_{n,m=1}^{F}\left[\frac{1}{2}\left(x^{(n)}+ip^{(n)}\right)\left(x^{(m)}-ip^{(m)}\right) - \gamma\delta_{nm}\right]|n\rangle\langle m| , \quad (56)$$

$$\hat{K}_{\text{ele}}^{-1}(\mathbf{x},\mathbf{p}) = \sum_{k,l=1}^{F}\left[\frac{1+F}{2(1+F\gamma)^2}\left(x^{(k)}+ip^{(k)}\right)\left(x^{(l)}-ip^{(l)}\right) - \frac{1-\gamma}{1+F\gamma}\delta_{kl}\right]|k\rangle\langle l| . \quad (57)$$

In this case, the normalization factor is $\bar{C}_{nm,kl}(t) \equiv 1$, with the invariant integral measure $d\mu(\mathbf{x},\mathbf{p}) = F d\mathbf{x} d\mathbf{p}$. The weight function is often set to a Dirac delta function $w(\gamma) = \delta(\gamma - \gamma_0)$ with an adjustable parameter $\gamma_0 \in (-1/F, +\infty)$ [319], although other reasonable choices for $w(\gamma)$ are possible[12]. We denote eqs (54)-(55) combined with eqs (56)-(57) as the covariant-covariant (cc) TCF. The term "covariant" implies that the mapping kernel and the inverse mapping kernel satisfy

$$\begin{aligned}\mathbf{U}\mathbf{K}_{\mathrm{ele}}(\mathbf{x},\mathbf{p})\mathbf{U}^\dagger &= \mathbf{K}_{\mathrm{ele}}\left(\mathrm{Re}\,\mathbf{U}(\mathbf{x}+i\mathbf{p}), \mathrm{Im}\,\mathbf{U}(\mathbf{x}+i\mathbf{p})\right), \\ \mathbf{U}\mathbf{K}_{\mathrm{ele}}^{-1}(\mathbf{x},\mathbf{p})\mathbf{U}^\dagger &= \mathbf{K}_{\mathrm{ele}}^{-1}\left(\mathrm{Re}\,\mathbf{U}(\mathbf{x}+i\mathbf{p}), \mathrm{Im}\,\mathbf{U}(\mathbf{x}+i\mathbf{p})\right)\end{aligned} \quad (58)$$

for arbitrary $F \times F$ unitary matrix $\mathbf{U}$. The cc TCF is always able to produce the exact population and coherence dynamics in the frozen nuclei limit for $F \geq 2$. The initial sampling of the variables $\{\mathbf{x},\mathbf{p}\}$ for electronic DOFs is according to the uniform distribution on one $(2F-1)$-dimensional sphere $\mathcal{S}(\mathbf{x},\mathbf{p};\gamma)$.

The cc TCF was employed by CMM in 2019 and by CMMcv in 2021 in refs [11, 318-320]. In spirit of the unified framework of mapping models with coordinate-momentum variables of ref [235] reported in 2016, the CMM methods on CPS were first proposed and derived in ref [318] in 2019 for nonadiabatic systems with general $F$ states ($F \geq 2$). The CPS with coordinate-momentum variables with the general value of $\gamma$ was first indicated by eqs (1), (5), (7), (19) and (28) of ref [318], and the CPS with action-angle variables for the $\gamma = 0$ case was first presented by eqs (A4)-(A5) of Appendix A of ref [318]. The strategy of using action-angle variables to establish or prove the CPS representations was first employed in Appendix A of ref [318], then in refs [323, 324], while the same strategy of using coordinate-momentum variables instead was presented in refs [12, 319]. Section S4 of the Supporting Information of this paper revisits the strategy of using action-angle variables of CPS of Appendix A of ref [318] only for *education purposes*. In ref [318], the constraint of eq (28) and the Hamiltonian model of either eq (7) or eq (19) inherently indicate the complex Stiefel manifolds U(*F*)/U(*F*-*r*) (with $1 \leq r < F$ )[325-328]. More specifically, in ref [318], the constraint of eq (28) and the Hamiltonian model of eq (7) with $\{\gamma_{nm} = \gamma \delta_{nm}; \forall n,m\}$ intrinsically imply the U(*F*)/U(*F*-1) complex Stiefel manifold, and the constraint of eq (28) and the Hamiltonian model of eq (19) with $\{\tilde{\gamma}_{nm} = \gamma \delta_{nm}; \forall n,m\}$ in principle leads to the U(*F*)/U(*F*-2) complex Stiefel manifold.

Following refs [235, 318, 352], ref [319] explicitly showed that the phase space parameter, $\gamma$, should lie in a *continuous* range $(-1/F, +\infty)$, and refs [11, 12, 320] first pointed out that the P, W and Q versions of Stratonovich phase space correspond to only three special cases of the $\mathrm{U}(F)/\mathrm{U}(F-1)$ constraint phase space used in CMM with $\gamma = 1$, $(\sqrt{F+1}-1)/F$ and 0. These three conventional versions of Stratonovich phase space[330, 331, 333, 338, 353] were used for studying nonadiabatic dynamics of composite systems for the $F = 2$ case in refs [337, 338], and later for the $F = 2$ case in ref [354], then for the multistate case ($F \geq 3$) in ref [204] in 2020. Their relations to the CMM/CPS approach first originally developed in refs [235, 318] for general *F*-state systems, were presented in ref [320] in 2021, and in ref [11], Appendix 3 of ref [12], ref [355], and ref [356]. For example, a special case for phase space parameter $\gamma$ in CMM[235, 318-320] where $\gamma = (\sqrt{F+1}-1)/F$ of the $\mathrm{U}(F)/\mathrm{U}(F-1)$ constraint phase space is used, is related to the W version of the spin mapping method in ref [204] for general $F \geq 3$ cases, where the conventional W version of the $\mathrm{SU}(F)/\mathrm{U}(F-1)$ Stratonovich phase space[331, 333, 353] is implemented. More details are presented in Appendix 3 of ref [12] on the relation as well as subtle difference between the $\mathrm{U}(F)/\mathrm{U}(F-1)$ constraint coordinate-momentum phase space with $2F$ variables and the $\mathrm{SU}(F)/\mathrm{U}(F-1)$

Stratonovich phase space[331, 333, 353] with $2F-2$ variables. The $\mathrm{U}(F)/\mathrm{U}(F-1)$ constraint coordinate-momentum phase space inherently includes a time-dependent global phase variable and naturally leads to the linear EOMs of quantum mechanics in the frozen nuclei limit. More recently, the twin-space representation with CMM/CPS has been proposed in ref [357].

Similar to CMMcv[320], when NaF is combined with the "cc" TCF (leading to the NaF-cc method[322]), it is essential to involve the commutator matrix in calculating nuclear force for satisfying the Born-Oppenheimer limit, where the initial condition reads[320]

$$\Gamma_{nm}(0) = \begin{cases} \dfrac{\left(x^{(n)}\right)^2 + \left(p^{(n)}\right)^2}{2} - \delta_{nj_0}, & n=m \\ 0, & n \neq m \end{cases}. \tag{59}$$

Here $j_0$ denotes the label of the initially occupied state. We have shown that $\gamma_0 \in \left[\left(\sqrt{F+1}-1\right)/F, 1/2\right]$ is recommended for NaF-cc[322], and we employ $\gamma_0 = 1/2$ for NaF-cc throughout this paper. (That is, the weight function is $w(\gamma) = \delta(\gamma - 1/2)$.)

As demonstrated in refs [325, 326], the mapping kernel as well as its inverse are not unique, while eqs (56)-(57) denote only a special choice (the covariant form). Several other kinds of mapping formalisms are discussed below as well as listed in ref [326].

**Triangle Window Functions:** We follow the formalism in refs [323, 324] for describing the TCF with TWFs. The mapping kernel is given by

$$\hat{K}_{\mathrm{ele}}(\mathbf{x},\mathbf{p}) = \sum_{n=1}^{F} W_n(\mathbf{x},\mathbf{p}) |n\rangle\langle n| \\ + \frac{3}{5} \sum_{n \neq m}^{F} \sum_{j=n,m} W_j(\mathbf{x},\mathbf{p}) \left(x^{(n)} + \mathrm{i}p^{(n)}\right)\left(x^{(m)} - \mathrm{i}p^{(m)}\right) |n\rangle\langle m| \tag{60}$$

with the TWF for the *n*-th state[195, 200, 324]

$$W_n(\mathbf{x},\mathbf{p}) = \frac{2(F^F-1)}{F \cdot F!}(2-e^{(n)})^{2-F} h(e^{(n)}-1)h(2-e^{(n)})\prod_{k \neq n}^{F} h(2-e^{(k)}-e^{(n)}) \; , \tag{61}$$

where $e^{(n)} = \left[(x^{(n)})^2 + (p^{(n)})^2\right]/2$. For the population-population and population-coherence TCFs (i.e., eq (55) with $n = m$), the corresponding inverse mapping kernel reads

$$\hat{K}_{\text{ele}}^{-1}(\mathbf{x},\mathbf{p}) = \sum_{k=1}^{F} h(e^{(k)}-1)\prod_{j \neq k}^{F} h(1-e^{(j)})|k\rangle\langle k| \\ + \sum_{k \neq l}^{F} \frac{1}{2}(x^{(k)}+\mathrm{i}p^{(k)})(x^{(l)}-\mathrm{i}p^{(l)})|k\rangle\langle l| \tag{62}$$

For the coherence-population and coherence-coherence TCFs (i.e., eq (55) with $n \neq m$), the inverse mapping kernel is set to be

$$\hat{K}_{\text{ele}}^{-1}(\mathbf{x},\mathbf{p}) = \sum_{k,l=1}^{F} \left[\frac{1}{2}(x^{(k)}+\mathrm{i}p^{(k)})(x^{(l)}-\mathrm{i}p^{(l)}) - \delta_{kl}/3\right]|k\rangle\langle l| \; . \tag{63}$$

(Note that in principle eq (63) can also be used for the population-population and population-coherence TCFs[326].) The weight function reads

$$w(\gamma) = \begin{cases} \dfrac{F^2}{F^F-1}(1+F\gamma)^{F-1}, & 0 \leq \gamma \leq \dfrac{F-1}{F} \\ 0, & \text{otherwise} \end{cases} \tag{64}$$

and $\mathrm{d}\mu(\mathbf{x},\mathbf{p}) = F\mathrm{d}\mathbf{x}\mathrm{d}\mathbf{p}$. Since $\gamma$ does not appear explicitly in the integrand, $\overline{Q}_{nm,kl}(\mathbf{x}_0,\mathbf{p}_0,\mathbf{x}_t,\mathbf{p}_t)$, for CPS with TWFs, it can be integrated over exactly, leading to[324]

$$\int_0^{(F-1)/F} \mathrm{d}\gamma w(\gamma) \int_{\mathcal{S}(\mathbf{x},\mathbf{p};\gamma)} \mathrm{d}\mu(\mathbf{x}_0,\mathbf{p}_0) \overline{Q}_{nm,kl}(\mathbf{x}_0,\mathbf{p}_0,\mathbf{x}_t,\mathbf{p}_t) \\ = \frac{F \cdot F!}{(2\pi)^F (F^F-1)} \int \mathrm{d}\mathbf{x}_0 \mathrm{d}\mathbf{p}_0 \overline{Q}_{nm,kl}(\mathbf{x}_0,\mathbf{p}_0,\mathbf{x}_t,\mathbf{p}_t) \tag{65}$$

Finally, the time-dependent normalization factor reads

$$\overline{C}_{nm,kl}(t) = \begin{cases} \dfrac{F \cdot F!}{(2\pi)^F (F^F - 1)} \sum_{j=1}^{F} \int d\mathbf{x}_0 d\mathbf{p}_0 \overline{Q}_{nn,jj}(\mathbf{x}_0, \mathbf{p}_0, \mathbf{x}_t, \mathbf{p}_t), & n = m \text{ and } k = l \\ 1, & \text{otherwise} \end{cases} \quad (66)$$

The TCF with TWFs for population dynamics (with eqs (60) and (62) for the $n = m$ and $k = l$ case) was first introduced by Cotton and Miller[195, 200], of which the $F = 2$ case belongs to the novel class of exact phase space representations of the pure two-state system in ref [323] and yields exact population dynamics in the frozen nuclei limit. Although it fails to produce exact population dynamics for $F \geq 3$ in the frozen nuclei limit, the TWFs ensure a positive semidefinite population and demonstrate reasonably good results for typical multistate systems[200, 324]. The above formalism for TCFs involving coherence terms (i.e., $n \neq m$ or $k \neq l$), which was first proposed in ref [324] by us, satisfies the frozen nuclei limit for arbitrary number of states. Reference [195] of Cotton and Miller provides a simple algorithm for efficient sampling or evaluation with TWFs, which is also described in details in our previous work[324].

Two possible approaches for combining NaF with TWFs have been proposed in ref [324]. The first approach employs $\mathbf{\Gamma} = 1/3$, following the recommended zero-point energy parameter 1/3 of Cotton and Miller[195, 200], and is referred to as NaF with TWFs (NaF-TW). The second method employs eq (59) as the initial condition of the commutator matrix, which is denoted as NaF with TWFs-2 (NaF-TW2). We have demonstrated that both NaF-TW and NaF-TW2 are comparable with NaF-cc and perform reasonably well for various benchmark gas-phase and condensed-phase systems[324].

**Hill Window Functions:** A novel positive semidefinite hill window functions (HWF) is proposed in ref [326] for general $F \geq 3$ cases, which can be utilized for constructing TCF formalisms. As an example, when choosing the mapping kernel as

$$\hat{K}_{\text{ele}}(\mathbf{x},\mathbf{p}) = \sum_{n=1}^{F} \prod_{k \neq n}^{F} h\left(\frac{(x^{(n)})^2 + (p^{(n)})^2}{2} - \frac{(x^{(k)})^2 + (p^{(k)})^2}{2}\right)|n\rangle\langle n| \\ + \sum_{n \neq m}^{F} \frac{(x^{(n)} + ip^{(n)})(x^{(m)} - ip^{(m)})}{2}|n\rangle\langle m|$$ (67)

the corresponding inverse mapping kernel for population-population and population-coherence TCFs (i.e., eq (55) with $n = m$) reads

$$\hat{K}_{\text{ele}}^{-1}(\mathbf{x},\mathbf{p}) = \sum_{k=1}^{F} \prod_{j \neq k}^{F} h\left(\frac{(x^{(k)})^2 + (p^{(k)})^2}{2} - \frac{(x^{(j)})^2 + (p^{(j)})^2}{2}\right) \\ \times \left(\frac{(x^{(k)})^2 + (p^{(k)})^2}{2} - \frac{(x^{(j)})^2 + (p^{(j)})^2}{2}\right)^{B(F)} |k\rangle\langle k| \\ + \sum_{k \neq l}^{F} \tilde{\gamma}_{\text{off}}(x^{(k)} + ip^{(k)})(x^{(l)} - ip^{(l)})|k\rangle\langle l|$$ (68)

with

$$B(F) = \frac{3}{7(F-1)} + \frac{60}{7(F+13)},$$ (69)

where the diagonal terms of the inverse mapping kernel eq (68) are HWFs. The weight function is $w(\gamma) = \delta\left(\gamma - (\sqrt{F+1} - 1)/F\right)$ and the invariant integral measure is $d\mu(\mathbf{x},\mathbf{p}) = F d\mathbf{x} d\mathbf{p}$. The (time-dependent) normalization factor reads

$$\overline{C}_{nm,kl}(t) = \begin{cases} \sum_{j=1}^{F} \int_{-1/F}^{+\infty} d\gamma\, w(\gamma) \int_{S(\mathbf{x},\mathbf{p};\gamma)} d\mu(\mathbf{x}_0,\mathbf{p}_0) \overline{Q}_{nn,jj}(\mathbf{x}_0,\mathbf{p}_0;\mathbf{x}_t,\mathbf{p}_t), & n=m, k=l \\ 1, & \text{otherwise} \end{cases}.$$ (70)

Similar to the TCF with TWFs, the TCF with HWFs is positive semidefinite for estimating electronic population. For the $F = 2$ case, the population-population TCF can be related to Appendix B of ref [294] in the frozen nuclei limit. In eq (68), parameter $\tilde{\gamma}_{\text{off}} = \dfrac{F-1}{2\sqrt{F+1}\sum_{k=2}^{F}k^{-1}}$ that appears in the population-coherence TCF can be related to ref [295]. The inverse mapping kernel for coherence-population and coherence-coherence TCFs (i.e., eq (55) with $n \neq m$) is identical to eq (57). Similar to NaF-TW and NaF-TW2 of ref [324], we also consider two approaches to use HWFs with NaF, namely NaF with HWFs (NaF-HW) and NaF with HWFs-2 (NaF-HW2). The initial condition of $\boldsymbol{\Gamma}$ for NaF-HW is defined as $\boldsymbol{\Gamma} = \left(\sqrt{F+1}-1\right)\mathbf{1}/F$, i.e., an invariant CPS parameter, and that for NaF-HW2 employs eq (59).

The initial sampling procedure for variables $\{\mathbf{x},\mathbf{p}\}$ for electronic DOFs with HWFs is listed below. For coherence-population and coherence-coherence TCFs, the initial condition of $\{\mathbf{x},\mathbf{p}\}$ is uniformly sampled on $\mathcal{S}(\mathbf{x},\mathbf{p};\gamma)$ where $\gamma = \left(\sqrt{F+1}-1\right)/F$. For population-population and population-coherence TCFs $D_{nn,kl}(t)$, the initial condition of $\{\mathbf{x},\mathbf{p}\}$ is uniformly sampled on

$$\mathcal{S}(\mathbf{x},\mathbf{p};\gamma)\prod_{k \neq n}^{F} h\left(\frac{\left(x^{(n)}\right)^2+\left(p^{(n)}\right)^2}{2} - \frac{\left(x^{(k)}\right)^2+\left(p^{(k)}\right)^2}{2}\right),$$

with the following sampling procedure for one phase point:

1. Uniformly sample a point on the $(2F-1)$-dimensional sphere $\mathcal{S}(\mathbf{x},\mathbf{p};\gamma)$.

2. If the phase variables of the point satisfy $\dfrac{\left(x^{(n)}\right)^2+\left(p^{(n)}\right)^2}{2} > \dfrac{\left(x^{(k)}\right)^2+\left(p^{(k)}\right)^2}{2}$ for any $k \neq n$, then accept the point as the initial condition. Otherwise repeat the procedure until one point is accepted.

**Covariant-Noncovariant:** When the covariant mapping kernel eq (56) is used as the mapping kernel $\hat{K}_{\text{ele}}(\mathbf{x},\mathbf{p})$ in eq (55), in principle any function of phase space variables (regardless of whether it is covariant or noncovariant) can be used as a matrix element of the inverse mapping kernel $\hat{K}_{\text{ele}}^{-1}(\mathbf{x},\mathbf{p})$, and the frozen nuclei limit is naturally satisfied as long as the mapping formalism satisfies the exact mapping condition[325, 326]. This statement was first proved in ref [325] by us in 2022 and then in ref [326]. When noncovariant phase space functions are used in the inverse mapping kernel, the TCFs are referred as the covariant-noncovariant (cx) TCFs in ref [326]. As an example, the inverse mapping kernel can be defined as

$$\hat{K}_{\text{ele}}^{-1}(\mathbf{x},\mathbf{p}) = \frac{1}{F}\left(\frac{1+F\gamma}{F\gamma}\right)^{F-1}\sum_{k=1}^{F} h\left(\frac{\left(x^{(k)}\right)^2 + \left(p^{(k)}\right)^2}{2} - 1\right)|k\rangle\langle k| \\ + \sum_{k\neq l}^{F} \frac{1+F}{2(1+F\gamma)^2}\left(x^{(k)} + ip^{(k)}\right)\left(x^{(l)} - ip^{(l)}\right)|k\rangle\langle l| \qquad (71)$$

with the weight function $w(\gamma) = \delta\left(\gamma - \left(\sqrt{F+1}-1\right)/F\right)$, and the elements of the corresponding normalization factor tensor are

$$\overline{C}_{nm,kl}(t) = \begin{cases} \sum_{j=1}^{F}\int_{0}^{+\infty} d\gamma\, w(\gamma)\int_{\mathcal{S}(\mathbf{x},\mathbf{p};\gamma)} d\mu(\mathbf{x}_0,\mathbf{p}_0)\overline{Q}_{nn,jj}(\mathbf{x}_0,\mathbf{p}_0;\mathbf{x}_t,\mathbf{p}_t), & n=m, k=l \\ 1, & \text{otherwise} \end{cases} \qquad (72)$$

The invariant integral measure for cx is $d\mu(\mathbf{x},\mathbf{p}) = F d\mathbf{x} d\mathbf{p}$. The cx TCFs satisfy the frozen nuclei limit. It is straightforward to implement NaF with the cx TCFs, which is denoted as NaF-cx. The initial condition of $\{\mathbf{x},\mathbf{p}\}$ is uniformly sampled on $\mathcal{S}(\mathbf{x},\mathbf{p};\gamma)$ with $\gamma = \left(\sqrt{F+1}-1\right)/F$. Similar to NaF-cc, the commutator matrix with the initial condition in eq (59) is also *critical* for NaF-cx to approach the correct Born-Oppenheimer limit.

Finally, when both nuclear and electronic DOFs are considered, the phase space counterpart of the quantum TCF

$$D(t) = \text{Tr}_{n,e}\left[\hat{\rho}_{\text{nuc}} \otimes |n\rangle\langle m| e^{i\hat{H}t}\left(\hat{B}_{\text{nuc}} \otimes |k\rangle\langle l|\right)e^{-i\hat{H}t}\right] \quad (73)$$

can be written as

$$D(t) \mapsto \frac{1}{\overline{C}_{nm,kl}(t)} \int d\mu(\mathbf{R},\mathbf{P}) \int d\gamma w(\gamma) \int_{\mathcal{S}(\mathbf{x},\mathbf{p},\Gamma;\gamma)} d\mu(\mathbf{x},\mathbf{p},\Gamma) \\ \times \rho_{\text{nuc}}(\mathbf{R},\mathbf{P}) \tilde{B}_{\text{nuc}}(\mathbf{R}_t,\mathbf{P}_t) \overline{Q}_{nm,kl}(\mathbf{x},\mathbf{p},\Gamma;\gamma;t) \quad (74)$$

where the phase space functions of nuclear operators are

$$\rho_{\text{nuc}}(\mathbf{R},\mathbf{P}) = \text{Tr}_n\left[\hat{\rho}_{\text{nuc}} \hat{K}_{\text{nuc}}(\mathbf{R},\mathbf{P})\right] \quad , \quad (75)$$

$$\tilde{B}_{\text{nuc}}(\mathbf{R},\mathbf{P}) = \text{Tr}_n\left[\hat{B}_{\text{nuc}} \hat{K}_{\text{nuc}}^{-1}(\mathbf{R},\mathbf{P})\right] \quad . \quad (76)$$

Here, when not otherwise specified, an operator of pure nuclear DOFs $\hat{B}_{\text{nuc}}$ is treated as $\hat{B}_{\text{nuc}} \otimes \sum_{k=1}^{F} |k\rangle\langle k|$. When Wigner phase space is employed for nuclear DOFs, the inverse mapping kernel $\hat{K}_{\text{nuc}}^{-1}(\mathbf{R},\mathbf{P})$ is identical to the mapping kernel defined in eq (18). In this paper, we utilize only Wigner phase space[6] for nuclear DOFs. We note that other quantum phase space representations for nuclear DOFs[10, 12, 15] may also be feasible for the NaF implementation.

**Table 1.** CPS Representations of the Electronic Time Correlation Function (by Using Equations (53)-(55)) in Sub-Section 2.4.

| Method | Elements of mapping kernel $K_{nm}$ | Elements of inverse mapping kernel $K_{kl}^{-1}$ | Weight function $w(\gamma)$ | Normalization factor $\bar{C}_{nm,kl}(t)$ | Initial Condition of $\Gamma$ |
|---|---|---|---|---|---|
| NaF-cc | eq (56) | eq (57) | $\delta(\gamma - \gamma_0)$; $\gamma_0 = 1/2$ | 1 | eq (59) |
| NaF-TW | eq (60) | eq (62) ($n=m$) <br> eq (63) ($n \neq m$) | eq (64) | eq (66) | $\frac{1}{3}\mathbf{1}$ |
| NaF-TW2 | eq (60) | eq (62) ($n=m$) <br> eq (63) ($n \neq m$) | eq (64) | eq (66) | eq (59) |
| NaF-HW | eq (67) | eq (68) ($n=m$) <br> eq (57) ($n \neq m$) | $\delta\left(\gamma - \frac{\sqrt{F+1}-1}{F}\right)$ | eq (70) | $\frac{\sqrt{F+1}-1}{F}\mathbf{1}$ |
| NaF-HW2 | eq (67) | eq (68) ($n=m$) <br> eq (57) ($n \neq m$) | $\delta\left(\gamma - \frac{\sqrt{F+1}-1}{F}\right)$ | eq (70) | eq (59) |
| NaF-cx | eq (56) | eq (71) | $\delta\left(\gamma - \frac{\sqrt{F+1}-1}{F}\right)$ | eq (72) | eq (59) |

## 3 Numerical Results of Model Systems

In this section, we apply the NaF methods outlined in Sub-Section 2.4 and listed in Table 1 to various typical model systems, including linear vibronic coupling models, one-dimensional nonadiabatic scattering models, system-bath systems with two or more states, and atom-in-cavity models. The results produced by NaF methods will be compared with the numerically exact results. Additionally, we employ three SH methods for comparison with NaF methods, namely, SH-1[253], SH-2[295] and SH-3[294]. Because SH-3 is applicable to only two-state systems, its results are but available for the two-state cases in the following benchmark tests. Unless otherwise specified, the TCFs are calculated in the diabatic representation in all cases where exact results are available only in the diabatic representation. While simulations using SH methods can only be performed in the adiabatic representation, NaF simulations can be performed in either the adiabatic or diabatic representation, leading to same numerically converged results (see Section S3 of the Supporting Information).

Our simulations are conducted on the new Sunway supercomputer. The heterogeneous platform is composed of millions of SW26010-Pro CPUs, each with 6 core groups (CGs) of a management processing element (MPE), 64 computing processing elements (CPEs) and 16 GB DDR4 memory. By implementing our nonadiabatic dynamics program with MPI and Athread, it can take full advantage of the super large-scale. The program package greatly enhances the efficiency of simulating long-time dynamics of nonadiabatic systems with numerous DOFs. Table 2 provides the parameters (the time-step size, the number of trajectories, etc.) for obtaining fully converged results of NaF methods for each benchmark model. We divide the trajectories into 20 groups to calculate the standard error of each physical property. The error bar for each result is plotted in the figure, although it is often smaller than the width of the line or the size of the marker.

Similar numbers of trajectories are used in SH methods for obtaining fully converged results. We note that the number of NaF trajectories or SH trajectories can be considerably decreased when a larger error bar is tolerated.

**Table 2**. The Number of Trajectories and the Time-Step Size for NaF Methods for Benchmark Models.

| Model | Time-step Size | Number of trajectories |
|---|---|---|
| Pyrazine LVCM | 0.01 fs | $\sim 10^5$ |
| Cr(CO)$_5$ LVCM | 0.01 fs | $\sim 10^5$ |
| Thymine LVCM | 0.01 fs | $\sim 10^5$ |
| Tully models | 0.01 fs | $\sim 10^5$ |
| 3-state photodissociation models | 0.01 fs | $\sim 10^6$ |
| 2-state photodissociation model | 0.01 fs | $6 \times 10^5$ |
| Spin-boson models | 0.01 au | $\sim 5 \times 10^5$ |
| FMO model | 0.5 fs | $\sim 10^6$ |
| Singlet fission model | 0.005 fs | $\sim 10^5$ |
| Atom-in-cavity models | 0.1 au | $\sim 10^5$ |
| One-dimensional Holstein model | 0.5 au | $\sim 10^6$ |

## 3.1 Linear Vibronic Coupling Models

The linear vibronic coupling model (LVCM) is a typical model for studying dynamics around molecular conical intersections. The Hamiltonian (in the diabatic representation) is $\hat{H} = \hat{H}_0 + \hat{H}_C$, with

$$\hat{H}_0 = \sum_{k=1}^{N_{\text{nuc}}} \frac{\omega_k}{2}\left(\hat{\bar{P}}_k^2 + \hat{\bar{R}}_k^2\right), \tag{77}$$

where $\hat{\bar{P}}_k$ and $\hat{\bar{R}}_k$ $(k=1,...,N)$ represent the dimensionless weighted normal-mode momentum and coordinate of the $k$-th nuclear DOF, respectively, and $\omega_k$ represents the corresponding vibrational frequency. The relation between $\{\bar{R}_k, \bar{P}_k\}$ and the corresponding canonical (mass-weighted) coordinate/momentum, $\{R_k, P_k\}$, in the diabatic representation reads $\bar{R}_k = \sqrt{\omega_k}R_k$ and $\bar{P}_k = P_k/\sqrt{\omega_k}$. The electron-nuclear coupling term is

$$\hat{H}_C = \sum_{n=1}^{F}\left(E_n + \sum_{k=1}^{N_{\text{nuc}}}\kappa_k^{(n)}\hat{\bar{R}}_k\right)|n\rangle\langle n| + \sum_{n\neq m}^{F}\left(\sum_{k=1}^{N_{\text{nuc}}}\lambda_k^{(nm)}\hat{\bar{R}}_k\right)|n\rangle\langle m|, \tag{78}$$

where $E_n$ $(n=1,...,F)$ represents the vertical excitation energy of the $n$-th state, while $\kappa_k^{(n)}$ and $\lambda_k^{(nm)}$ denote the linear coupling terms of the $k$-th nuclear DOF for the corresponding diagonal and off-diagonal elements, respectively.

Two typical 2-state LVCMs for pyrazine—the 3-mode model of ref [358] and the 24-mode model of ref [359] are considered. The 3-state 2-mode LVCM of ref [360], which describes the dynamics of Cr(CO)$_5$ through a Jahn–Teller conical intersection after photodissociation, is also tested. The parameters and more details of these three models are described in refs [358-360] and also summarized in our previous work[322]. Both electronic population and nuclear properties are investigated and compared with the numerically exact results produced by MCTDH[134]. The Heidelberg MCTDH package (V8.5)[361] is employed for producing the exact data of the LVCMs for pyrazine, while the benchmark data of the Cr(CO)$_5$ model are directly obtained from ref [360].

In addition, a 7-state 39-mode LVCM of Thymine[362] is investigated in this paper. This model, proposed by Improta, Santoro and their co-workers, is established from CAM-B3LYP calculations[362]. We only consider the case where the $\pi\pi^*2$ state is initially occupied, with each nuclear mode in its corresponding vibrational ground state. The exact results provided by ML-MCTDH[136, 137] are available in ref [362]. Although this LVCM includes seven states, we focus on three of them: the n$o\pi^*1$ state, the $\pi\pi^*1$ state and the $\pi\pi^*2$ state. These three states exhibit the most significant population transfer among all the states.

Figures 1-3 present the results of the LVCMs for pyrazine and Cr(CO)$_5$, demonstrating both the electronic population and nuclear properties. All SH methods and NaF methods produce similar reasonable descriptions in both electronic and nuclear dynamics of the LVCMs for these two molecules. We also show the results of CMM[235, 318, 319] using the smallest and largest values of $\gamma$ in the region $\left[(\sqrt{F+1}-1)/F, 1/2\right]$ recommended in refs [11, 320]. Although CMM often demonstrates superior performance in various benchmark condensed-phase models, it may produce unphysical negative populations and performs poorly for nuclear dynamics of LVCMs, of which such drawbacks have already been demonstrated in several (gas-phase) models in our previous works, e.g., in Figures 1 and 3 of ref [320] and Figure 7 of ref [12]. It is evident from Panel (a4) of Figure 1 for pyrazine and from Panel (i) of Figure 4 for Thymine that $\gamma = (\sqrt{F+1}-1)/F$ should *not* be the best choice for phase space parameter $\gamma$ for CMM for general real molecular systems. Figure 4 illustrates population dynamics of the LVCM for Thymine where 39 modes are included. Neither SH nor CMM methods offer reasonable predictions for this model. In contrast, NaF methods consistently demonstrate better performance in reasonably describing the population transfer of this LVCM model. Figure 4 for Thymine implies that the nonadiabatic

(nuclear) force can play an important role around CI regions (for LVCMs) for real large molecular systems.

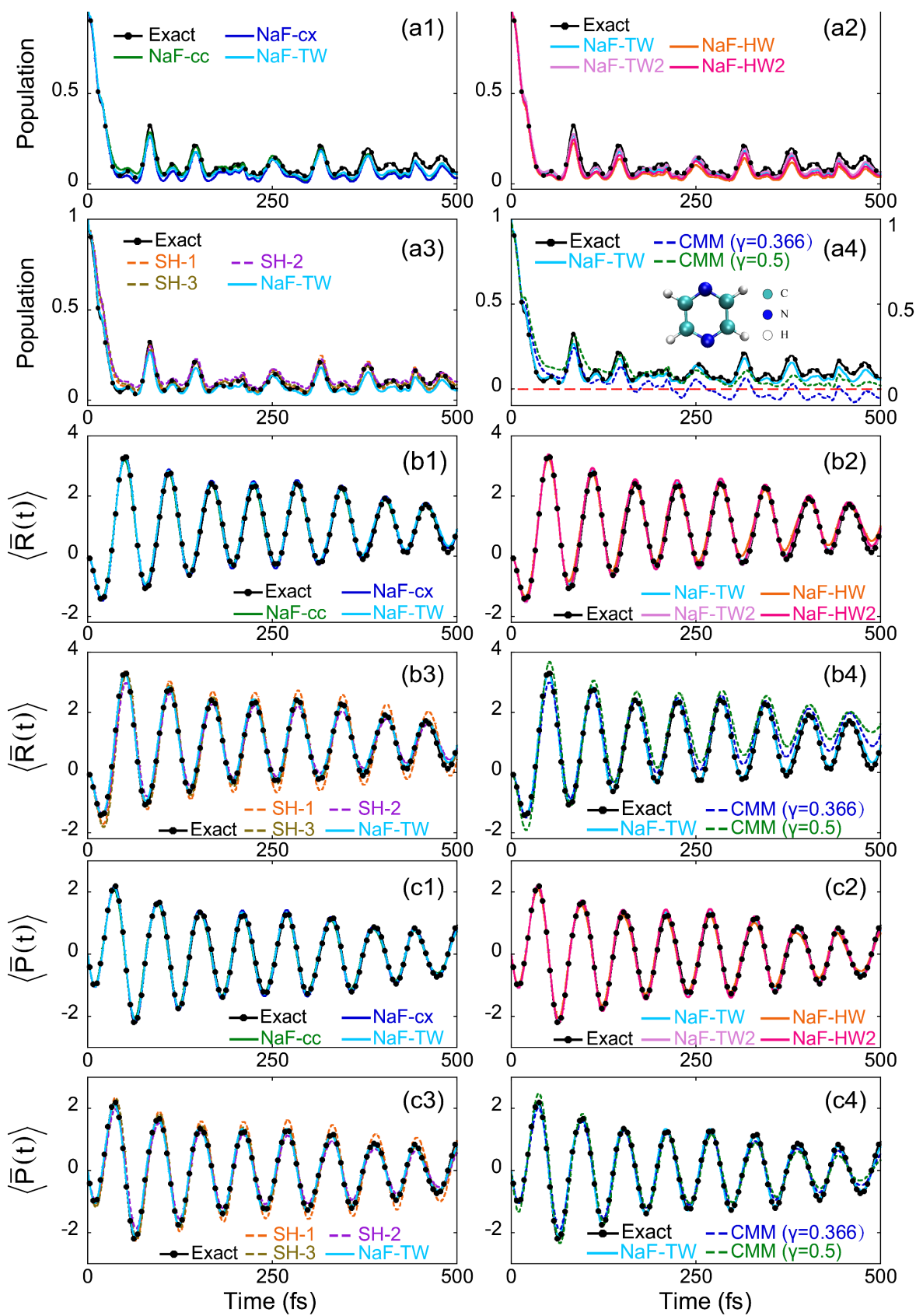

**Figure 1.** Results of the 24-mode LVCM for pyrazine. Panels (a1)-(a4): Population of the state 2; Panels (b1)-(b4): The average dimensionless coordinate $\langle \bar{R}(t) \rangle$ of the nuclear normal mode $v_{6a}$; Panels (c1)-(c4): The average dimensionless momentum $\langle \bar{P}(t) \rangle$ of the nuclear normal mode $v_{6a}$. In panels (a1), (b1) and (c1), the green, blue and cyan solid lines represent the results of NaF-cc, NaF-cx and NaF-TW, respectively. In panels (a2), (b2) and (c2), the cyan, pink, orange and magenta solid lines represent the results of NaF-TW, NaF-TW2, NaF-HW and NaF-HW2, respectively. In panels (a3), (b3) and (c3), the orange dashed line, purple dashed line, brown dashed line and cyan solid line denote the results of SH-1, SH-2, SH-3 and NaF-TW, respectively. In panels (a4), (b4) and (c4), the blue dashed line, green dashed line and cyan solid line denote the results of CMM (γ=0.366), CMM (γ=0.5) and NaF-TW, respectively. The numerically exact results produced by MCTDH[361] are demonstrated by black solid lines with black points in each panel.

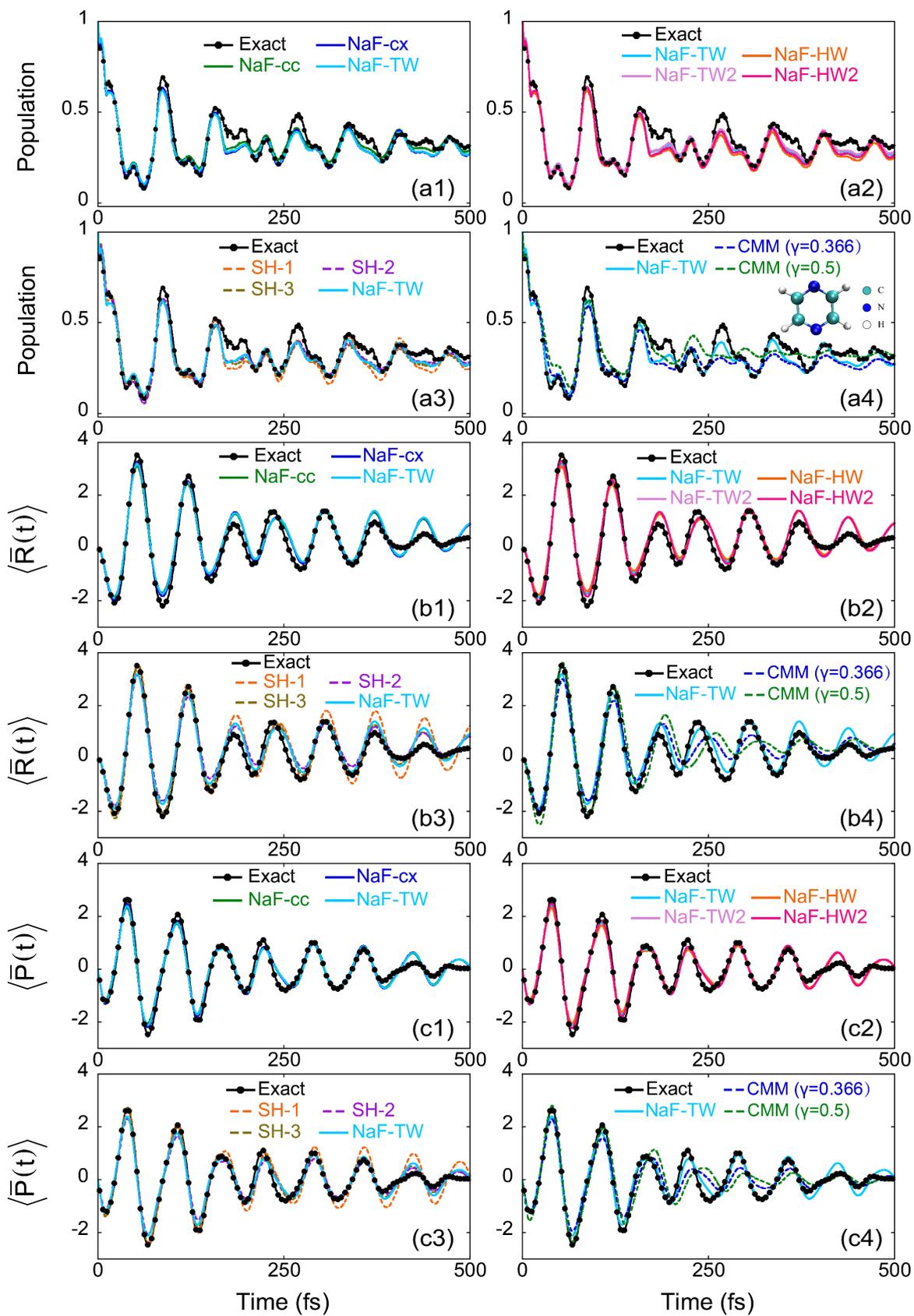

**Figure 2.** Results of the 3-mode LVCM for pyrazine. Panels (a1)-(a4): Population of the state 2; Panels (b1)-(b4): The average dimensionless coordinate $\langle \bar{R}(t) \rangle$ of the nuclear normal mode $v_{6a}$; Panels (c1)-(c4): The average dimensionless momentum $\langle \bar{P}(t) \rangle$ of the nuclear normal mode $v_{6a}$. In panels (a1), (b1) and (c1), the green, blue and cyan solid lines represent the results of NaF-cc, NaF-cx and NaF-TW, respectively. In panels (a2), (b2) and (c2), the cyan, pink, orange and magenta solid lines represent the results of NaF-TW, NaF-TW2, NaF-HW and NaF-HW2, respectively. In panels (a3), (b3) and (c3), the orange dashed line, purple dashed line, brown dashed line, and cyan solid line denote the results of SH-1, SH-2, SH-3 and NaF-TW, respectively. In panels (a4), (b4) and (c4), the blue dashed line, green dashed line and cyan solid line denote the results of CMM (γ=0.366), CMM (γ=0.5) and NaF-TW, respectively. The numerically exact results produced by MCTDH[361] are demonstrated by black solid lines with black points in each panel.

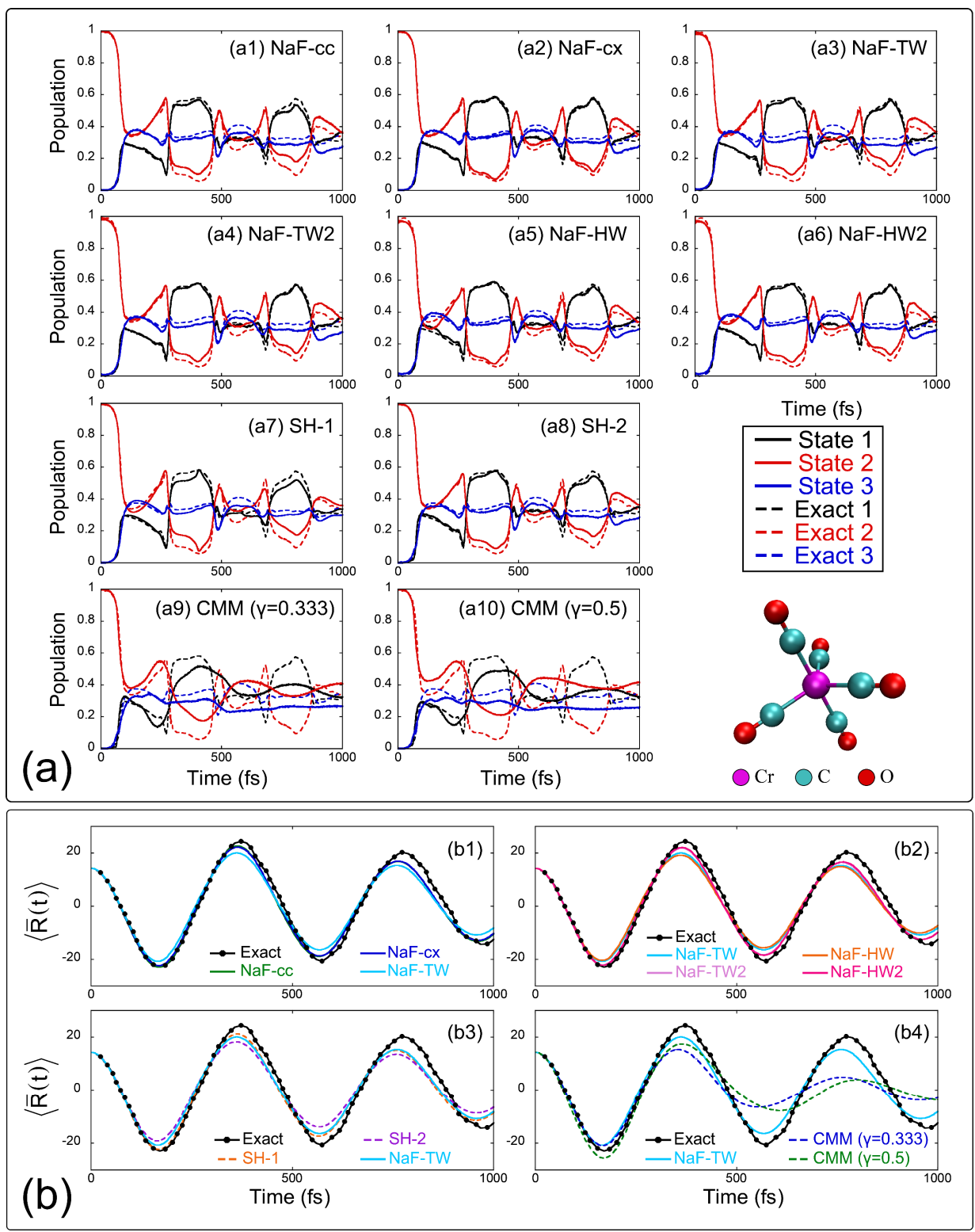

**Figure 3.** Results of the 2-mode LVCM for the $Cr(CO)_5$ molecule. Panels (a1)-(a10): Population dynamics, where the black, red and blue solid lines represent the population of states 1-3,

respectively, and the numerically exact results (taken from ref [360]) are plotted by dashed lines with corresponding colors. Panel (a1): NaF-cc; Panel (a2): NaF-cx; Panel (a3): NaF-TW; Panel (a4): NaF-TW2; Panel (a5): NaF-HW; Panel (a6): NaF-HW2; Panel (a7): SH-1; Panel (a8): SH-2; Panel (a9): CMM ($\gamma$=0.333); Panel (a10): CMM ($\gamma$=0.5). Panels (b1)-(b4): The average dimensionless coordinate $\langle \bar{R}(t) \rangle$ of the second nuclear normal mode. In panel (b1), the green, blue and cyan solid lines represent the results of NaF-cc, NaF-cx and NaF-TW, respectively. In panel (b2), the cyan, pink, orange and magenta solid lines represent the results of NaF-TW, NaF-TW2, NaF-HW and NaF-HW2, respectively. In panel (b3), the orange dashed line, purple dashed line and cyan solid line denote the results of SH-1, SH-2 and NaF-TW, respectively. In panel (b4), the blue dashed line, green dashed line and cyan solid line denote the results of CMM ($\gamma$=0.333), CMM ($\gamma$=0.5) and NaF-TW, respectively. Note that SH-3 is not applicable for this 3-state model. The numerically exact results produced by MCTDH (taken from ref [360]) are demonstrated by black solid lines with black points in panels (b1)-(b4).

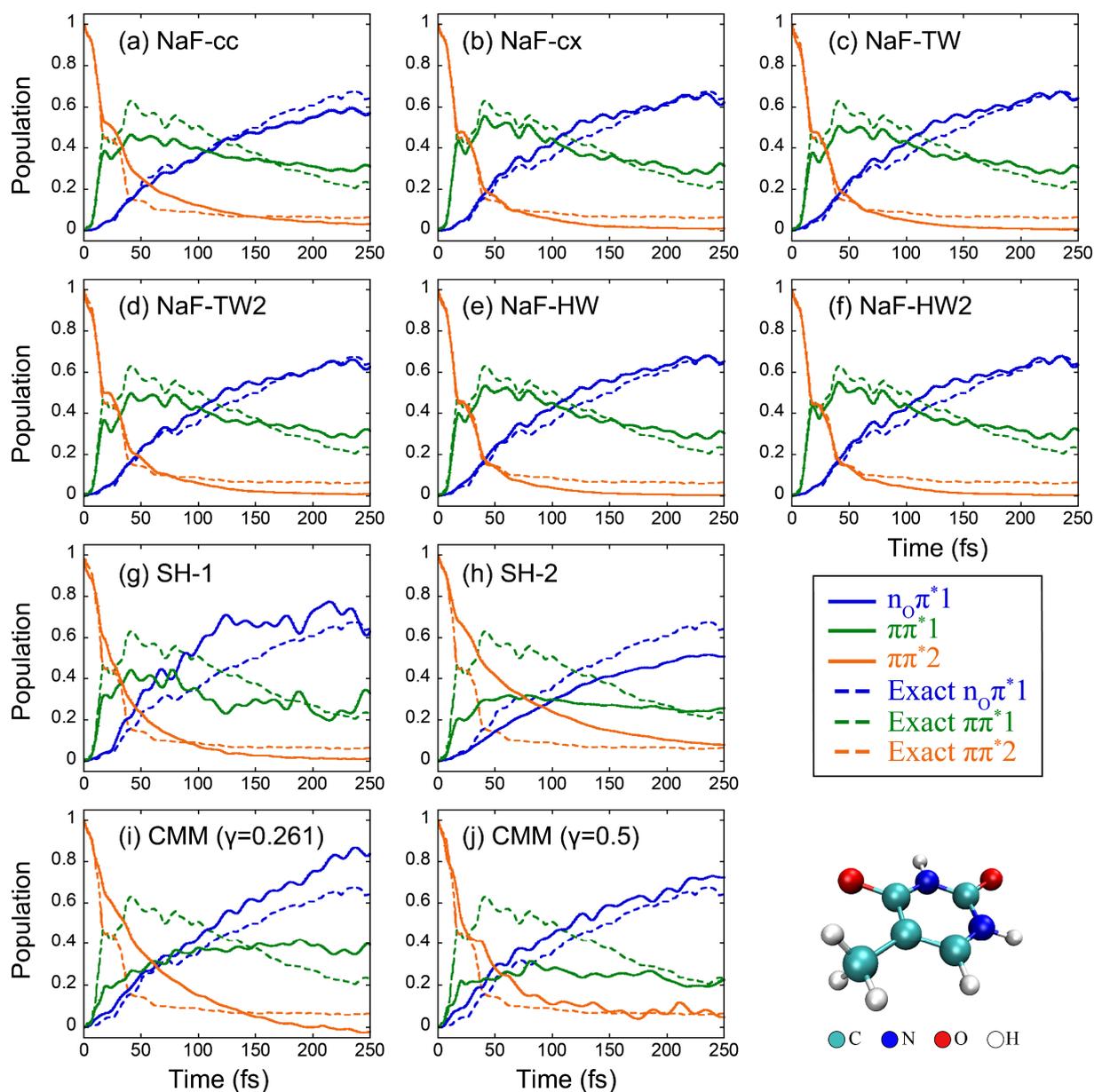

**Figure 4.** Results of the 39-mode LVCM for the Thymine parametrized from the CAM-B3LYP calculations. In each panel, population dynamics of 3 states are demonstrated, where the blue, green and orange solid lines represent the population of the $n_O\pi^*1$ state, the $\pi\pi^*1$ state and the $\pi\pi^*2$ state, respectively. Panel (a): NaF-cc; Panel (b): NaF-cx; Panel (c): NaF-TW; Panel (d): NaF-TW2; (e): NaF-HW; Panel (f): NaF-HW2; Panel (g): SH-1; Panel (h): SH-2; Panel (i): CMM (γ=0.261); Panel (j): CMM (γ=0.5). Note that SH-3 is not applicable for this 7-state model. The numerically

exact results produced by ML-MCTDH (taken from ref [362]) are demonstrated by dashed lines with corresponding colors in each panel.

Although both CMM(cv) and NaF employ the CPS formulation, Figures 1-4 show that NaF is superior to CMM(cv) for LVCMs of real molecular systems. Because CMM(cv) employs Ehrenfest-like dynamics, the comparison in Figures 1-4 agrees with the conclusion of Refs [322, 324] that NaF performs better than Ehrenfest-like dynamics when both nuclear and electronic motion is considered in the asymptotic region where the nonadiabatic coupling disappears. We then only focus on NaF with the CPS formulation in the following benchmark tests.

### 3.2 Nonadiabatic Scattering Processes in Gas-Phase

We demonstrate the performance of NaF methods on gas-phase scattering problems with one nuclear DOF. We first consider Tully's three scattering models with one nuclear DOF and two electronic states[253], namely, the single avoided crossing (SAC) model, the dual avoided crossing (DAC) model and the extended coupling region (ECR) model. The potential energy operator in the diabatic representation is $\hat{V}(R) = V_{11}(R)|1\rangle\langle 1| + V_{22}(R)|2\rangle\langle 2| + V_{12}(R)(|1\rangle\langle 2| + |2\rangle\langle 1|)$, where

$$\begin{aligned} V_{11}(R) &= A(1-e^{-B|R|})\operatorname{sgn}(R) \\ V_{22}(R) &= -V_{11}(R) \\ V_{12}(R) &= Ce^{-DR^2} \end{aligned}, \qquad (79)$$

for the SAC model with $A = 0.01$, $B = 1.6$, $C = 0.005$ and $D = 1.0$ (all in atomic units, the same below),

$$\begin{aligned} V_{11}(R) &= 0 \\ V_{22}(R) &= -Ae^{-BR^2} + E_0 \\ V_{12}(R) &= Ce^{-DR^2} \end{aligned} \qquad (80)$$

for the DAC model with $A = 0.1$, $B = 0.28$, $C = 0.015$, $D = 0.06$ and $E_0 = 0.05$, and

$$\begin{aligned} V_{11}(R) &= +E_0 \\ V_{22}(R) &= -E_0 \\ V_{12}(R) &= C\left[e^{BR}h(-R) + (2 - e^{-BR})h(R)\right] \end{aligned} \qquad (81)$$

for the ECR model with $B = 0.9$, $C = 0.1$ and $E_0 = -0.0006$, respectively. The system is initially in the electronic ground state in the adiabatic representation with the nuclear mass $M = 2000$ au and the nuclear wave function

$$\psi(R) \propto e^{-\alpha(R-R_0)^2/2 + iP_0(R-R_0)} . \qquad (82)$$

The corresponding Wigner distribution of eq (82) reads

$$\rho_W(R, P) \propto e^{-\alpha(R-R_0)^2 - (P-P_0)^2/\alpha} , \qquad (83)$$

where the center of the wave function, $R_0$, is set to -3.8, -10, and -13 au for SAC, DAC, and ECR models, respectively. The width parameter $\alpha = 1$. Initial momentum $P_0$ is adjustable. The scattering probabilities of each channel in the adiabatic representation are investigated. For these three models, the TCFs are directly expressed and calculated in the adiabatic representation. The numerically exact results are produced by using Colbert and Miller's version of the discrete variable representation (DVR) of ref [363].

Figures 5-6 illustrate the scattering probability in the adiabatic representation for all Tully models versus $P_0$. SH-1 performs reasonably for the SAC and DAC models. Although SH-2 and SH-3 are consistent with SH-1 in the large momentum region, they perform worse than SH-1 in the relatively small momentum region. For example, both SH-2 and SH-3 underestimate the transmission probability of the electronic ground state in the SAC and DAC models for small $P_0$. Additionally, SH-2 even produces negative transmission probabilities for the electronic excited state. All NaF methods yield relatively more accurate results than SH-2 for the SAC and DAC models. For the more challenging ECR model, the transmission probabilities obtained from SH-1 and SH-3 show overall good agreement with numerically exact results. SH-2 and all NaF methods underestimate the transmission probabilities on the electronic ground state in the intermediate momentum region. NaF-cc also significantly overestimates the transmission probability on the electronic excited state in the intermediate momentum region, while SH-2, NaF-cc and NaF-cx exhibit negative probability issues. NaF-TW, NaF-TW2, NaF-HW and NaF-HW2 yield more accurate results than other NaF methods due to their positive semidefinite TCFs for population dynamics.

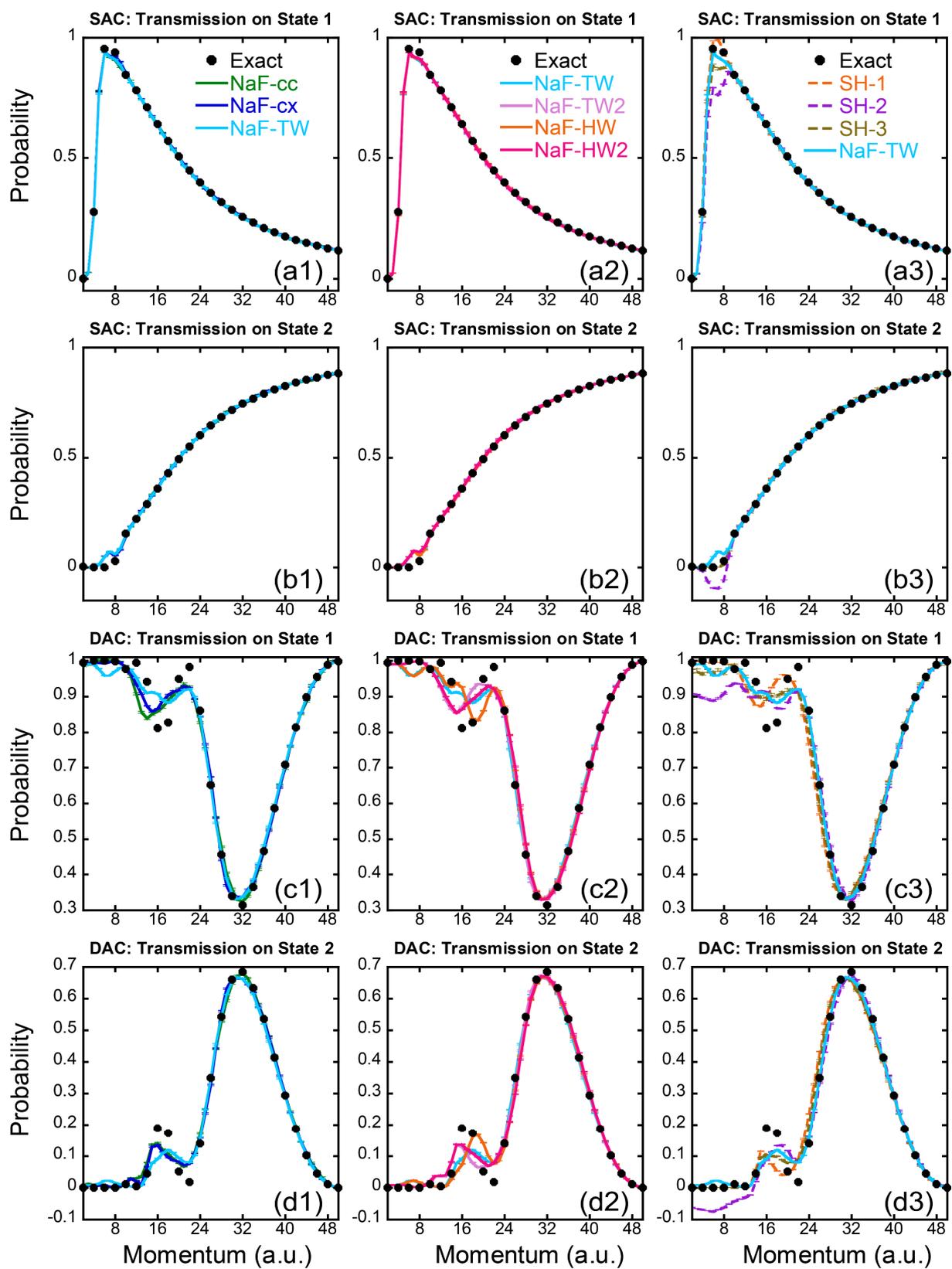

**Figure 5.** Results of the SAC and DAC models. Panels (a1) and (b1): transmission probabilities on the adiabatic ground and excited state of the SAC model, respectively; Panels (c1) and (d1): transmission probabilities on the adiabatic ground and excited state of the DAC model, respectively; In panels (a1), (b1), (c1) and (d1), the green, blue, and cyan solid lines represent the results of NaF-cc, NaF-cx and NaF-TW, respectively. Panels (a2), (b2), (c2) and (d2) are similar to panels (a1), (b1), (c1) and (d1), respectively, but the cyan, pink, orange and magenta solid lines denote the results of NaF-TW, NaF-TW2, NaF-HW and NaF-HW2, respectively. Panels (a3), (b3), (c3) and (d3) are similar to panels (a1), (b1), (c1) and (d1), respectively, but the orange dashed lines, purple dashed lines, brown dashed lines and cyan solid lines denote the results of SH-1, SH-2, SH-3 and NaF-TW, respectively. The numerically exact results produced by DVR[363] are demonstrated by black points in each panel.

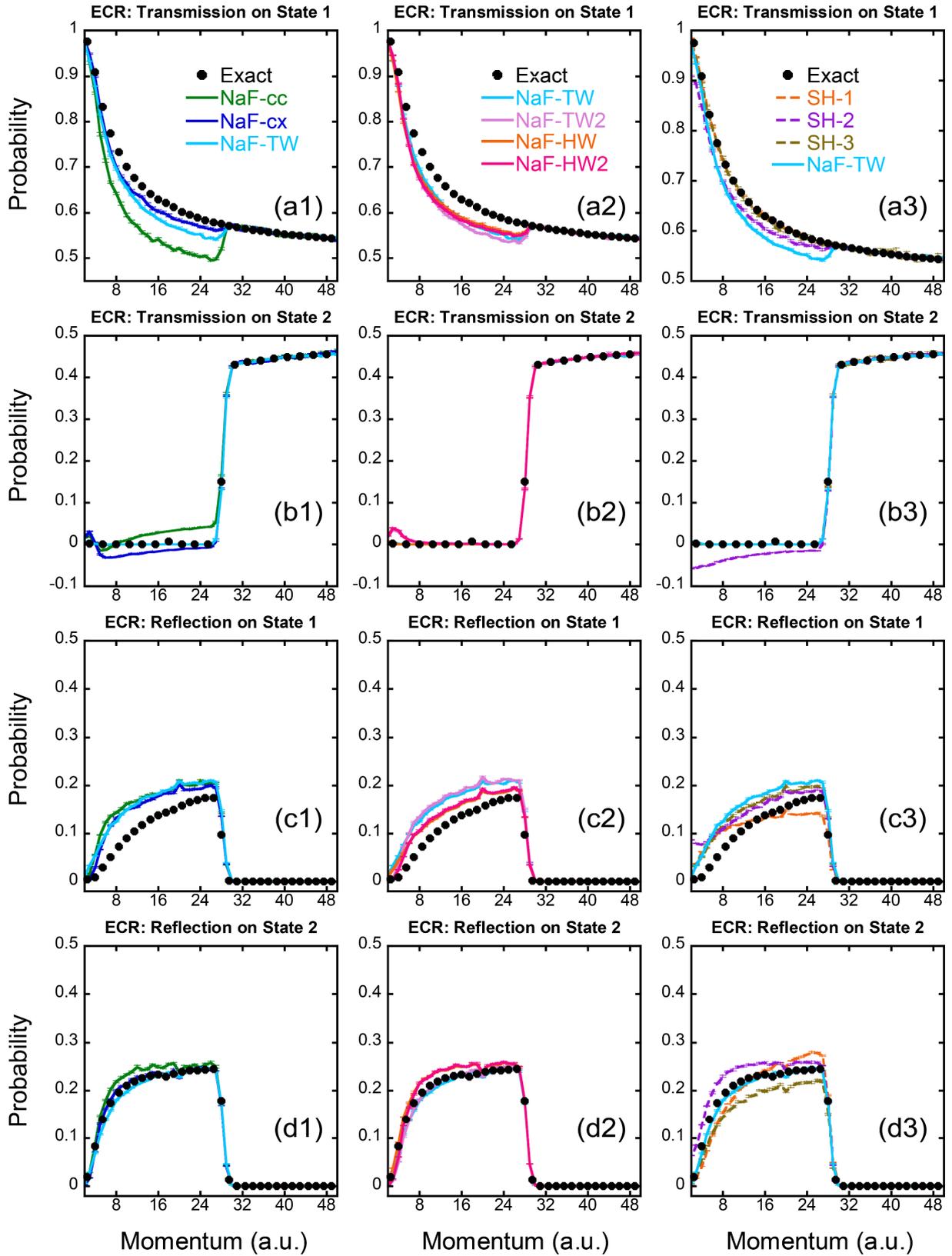

**Figure 6.** Results of the ECR models. Panels (a1) and (b1): transmission probabilities on the adiabatic ground and excited state, respectively; Panels (c1) and (d1): reflection probabilities on the adiabatic ground and excited state, respectively. In panels (a1), (b1), (c1) and (d1), the green, blue, and cyan solid lines represent the results of NaF-cc, NaF-cx and NaF-TW, respectively. Panels (a2), (b2), (c2) and (d2) are similar to panels (a1), (b1), (c1) and (d1), respectively, but the cyan, pink, orange and magenta solid lines denote the results of NaF-TW, NaF-TW2, NaF-HW and NaF-HW2, respectively. Panels (a3), (b3), (c3) and (d3) are similar to panels (a1), (b1), (c1) and (d1), respectively, but the orange dashed lines, purple dashed lines, brown dashed lines and cyan solid lines denote the results of SH-1, SH-2, SH-3 and NaF-TW, respectively. The numerically exact results produced by DVR[363] are demonstrated by black points in each panel.

In addition, we employ three anharmonic photodissociation models with three electronic states, developed by Miller and co-workers[185]. The diagonal elements of the potential matrix in the diabatic representation are Morse potential functions:

$$V_{nn}(R) = D_n \left[1 - e^{-\beta_n(R-R_n)}\right]^2 + C_n, \quad n = 1, 2, 3, \tag{84}$$

while the diabatic coupling elements are Gaussian functions:

$$V_{nm}(R) = V_{mn}(R) = A_{nm} e^{-\alpha_{nm}(R-R_{nm})^2}, \quad n, m = 1, 2, 3; \text{ and } n \neq m. \tag{85}$$

All parameters for these three models are provided in ref [185] and are also summarized in our pervious works[320, 322]. We present the results of only Model 2 in the main text, leaving the other two models in Section S7 of the Supporting Information. The initial values of the nuclear DOF with mass $M = 20000$ au are sampled from the Wigner distribution function

$$\rho_W(R,P) \propto e^{-M\omega(R-R_e)^2 - P^2/M\omega} \quad . \tag{86}$$

Here $\omega = 0.005$ au and $R_e = 3.3$ au for Model 2. The electronic DOFs are initially excited on the first electronic state. The numerically exact results are produced by DVR[363].

Panels (a)-(h) of Figure 7 present the electronic population dynamics of Model 2 of the 3-state photodissociation models. Interestingly, population dynamics generated by SH-2, illustrated in panel (h) of Figure 7, exhibit unreasonable oscillations. These oscillations are due to the unphysical momentum reversal for the frustrated hops in the SH-2 algorithm[295]. This can also be inferred from the nuclear properties. Panels (i)-(k) of Figure 7 illustrate the nuclear momentum distribution in the asymptotic region ($t$=200 fs) (See Sub-Section S1-E of the Supporting Information[340] of ref [322] for further details), where panel (k) shows that the nuclear momentum distribution of SH-2 has a greater proportion falling into the negative momentum region compared to those of SH-1 and NaF methods, indicating that the momentum reversal procedure affects the direction of nuclear momentum even in the long-time region. Note that SH-1 and all NaF methods shown here do not involve any momentum reversal, which consistently provide accurate descriptions for electronic population dynamics of this model. (The algorithm of SH-1 utilized in this work can be referred to Section S7 of the Supporting Information[340] of ref [322].) The differences in performance of all NaF methods are marginal for the other two models, but all of them slightly outperform SH-1 and SH-2, as demonstrated in Section S7 of the Supporting Information.

We further test a two-state anharmonic photodissociation model. The diagonal elements of the diabatic potential energy matrix are identical to states 1 and 3 of Model 2 of the 3-state photodissociation model, while the off-diagonal elements follow eq (85) with parameters $A_{12}=A_{21}=0.005$ au, $R_{12}=R_{21}=3.34$ au and $\alpha_{12}=\alpha_{21}=8$ au. All other computational settings remain the same as those of Model 2 of the three-state photodissociation model. Figure 8 illustrates population dynamics of the first diabatic state and the nuclear momentum distribution at 200 fs of the 2-state photodissociation model. As shown in panels (h) and (i) of Figure 8, both population dynamics of SH-2 and that of SH-3 exhibit artificial oscillations, indicating that the reversal of momentum in the frustrated hopping event in the two SH algorithms leads to unphysical behavior for the two-state benchmark model. As demonstrated in panel (g) of Figure 8, SH-1 provides physically correct population transfer behavior, but the results considerably deviate from the exact data. In contrast, all NaF methods consistently produce more accurate results for population dynamics, as illustrated in panels (a)-(f) of Figure 8. All NaF methods and SH methods generate comparable reasonable nuclear momentum distribution in the asymptotic region.

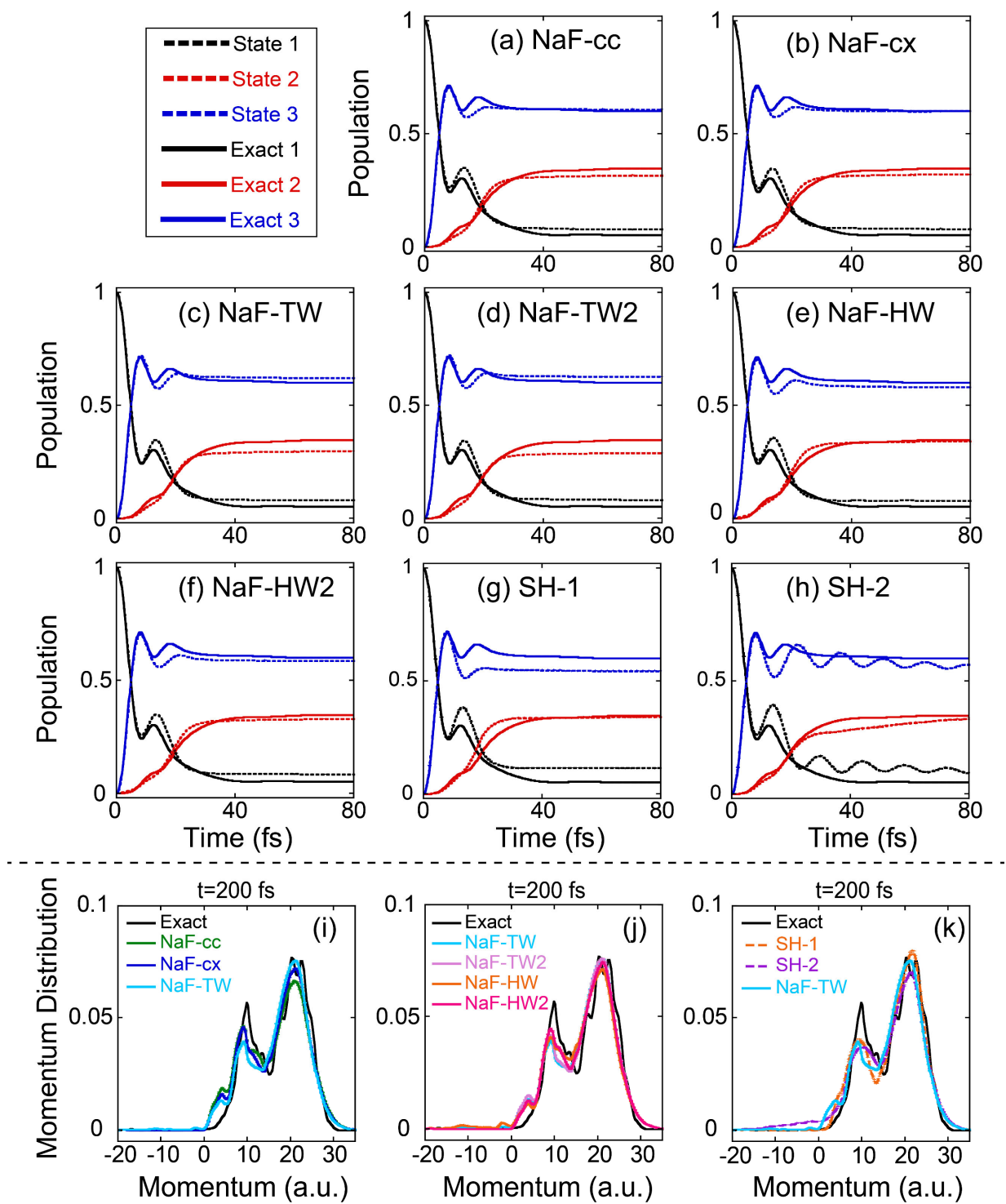

**Figure 7.** Results of Model 2 of the 3-state photodissociation models. In panels (a)-(h), the black, red and blue dashed lines represent population dynamics of states 1-3, respectively. Panel (a): NaF-cc; Panel (b): NaF-cx; Panel (c): NaF-TW; Panel (d): NaF-TW2; Panel (e): NaF-HW; Panel (f):

NaF-HW2; Panel (g): SH-1; Panel (h): SH-2. Note that SH-3 is not applicable for this 3-state model. The numerically exact results produced by DVR[363] are demonstrated by solid lines with corresponding colors in panels (a)-(h). Panels (i)-(k) illustrate the nuclear momentum distribution at 200 fs. The green, blue, and cyan solid lines in panel (i) represent the results of NaF-cc, NaF-cx and NaF-TW, respectively. The cyan, pink, orange and magenta solid lines in panel (j) denote the results of NaF-TW, NaF-TW2, NaF-HW and NaF-HW2, respectively. The orange dashed lines, purple dashed lines and cyan solid lines in panel (k) denote the results of SH-1, SH-2 and NaF-TW, respectively. The black solid lines in panels (i)-(k) denote the numerically exact results produced by DVR[363].

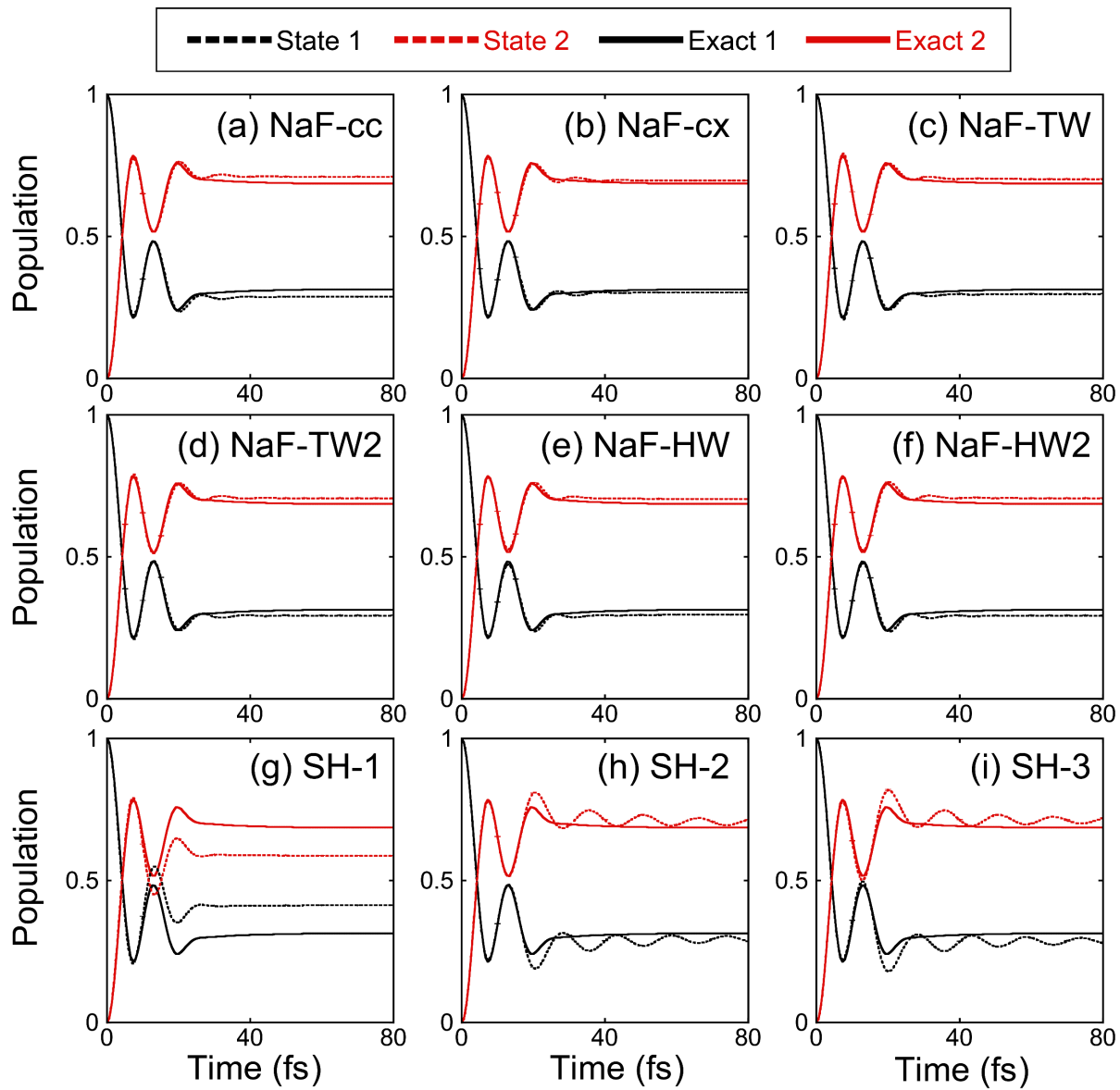
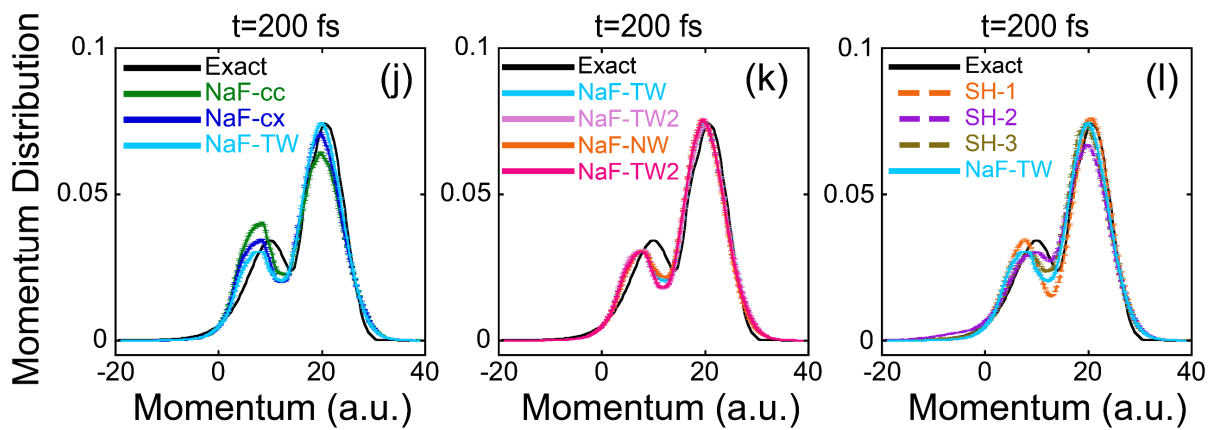

**Figure 8.** Results of the 2-state photodissociation model. In panels (a)-(i), the black and red dashed lines represent population dynamics of states 1 and 2, respectively. Panel (a): NaF-cc; Panel (b): NaF-cx; Panel (c): NaF-TW; Panel (d): NaF-TW2; Panel (e): NaF-HW; Panel (f): NaF-HW2; Panel (g): SH-1; Panel (h): SH-2; Panel (i): SH-3. The numerically exact results produced by DVR[363] are demonstrated by solid lines with corresponding colors in panels (a)-(i). Panels (j)-(l) illustrate the nuclear momentum distribution at 200 fs. The green, blue, and cyan solid lines in panel (j) represent the results of NaF-cc, NaF-cx and NaF-TW, respectively. The cyan, pink, orange and magenta solid lines in panel (k) denote the results of NaF-TW, NaF-TW2, NaF-HW and NaF-HW2, respectively. The orange dashed lines, purple dashed lines, brown dashed lines and cyan solid lines in panel (l) denote the results of SH-1, SH-2, SH-3 and NaF-TW, respectively. The black solid lines in panels (j)-(l) denote the numerically exact results produced by DVR[363].

### 3.3 System-Harmonic Bath Models

System-harmonic bath models are widely employed to study condensed-phase dissipative processes, quantum phase transitions, quantum thermodynamics, heat/energy transport/relaxation properties, and electron/proton transfer processes in physics, chemical, biological, materials, quantum computation and quantum information systems[208, 213, 223, 364-373]. These models also serve as benchmark problems for testing nonadiabatic dynamics methods[11, 12, 191, 195, 200, 209, 318, 319, 322, 324, 374-377], because numerically exact results are often available by real-time path integral methods[108-110, 114-116, 120], (ML-)MCTDH[134-137], TD-DMRG[151-160], HEOM[122-127], *etc*. The Hamiltonian operator can be divided into three parts: the system part, the harmonic bath, and the bilinear coupling term:

$$\hat{H} = \hat{H}_s + \hat{H}_b + \hat{H}_{s-b} \quad . \tag{87}$$

A typical example is the 2-state spin-boson model[366], whose Hamiltonian is of the form:

$$\hat{H}_s^{\text{(spin-boson)}} = \varepsilon \hat{\sigma}_z + \Delta \hat{\sigma}_x \ , \tag{88}$$

$$\hat{H}_b^{\text{(spin-boson)}} = \sum_{j=1}^{N_{\text{nuc}}} \frac{1}{2} \left( \hat{P}_j^2 + \omega_j^2 \hat{R}_j^2 \right) \ , \tag{89}$$

$$\hat{H}_{s-b}^{\text{(spin-boson)}} = \sum_{j=1}^{N_{\text{nuc}}} c_j \hat{R}_j \hat{\sigma}_z \ , \tag{90}$$

where $\hat{\sigma}_z = |1\rangle\langle 1| - |2\rangle\langle 2|$ and $\hat{\sigma}_x = |1\rangle\langle 2| + |2\rangle\langle 1|$ are the corresponding components of the Pauli operators. Another type is the site-exciton model[378], which reads

$$\hat{H}_s^{\text{(site-exciton)}} = \sum_{n,m=1}^{F} h_{nm} |n\rangle\langle m| \ , \tag{91}$$

$$\hat{H}_b^{\text{(site-exciton)}} = \sum_{n=1}^{F} \sum_{j=1}^{N_b} \frac{1}{2} \left( \hat{P}_{nj}^2 + \omega_j^2 \hat{R}_{nj}^2 \right) \ , \tag{92}$$

$$\hat{H}_{s-b}^{\text{(site-exciton)}} = \sum_{n=1}^{F} \sum_{j=1}^{N_b} c_j \hat{R}_{nj} |n\rangle\langle n| \ , \tag{93}$$

where $N_b = N_{\text{nuc}} / F$ represents the number of bath modes on each site/state.

The frequencies, $\{\omega_j\}$, and coefficients, $\{c_j\}$, of bath DOFs are often obtained by discretizing the spectral density as first suggested by Makri in ref [379]. For the Ohmic spectral density $J(\omega) = \frac{\pi}{2} \alpha \omega e^{-\omega/\omega_c}$ with Kondo parameter $\alpha$ and cutoff frequency $\omega_c$, the discretization scheme is often presented as[379, 380]

$$\begin{cases} \omega_j = -\omega_c \ln(1 - j/(1+N_b)) \\ c_j = \omega_j \sqrt{\alpha \omega_c / (1+N_b)} \end{cases}, \ j = 1, \cdots, N_b \ , \tag{94}$$

Similarly, the Debye spectral density $J(\omega) = \dfrac{2\lambda\omega\omega_c}{\omega^2 + \omega_c^2}$ with reorganization energy $\lambda$ and characteristic frequency $\omega_c$ is discretized as[135, 374, 381]

$$\begin{cases} \omega_j = \omega_c \tan\left(\dfrac{\pi}{2}\left(1 - \dfrac{j}{1+N_b}\right)\right), \\ c_j = \omega_j \sqrt{2\lambda/(1+N_b)} \end{cases} \quad j = 1,\cdots,N_b \ . \tag{95}$$

More recent progress on obtaining the harmonic-bath parameters from molecular dynamics simulations has been discussed in ref [382]. The initial density is often set as the tensor product between the density of system and the thermal equilibrium density of bath DOFs:

$$\hat{\rho}(0) = \hat{\rho}_s \otimes e^{-\beta \hat{H}_b} / \mathrm{Tr}\left[e^{-\beta \hat{H}_b}\right] \ . \tag{96}$$

We first utilize four spin-boson models with Ohmic spectral density at low temperature $\beta = 5$. The values of the Kondo parameter and the cutoff frequency are listed in $\alpha \in \{0.1, 0.4\}$ and $\omega_c \in \{1, 2.5\}$, and the energy bias and the coupling are set to $\varepsilon = \Delta = 1$. The system is initially in state $|1\rangle$ (that is, $\hat{\rho}_s = |1\rangle\langle 1|$). 300 bath DOFs are utilized for each spin-boson model. These models have been tested in refs [12, 319, 322, 324], where the numerically exact results produced by extended hierarchy equations of motion (eHEOM)[122-127, 383] are available. In principle, one can also produce numerically exact data from other benchmark methods. Figures 9-10 illustrate electronic population and coherence dynamics of all spin-boson models, respectively. The results generated by SH methods deviate from the exact data since relatively short time, especially for the models with higher cutoff frequencies and stronger system-bath coupling. In comparison, the data produced by NaF methods exhibit better agreements with the numerically exact results in all cases.

We also consider two multi-state site-exciton models with Debye spectral density, which are derived from real chemical systems. The first model is the 7-state Fenna-Matthews-Olson (FMO) model, which describes the exciton energy transfer processes within the light-harvesting complex in green sulfur bacteria[81, 189, 384-389]. The system Hamiltonian is given by:

$$\hat{H}_s = \begin{pmatrix} 12410 & -87.7 & 5.5 & -5.9 & 6.7 & -13.7 & -9.9 \\ -87.7 & 12530 & 30.8 & 8.2 & 0.7 & 11.8 & 4.3 \\ 5.5 & 30.8 & 12210 & -53.5 & -2.2 & -9.6 & 6.0 \\ -5.9 & 8.2 & -53.5 & 12320 & -70.7 & -17.0 & -63.3 \\ 6.7 & 0.7 & -2.2 & -70.7 & 12480 & 81.1 & -1.3 \\ -13.7 & 11.8 & -9.6 & -17.0 & 81.1 & 12630 & 39.7 \\ -9.9 & 4.3 & 6.0 & -63.3 & -1.3 & 39.7 & 12440 \end{pmatrix} \text{cm}^{-1}. \quad (97)$$

The parameters of the spectral density are $\lambda = 35 \text{ cm}^{-1}$ and $\omega_c = 106.14 \text{ cm}^{-1}$. We employ 50 bath modes for each state. The first state of the system is initially occupied, with temperature set at $T=77$ K. Numerically exact data are obtained from TD-DMRG calculations[151-160], using the same discretization scheme to establish a Hamiltonian operator for the FMO model, which are also employed by trajectory-based methods. This ensures that all methods are fairly compared for a well-defined quantum mechanical Hamiltonian operator. The second case is a 3-state model used to describe the singlet fission (SF) of pentacene[214, 390]. The system Hamiltonian is:

$$\hat{H}_S = \begin{pmatrix} 0.2 & -0.05 & 0 \\ -0.05 & 0.3 & -0.05 \\ 0 & -0.05 & 0 \end{pmatrix} \text{eV}, \quad (98)$$

where the three states represent the high-energy singlet state (S1), the charge-transfer state (CT), and the multi-exciton state (TT), respectively. The parameters of the Debye spectral density are $\lambda = 0.1$ eV and $\omega_c = 0.18$ eV. 200 bath modes are utilized for each state. The system is initially located at the S1 state, and the temperature is set at 300 K. The numerically exact data for the SF model are obtained from HEOM[122-126].

Figures 11-12 present the results of electronic population and coherence dynamics of the 7-state FMO model, respectively. We evolve the trajectories up to 10000 fs, approaching the long-time steady-state region. As shown in Figure 11(g) and Figure 12(g), SH-1 is not able to produce reasonable results for this model. Interestingly, although SH-2 outperforms SH-1 in the first 1000 fs, the long-time predictions by SH-2 are *not* reasonable. SH-2 generates the unphysical attenuation rather than the plateau for the population of the third state, as well as for the coherence term $|\rho_{34}|$ after 2000 fs. This indicates that SH-2 does not rationally describe the steady-state region in the long-time limit. We note that the SH-2 results shown in Figures 11-12 differ from those presented in ref [295]. This discrepancy arises because ref [295] used the classical Boltzmann distribution instead of the Wigner distribution for sampling the initial values of nuclear DOFs, thus excluding any nuclear quantum effects (for bath modes) during the simulations. TD-DMRG suggests that quantum effects in nuclear dynamics are not negligible for the FMO model at 77 K. Even when only electronic dynamics is investigated, nuclear (bath) modes should be treated quantum mechanically to predict time-dependent electronic properties for the *right* reason. Compared to the exact data produced by TD-DMRG using the identical Hamiltonian operator (with the same number of bath modes), SH-2 fails to describe the correct long-time asymptotic behavior in the steady-state region when nuclear quantum effects are not neglected. In comparison, all NaF methods, with nuclear (bath) DOFs consistently sampled from the Wigner distribution, produce overall reasonable results and outperform the SH methods. While NaF-cc and NaF-cx yield more accurate population dynamics in the first 1000 fs, NaF-TW, NaF-TW2, NaF-HW, and NaF-HW2 predict more reasonable long-time population dynamics. In addition, NaF-TW, NaF-TW2, NaF-

HW, and NaF-HW2 exhibit better performance than other NaF methods in describing the coherence dynamics, especially in the long-time region (after 1000 fs).

Figures 11-12 imply the importance of nuclear nonadiabatic force for correctly describing the quantum mechanical behavior of both electronic and nuclear motion in the model system. Since the mixed quantum-classical limit (where nuclear DOFs are treated classically and electronic DOFs are treated quantum mechanically) is not defined without any ambiguity for nonadiabatic systems where nuclear motion and electronic motion are coupled, *caution should often be taken* when nuclear DOFs are described by classical mechanics in either thermodynamics or nonadiabatic transition dynamics. For instance, when the number of path integral beads becomes one in the adiabatic representation for coupled multi-electronic-state systems, nuclear DOFs are treated classically, but such a mixed quantum-classical limit does not always leads to consistent results for electronic thermodynamic properties[376].

The results of 7-state FMO model at zero temperature (0 K) are presented in Section S7 of the Supporting Information, which provides a significant challenging test for trajectory-based nonadiabatic dynamics methods. When the initial nuclear distribution is described by classical mechanics for this case, it does not yield any meaningful results for NaF, SH, or other independent-trajectory-based nonadiabatic dynamics methods. Even in such a challenging case, NaF methods also lead to reasonably good results (in comparison to the TD-DMRG results) and consistently outperform SH methods. The zero-temperature benchmark test case of the effective Hamiltonian in the discretization scheme of the FMO model in Section S7 of the Supporting Information is also *heuristic*. The initial nuclear condition of many practical *ab initio* photo-dynamics simulations is set to be the ground vibrational state of an electronic state (e.g., often the ground adiabatic electronic state or a diabatic state) of the real multi-dimensional molecular system before photo-excitation. Such an initial condition is equivalent to the zero-temperature limit for nuclear DOFs on a single adiabatic or diabatic state PES of a complex molecular system.

Figure 13 illustrates electronic population dynamics of the SF model. Similar to the FMO model, SH-2 also produces noticeable attenuation of the population dynamics of the TT state in the long-time dynamics region, in comparison to exact results. In contrast to SH methods, NaF methods produce overall more reasonable results.

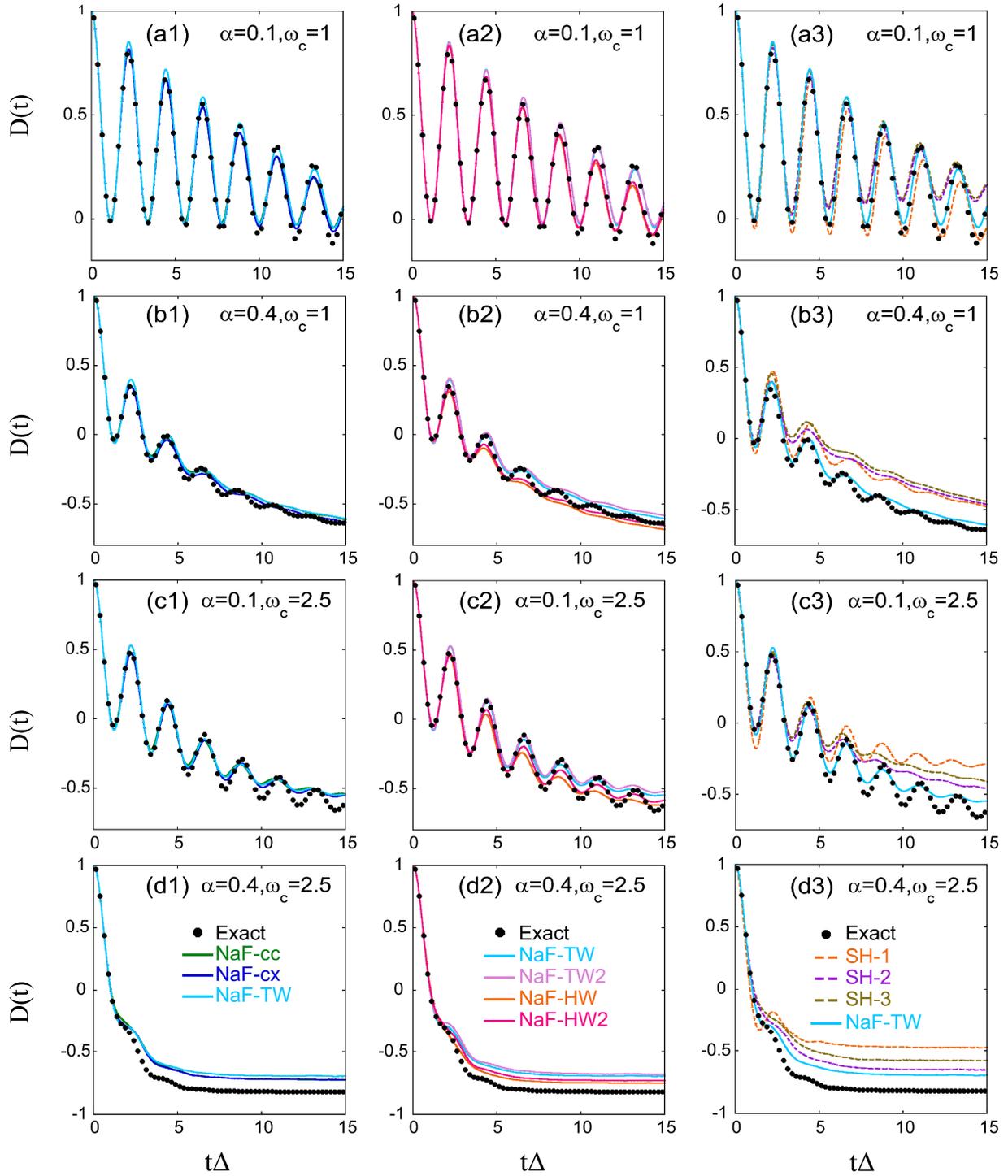

**Figure 9.** Results of the population difference $D(t) = \rho_{11}(t) - \rho_{22}(t)$ of spin-boson models with Ohmic spectral density at $\beta = 5$. The first to fourth rows illustrate the models with parameters

$\{\alpha=0.1, \omega_c=1\}$, $\{\alpha=0.4, \omega_c=1\}$, $\{\alpha=0.1, \omega_c=2.5\}$ and $\{\alpha=0.4, \omega_c=2.5\}$, respectively. In panels (a1), (b1), (c1) and (d1), the green, blue and cyan solid lines represent the results produced by NaF-cc, NaF-cx and NaF-TW, respectively. Panels (a2), (b2), (c2) and (d2), the cyan, pink, orange and magenta solid lines denote the results produced by NaF-TW, NaF-TW2, NaF-HW and NaF-HW2, respectively. In panels (a3), (b3), (c3) and (d3), the orange dashed lines, purple dashed lines, brown dashed lines and cyan solid lines denote the results produced by SH-1, SH-2, SH-3 and NaF-TW, respectively. The numerically exact results produced by eHEOM[122-127, 383] are demonstrated by black points in each panel.

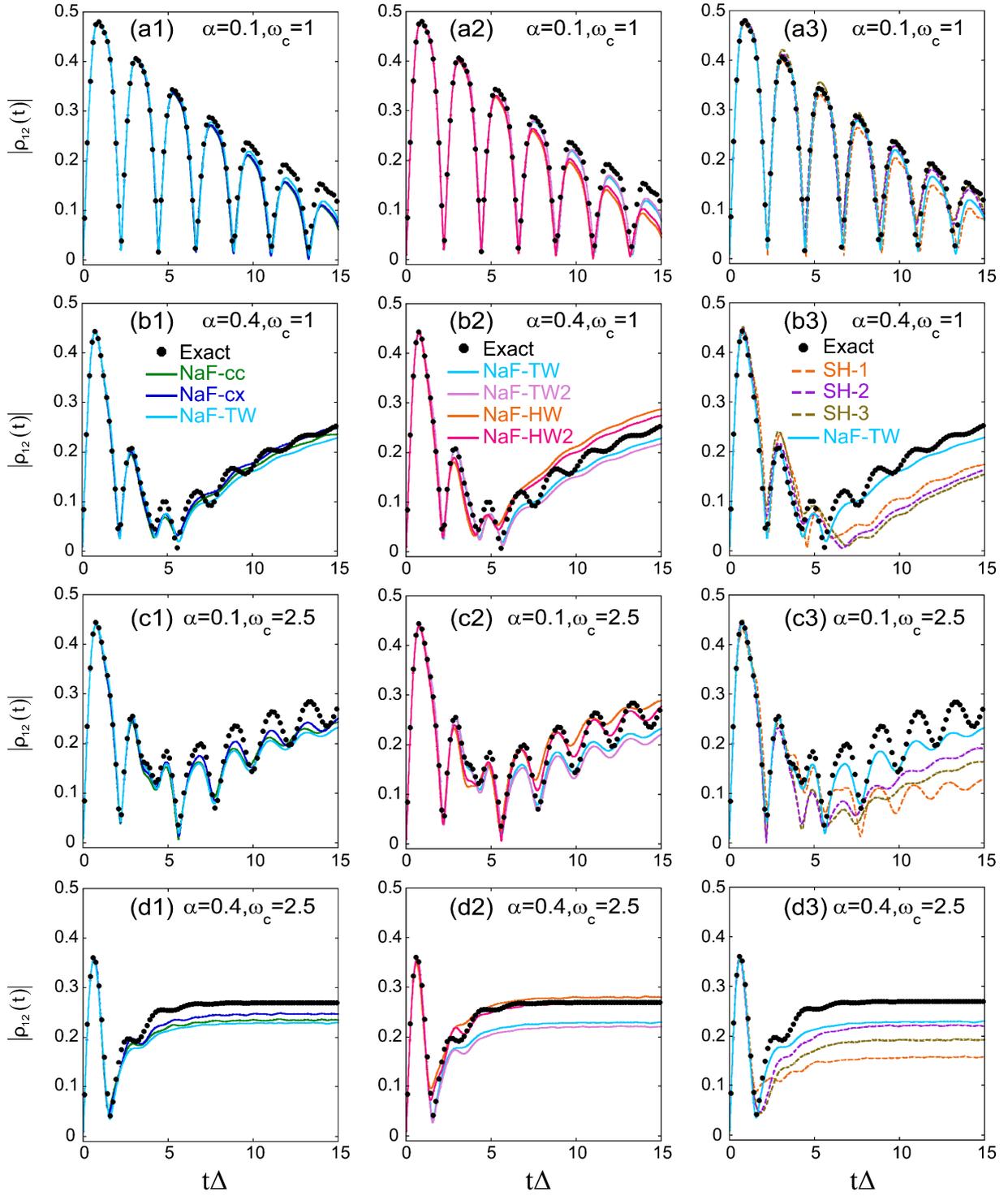

**Figure 10.** Similar to Figure 9, but illustrates the results of the modulus of the off-diagonal term $|\rho_{12}(t)|$.

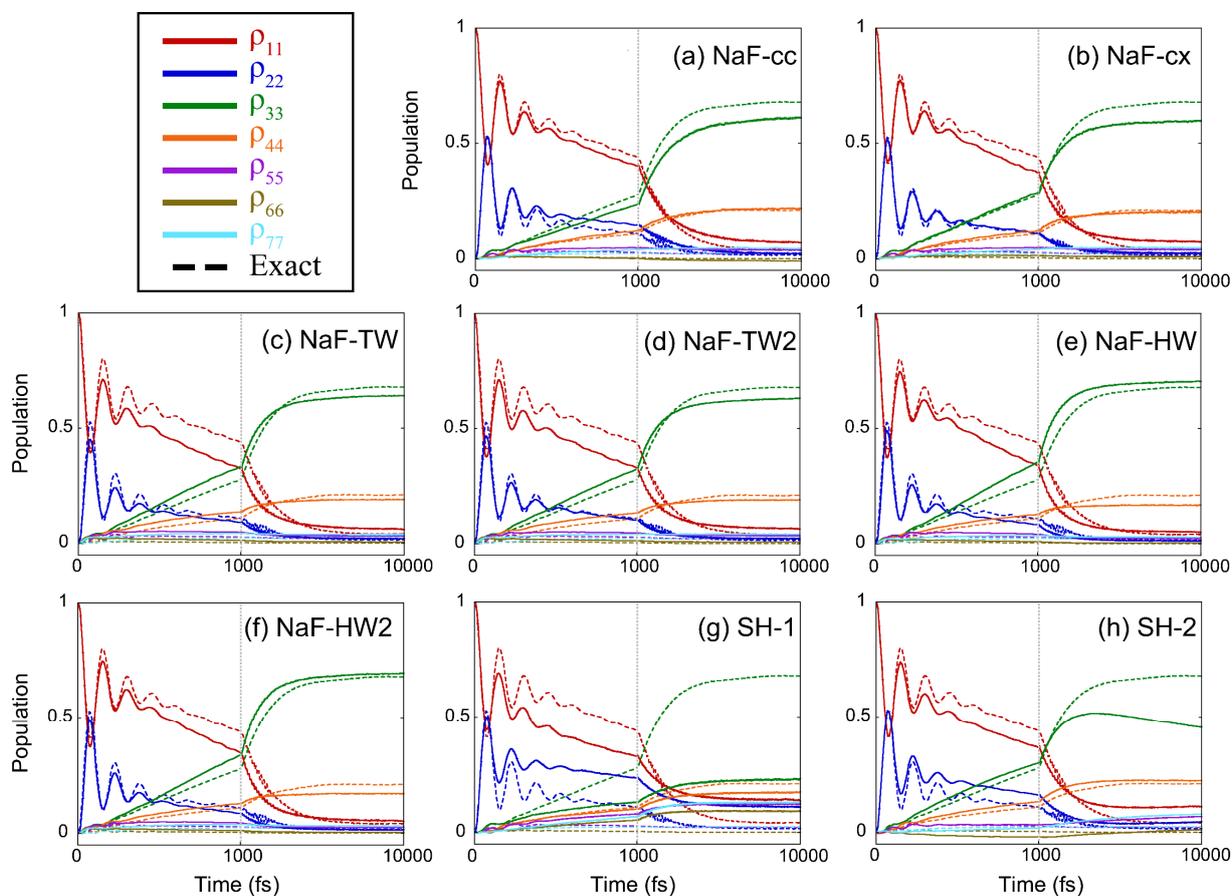

**Figure 11.** Results of population dynamics of the 7-state FMO model at 77 K. 50 nuclear (bath) modes in the discretization scheme are employed for each state in the simulations. In each panel, the red, blue, green, orange, purple, brown and cyan solid lines represent the population of states 1-7, respectively. Panel (a): NaF-cc; Panel (b): NaF-cx; Panel (c): NaF-TW; Panel (d): NaF-TW2; Panel (e): NaF-HW; Panel (f): NaF-HW2; Panel (g): SH-1; Panel (h): SH-2. Note that SH-3 is not applicable for this 7-state model. The numerically exact results produced by TD-DMRG[151-160] for the same effective Hamiltonian in the discretization scheme are demonstrated by dashed lines with corresponding colors in each panel.

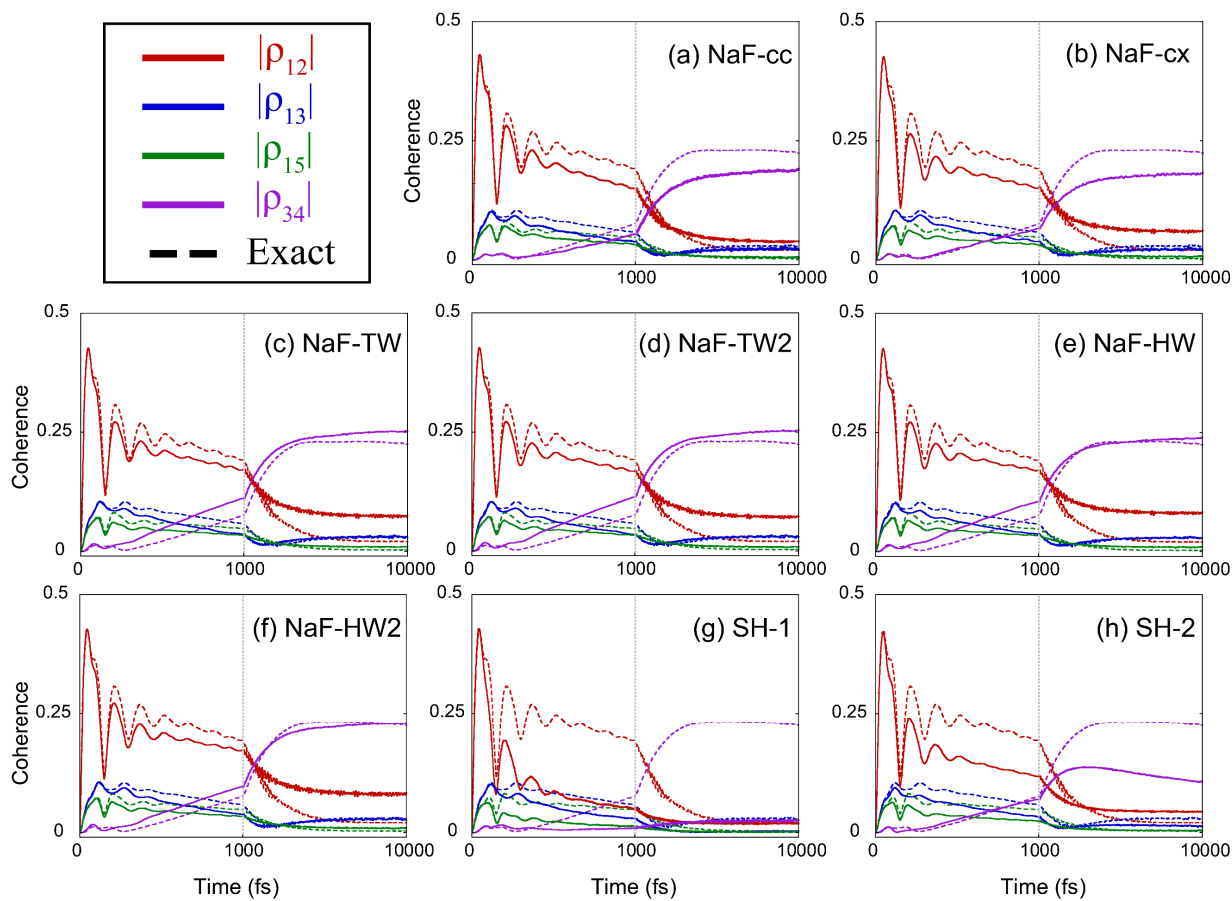

**Figure 12.** Results of the coherence dynamics of the 7-state FMO model at 77 K. 50 nuclear (bath) modes in the discretization scheme are employed for each state in the simulations. In each panel, the red, blue, green, and purple solid lines represent the moduli of the off-diagonal terms $\rho_{12}$, $\rho_{13}$, $\rho_{15}$ and $\rho_{34}$, respectively. Panel (a): NaF-cc; Panel (b): NaF-cx; Panel (c): NaF-TW; Panel (d): NaF-TW2; Panel (e): NaF-HW; Panel (f): NaF-HW2; Panel (g): SH-1; Panel (h): SH-2. Note that SH-3 is not applicable for this 7-state model. The numerically exact results produced by TD-DMRG[151-160] for the same effective Hamiltonian in the discretization scheme are demonstrated by dashed lines with corresponding colors in each panel.

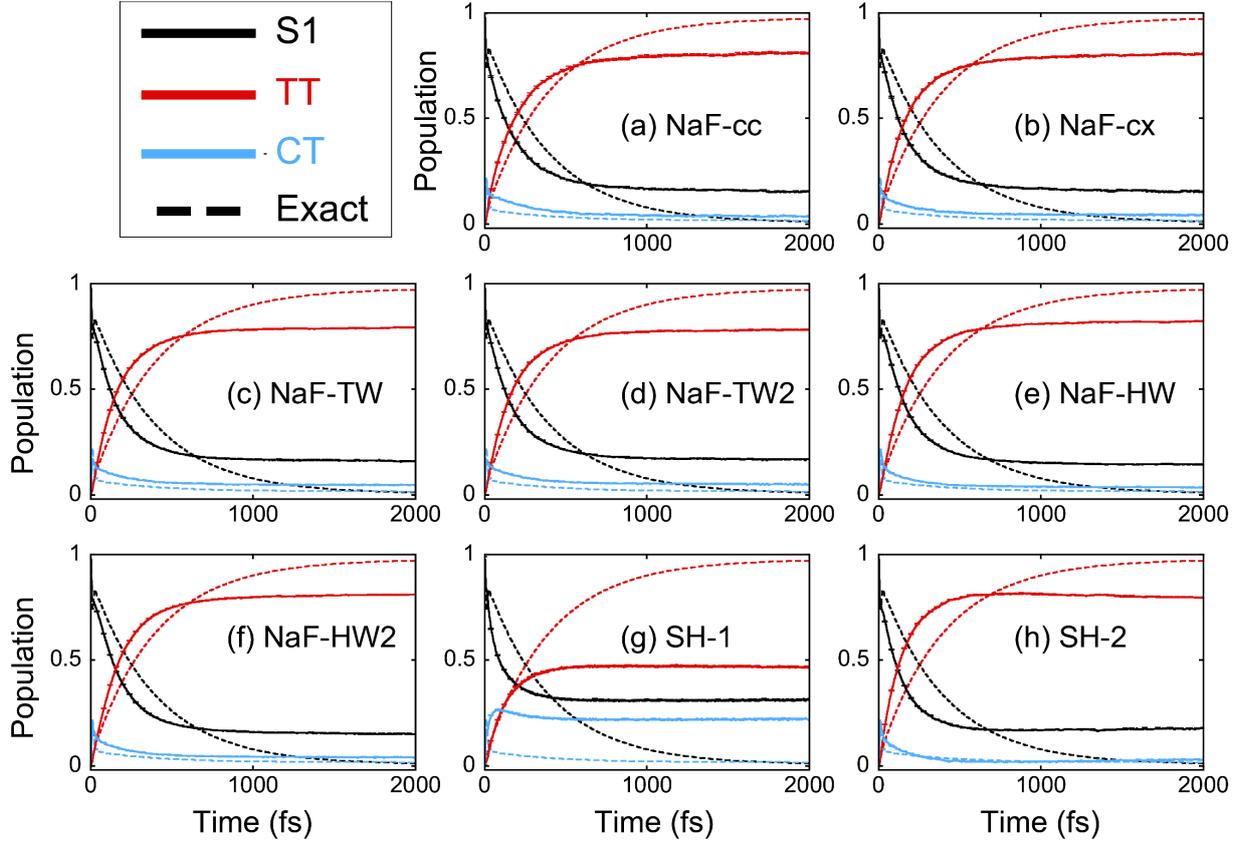

**Figure 13.** Population dynamics of the SF model. 200 nuclear modes are employed for each state in the simulations. In each panel, the black, red and cyan solid lines represent the population of the S1, TT and CT state, respectively. Panel (a): NaF-cc; Panel (b): NaF-cx; Panel (c): NaF-TW; Panel (d): NaF-TW2; Panel (e): NaF-HW; Panel (f): NaF-HW2; Panel (g): SH-1; Panel (h): SH-2. Note that SH-3 is not applicable for this 3-state model. The numerically exact results produced by HEOM[122-126] are demonstrated by dashed lines with corresponding colors in each panel.

### 3.4 Cavity Quantum Electrodynamics Processes

We employ the atom-in-cavity models in refs [391, 392] to test NaF methods on cavity quantum electrodynamics (cQED) processes[211, 216, 391-411]. Similar to the system-bath models, the total

Hamiltonian of the atom-in-cavity models can be decomposed into three parts. The atomic part is expensed by its eigenstates:

$$\hat{H}_a = \sum_{n=1}^{F} \varepsilon_n |n\rangle\langle n|, \qquad (99)$$

where $\varepsilon_n$ denotes the $n$-th corresponding atomic energy level. The optical field part reads

$$\hat{H}_p = \sum_{j=1}^{N_{nuc}} \omega_j \left(\hat{a}_j^\dagger \hat{a}_j + 1/2\right), \qquad (100)$$

where $\hat{a}_j^\dagger$ ($\hat{a}_j$) denotes the creation (annihilation) operator of the $j$-th optical field mode, and $\omega_j$ represents the corresponding photonic frequency. The coupling term is represented using the dipole approximation as

$$\hat{H}_c = \sum_{j=1}^{N_{nuc}} \sum_{n \neq m}^{F} \sqrt{\frac{\omega_j}{2}} \left(\hat{a}_j^\dagger + \hat{a}_j\right) \lambda_j(r_0) \mu_{nm} |n\rangle\langle m|, \qquad (101)$$

where $\mu_{nm}$ denotes the transitional dipole moment between the atomic eigenstates $|n\rangle$ and $|m\rangle$, and $\lambda_j(r_0) = \sqrt{2/\varepsilon_0 L} \sin(j\pi r_0/L)$ $(j=1,\cdots,N)$ denotes the atom-optical field interaction. Here $L = 2.362 \times 10^5$ au, $\varepsilon_0$ and $r_0 = L/2$ denote the volume length of the cavity, the vacuum permittivity, and the location of the atom, respectively. We utilize $N_{nuc} = 400$ optical field modes with the frequency $\omega_j = j\pi c/L$, where $c = 137.036$ au denotes the light speed in vacuum.

Two models are considered in this work, one is the two-state model with the atomic energy levels $\varepsilon_1 = -0.6738$, $\varepsilon_2 = -0.2798$, and the dipole moments $\mu_{12} = -1.034$. The other model is the three-state model, which extends from the two-state model with an additional atomic energy level $\varepsilon_3 = -0.1547$ and dipole moment $\mu_{23} = -2.536$ (all in atomic units). The system is initially located at the highest atomic eigenstate, while each optical field mode is in the corresponding optical vacuum state (ground state).

When trajectory-based methods are employed for simulating these two models, it is more convenient to recast the optical field modes by their canonical coordinates and momentum in the diabatic representation:

$$\hat{R}_j = \sqrt{\frac{1}{2\omega_j}}\left(\hat{a}_j^\dagger + \hat{a}_j\right), \hat{P}_j = i\sqrt{\frac{\omega_j}{2}}\left(\hat{a}_j^\dagger - \hat{a}_j\right) \qquad (102)$$

and treating $\{\hat{R}_j, \hat{P}_j\}$ as continuous nuclear DOFs. This implies that one can map the optical field modes onto the Wigner phase space and employ trajectory-based methods to study the evolution. Geva and co-workers have applied the Meyer-Miller mapping model approach to study the same systems[211]. Some more tests have also been performed in refs [11, 12, 320, 322, 324]. Here we compare the results obtained by the NaF and SH methods with the numerically exact results yielded by truncated configuration interaction in refs [391, 392].

Figures 14-15 illustrate the results of all atom-in-cavity models. All SH methods exhibit poor performance in both the short-time spontaneous emission process and the re-absorption/re-emission process at around 1800 au. In contrast, all NaF methods offer a much better description of spontaneous emission as well as the re-absorption/re-emission processes, and predict more accurate population results in the plateau region.

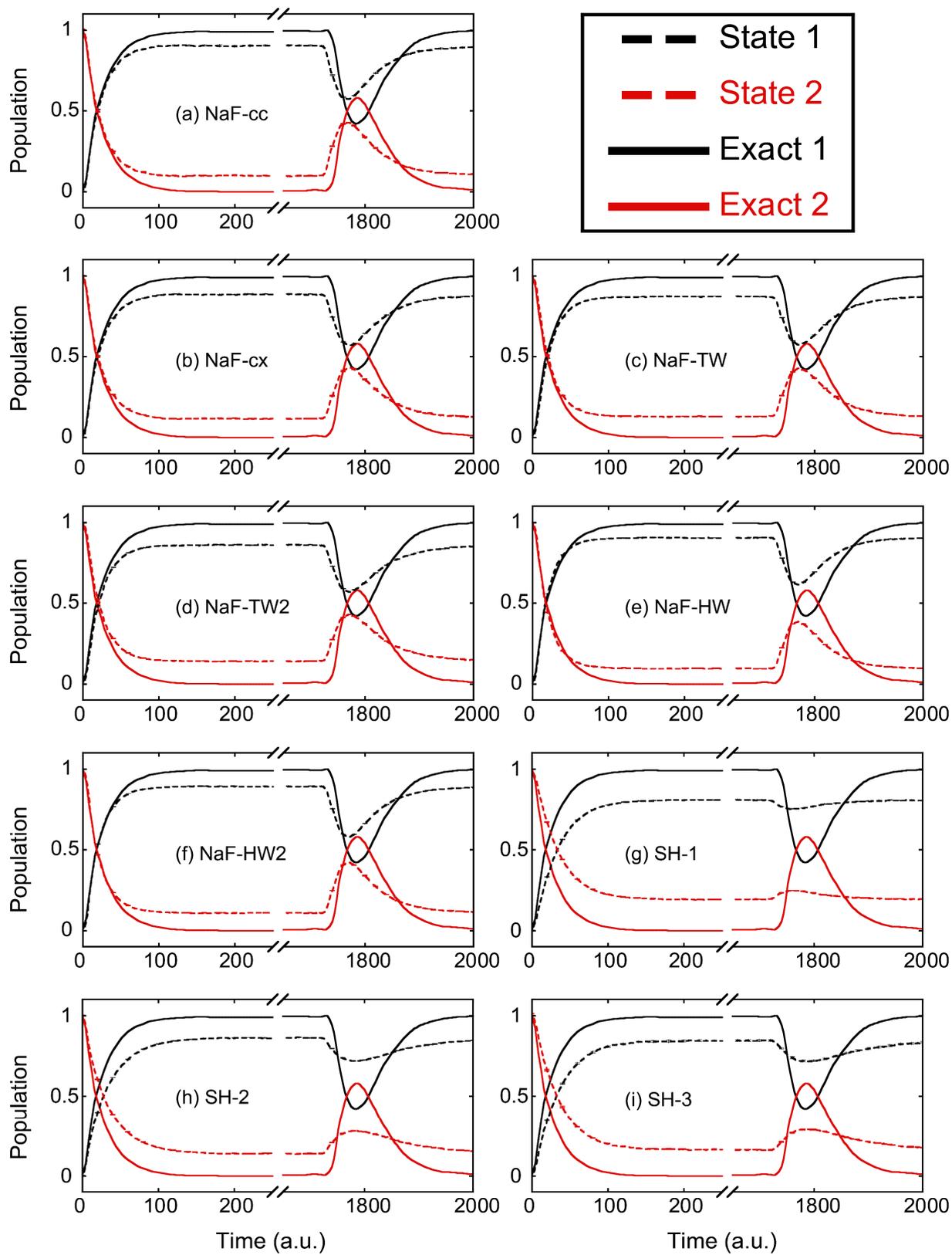

**Figure 14.** Population dynamics of the 2-state atom-in-cavity model with 400 standing wave modes. In each panel, the black and red dashed lines represent the population of the atomic ground and excited state, respectively. Panel (a): NaF-cc; Panel (b): NaF-cx; Panel (c): NaF-TW; Panel (d): NaF-TW2; Panel (e): NaF-HW; Panel (f): NaF-HW2; Panel (g): SH-1; Panel (h): SH-2; Panel (i): SH-3. The numerically exact results produced by truncated configuration interaction (taken from refs [391, 392]) are demonstrated by solid lines with corresponding colors in each panel.

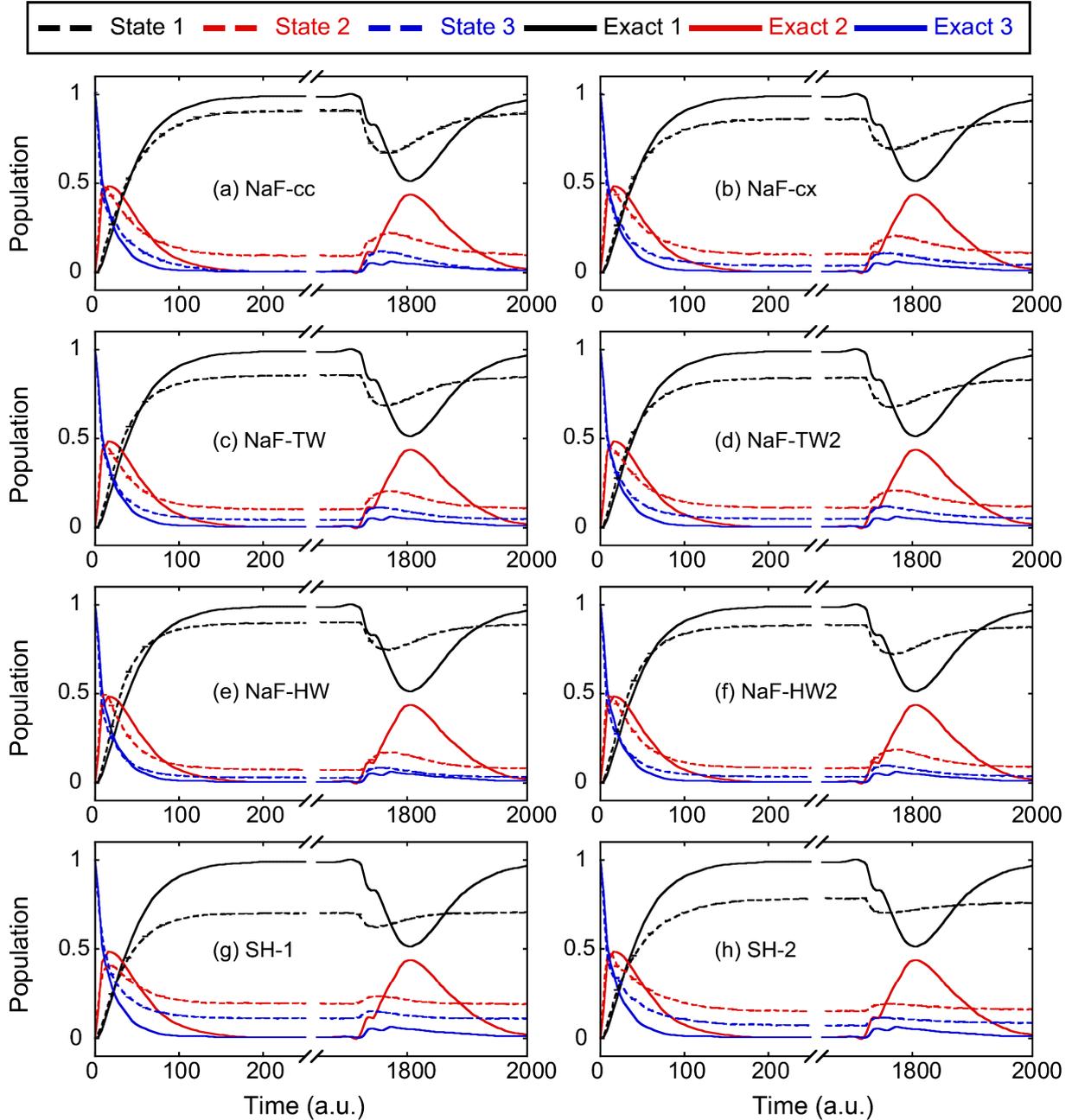

**Figure 15.** Population dynamics of the 3-state atom-in-cavity model with 400 standing wave modes. In each panel, the black, red and blue dashed lines represent the population of the first, second and third atomic state, respectively. Panel (a): NaF-cc; Panel (b): NaF-cx; Panel (c): NaF-TW; Panel (d): NaF-TW2; Panel (e): NaF-HW; Panel (f): NaF-HW2; Panel (g): SH-1; Panel (h): SH-2. Note that SH-3 is not applicable for this 3-state model. The numerically exact results

produced by truncated configuration interaction (taken from refs [391, 392]) are demonstrated by solid lines with corresponding colors in each panel.

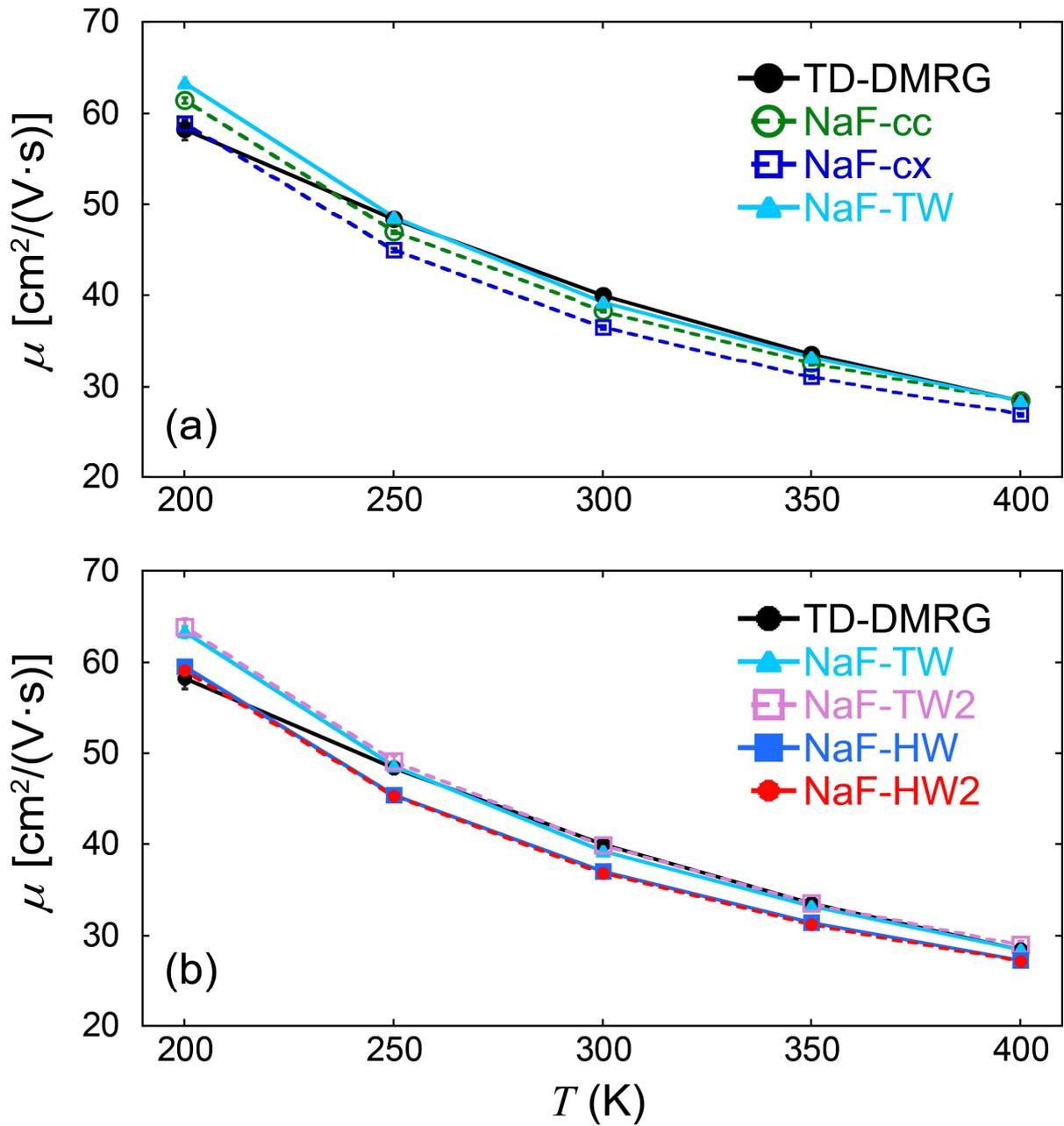

**Figure 16**. Benchmark results of $\mu$ of the one-dimensional Holstein model in ref [412] as functions of temperature. In panel (a), the green hollow circles with green dashed lines, blue hollow squares

with blue dashed lines and cyan triangles with cyan solid lines represent the results of NaF-cc; NaF-cx and NaF-TW, respectively. In panel (b), the cyan triangles with cyan solid lines, pink hollow squares with pink dashed lines, blue squares with blue solid lines, red points with red dashed lines represent the results of NaF-TW, NaF-TW2, NaF-HW and NaF-HW2, respectively. In each panel, black points with black solid lines denote the results of TD-DMRG. (Please see more details in Section S5 of the Supporting Information.) Figure S2(c) of the Supporting Information shows that, when the initial nuclear condition is sampled by classical mechanics, the results significantly deviate from the TD-DMRG results. It suggests that nuclear quantum effects are important for this system. Consistent definitions of the electronic coherence-coherence TCF for multistate systems, which satisfy the frozen nuclei limit, are not available for these SH methods in the literature[253, 294, 295]. So it will be fair to compare SH methods to TD-DMRG and NaF methods in the future.

# 4 Conclusions

We have recently developed NaF, a conceptually novel nonadiabatic dynamics approach with independent trajectories on quantum phase space with coordinate-momentum variables, where CPS is used for discrete electronic-state DOFs and Wigner phase space is employed for nuclear DOFs. In the paper we develop an exact integrator for updating the nuclear kinematic momentum from the contribution of the effective nonadiabatic force for a finite time-step, which is included in the efficient algorithm for NaF in the adiabatic representation described in Section 2.3. When the diabatic representation is available, two additional efficient algorithms that take advantage of the diabatic representation are proposed in Sections S2-S3 of the Supporting Information. We implement these efficient NaF algorithms with MPI and Athread for the heterogeneous platform.

We show a few expressions of (electronic) TCFs[195, 318, 319, 324, 326] in quantum phase space representation, then apply them to NaF. These NaF methods are employed to investigate a suite of benchmark model systems, ranging from low-dimensional gas-phase models to condensed-phase complex systems, and are compared with numerically exact approaches and three SH methods. NaF methods are robust and competent in presenting reasonable descriptions for both electronic and nuclear motion in the nonadiabatic coupling region as well as the asymptotic region. As demonstrated in Figures 1-4 for the LVCMs for molecular CIs, Figures 5-6 for one-dimensional scattering models of Tully, and Figure 7 for one-dimensional three-state photodisassociation models of Miller and co-workers and Figure 8 for a two-state anharmonic model, the overall performance of NaF methods is comparable to that of SH methods for gas-phase systems. More importantly, NaF methods are superior to SH methods in high-dimensional (condensed-phase) systems where the nonadiabatic coupling region is wide or the states stay coupled all the time, as shown in Figures 9-10 for spin-boson models, Figures 11-12 for the 7-site FMO model, Figure 13 for the three-state SF model, and Figures 14-15 for cavity-induced chemical processes. The benchmark tests shown in the present paper as well as in ref [324] and Section S7 of the Supporting Information[340] of ref [322] imply that, although the phase space expressions of (electronic) TCFs may also be used with various surface hopping or Ehrenfest-like dynamics methods, NaF should always be highly recommended for its superiority.

The suite of benchmark model tests also indicates that the performance of NaF is relatively insensitive to the phase space expression of electronic TCFs. In addition, Figure 16 investigates the Holstein model for the carrier mobility of organic semiconductors, which one of us already studied in ref [412] in 2020. The isomorphism of ref [317] suggests that the Holstein model can be studied by NaF and other nonadiabatic dynamics methods. Figure 16 compares the NaF results related to the electronic coherence-coherence TCF and those benchmark results produced by TD-DMRG, for the same initial condition in quantum mechanics. Figure 16 shows that NaF yields reasonable results in a wide temperature regime. In addition, as shown in Figure S4 of Section S3 of the Supporting Information[351] of ref [324], NaF methods are capable of describing the electron transfer rate that is also related to the electronic coherence-coherence TCF, which reproduce the Marcus theory[413-415] in its valid region. Among all the NaF methods, NaF-TW is recommended due to its positive semidefinite TCF of population dynamics and the overall reasonable accuracy for various model systems. It is expected that NaF-TW can be performed for simulations of nonadiabatic transition processes of real molecular systems with *ab initio* on-the-fly calculations[97, 416-427] or machine learning approaches[237, 428-433]. It will be intriguing to use NaF to study electronically (or vibrationally) nonadiabatic processes involving numerous DOFs, such as electron/hole/positron/proton/hydrogen transfer and matter/energy transport in bulk or at interfaces in complex systems, especially where the nonadiabatic coupling region is broad, in chemistry, physics, materials, biology, and so forth. A simulation package for NaF and other nonadiabatic molecular dynamics methods will be reported by us soon[434].

We note that, despite its overall robust superior performance for various nonadiabatic/composite systems, NaF is not adequate to describe deep quantum tunneling effects or true quantum recurrence/coherence/resonance effects, as implied in Figure S7 of Section S8 of the Supporting Information. This is because NaF only involves the independent trajectory without the phase. When the NaF trajectory with its associated phase is included in the semiclassical analogy approach or time-dependent multiconfiguration approach, it will offer potential tools for faithfully describing these more challenging quantum mechanical effects with more computational effort, although it is expected that such effects are often quenched in large molecular systems.

Finally, we suggest that the suite of gas phase and condensed phase benchmark models in the main text and the Supporting Information of the paper should be important for testing various developed or new practical nonadiabatic molecular dynamics methods in the community. Because numerically exact results are available for these benchmark models that represent various regions (or limits) of nonadiabatic transition processes, it is clean and clear to illustrate whether both electronic motion and nuclear quantum effects are reasonably described by the practical method that is designed for *ab initio* nonadiabatic quantum molecular dynamics. It will also be useful to include more typical representative benchmark tests in the suite. Such a suite will help develop consistent practical approaches that faithfully capture the main features of the **quantum mechanical behavior of both electrons and nuclei** in nonadiabatic transition processes in real complex molecular systems.

ASSOCIATED CONTENT

**Supporting Information**.

Supporting Information includes ten sections: Numerical integrator for effective nonadiabatic force of NaF; Further details of the numerical integrator of NaF; Nuclear force of NaF in the diabatic representation; Derivation of the covariant-covariant TCF with action-angle variables; Simulation details of one-dimensional Holstein model; Details of TD-DMRG simulations; Additional numerical results for the 3-state photodissociation models and for the FMO model at zero temperature; Transition path flight time simulation; Relation between the EOMs of NaF and the exact EOMs in the generalized coordinate-momentum phase space formulation of quantum mechanics; Electronic TCF of the weighted mapping model with the Born-Oppenheimer limit.

AUTHOR INFORMATION
**Corresponding Author**

*E-mail: jianliupku@pku.edu.cn

**ORCID**

Baihua Wu: 0000-0002-1256-6859

Bingqi Li: 0000-0001-7273-8299

Xin He: 0000-0002-5189-7204

Xiangsong Cheng: 0000-0001-8793-5092

Jiajun Ren: 0000-0002-1508-4943

Jian Liu: 0000-0002-2906-5858

Author Contributions

Baihua Wu, Bingqi Li and Xin He contributed equally to this work.

Funding Sources

National Science Fund for Distinguished Young Scholars (Grant No. 22225304)

Notes

The authors declare no competing financial interest.

ACKNOWLEDGMENT

We thank Youhao Shang, Bill Miller, Wolfgang Domcke, Eli Pollak, Haocheng Lu, and Yu Zhai for helpful discussions. We also thank Martha Yaghoubi Jouybari and Fabrizio Santoro for providing the parameters of the LVCM for Thymine in ref [362]. This work was supported by the National Science Fund for Distinguished Young Scholars Grant No. 22225304. We acknowledge the High-performance Computing Platform of Peking University, Beijing PARATERA Tech Co., Ltd., and Guangzhou Supercomputer Center for providing computational resources. We also thank the Laoshan Laboratory (LSKJ202300305) for providing the computational resources of the new Sunway platform and for the technical support.

# ABBREVIATIONS AND NOTATIONS

| Abbreviation and Notations | Complete Terminology |
|:---:|:---:|
| BO | Born-Oppenheimer |
| $\overline{C}_{nm,kl}(t)$ | time-dependent normalization factor tensor |
| cc | covariant-covariant |
| CMM | classical mapping model |
| CMMcv | classical mapping model with commutator variables |
| CPS | constraint coordinate-momentum phase space |
| cx | covariant-noncovariant |
| $d\mu(\mathbf{x},\mathbf{p})$ | integral measure of electronic phase space variables |
| $D_{nm,kl}(t)$ | time correlation function of electronic degrees of freedom |
| $D_{nm,kl}(t)$ ($n=m, k=l$) | (electronic) population-population time correlation function |
| $D_{nm,kl}(t)$ ($n=m, k\neq l$) | (electronic) population-coherence time correlation function |
| $D_{nm,kl}(t)$ ($n\neq m, k=l$) | (electronic) coherence-population time correlation function |
| $D_{nm,kl}(t)$ ($n\neq m, k\neq l$) | (electronic) coherence-coherence time correlation function |
| DAC | dual avoided crossing |
| DVR | discrete variable representation |
| DOF | degree of freedom |
| ECR | extended coupling region |
| eHEOM | extended hierarchy equations of motion |
| EOM | equation of motion |
| FMO | Fenna-Matthews-Olson |
| **g** | vector $\mathbf{x} + i\mathbf{p}$ |
| $\gamma$ | parameter of constraint coordinate-momentum phase space |
| $\boldsymbol{\Gamma}$ | commutator matrix of electronic degrees of freedom |
| $H_{\text{NaF}}(\mathbf{R}, \mathbf{P}, \tilde{\boldsymbol{\rho}})$ | mapping energy on quantum phase space of nonadiabatic field |
| HEOM | hierarchy equations of motion |
| HWF | hill window function |
| $\hat{K}_{\text{ele}}(\mathbf{x}, \mathbf{p}, \boldsymbol{\Gamma})$ | mapping kernel of electronic degrees of freedom |
| $\hat{K}_{\text{ele}}^{-1}(\mathbf{x}, \mathbf{p}, \boldsymbol{\Gamma})$ | inverse mapping kernel of electronic degrees of freedom |
| $\hat{K}_{\text{nuc}}(\mathbf{R}, \mathbf{P})$ | mapping kernel of nuclear degrees of freedom |
| $\hat{K}_{\text{nuc}}^{-1}(\mathbf{R}, \mathbf{P})$ | inverse mapping kernel of nuclear degrees of freedom |
| LSC-IVR | linearized semiclassical initial value representation |
| LVCM | linear vibronic coupling model |
| MCTDH | multiconfiguration time-dependent Hartree |
| ML-MCTDH | multilayer multiconfiguration time-dependent Hartree |
| MM | Meyer-Miller |
| NaF | nonadiabatic field |
| NaF-cc | nonadiabatic field with covariant-covariant time correlation functions |
| NaF-cx | nonadiabatic field with covariant-noncovariant time correlation functions |
| NaF-HW | nonadiabatic field with hill window functions |
| NaF-HW2 | nonadiabatic field with hill window functions-2 |
| NaF-TW | nonadiabatic field with triangle window functions |
| NaF-TW2 | nonadiabatic field with triangle window functions-2 |
| PES | potential energy surface |
| $\bar{Q}_{nm,kl}(\mathbf{x},\mathbf{p},\mathbf{\Gamma};\boldsymbol{\gamma};t)$ | integrand of electronic time correlation function in constraint coordinate-momentum phase space representation |
| $\tilde{\boldsymbol{\rho}}(\tilde{\mathbf{x}},\tilde{\mathbf{p}},\tilde{\mathbf{\Gamma}})$ | effective electronic density matrix in the adiabatic representation |
| RHS | right-hand side |
| $\mathcal{S}(\mathbf{x},\mathbf{p};\gamma)$ | manifold of constraint coordinate-momentum phase space |
| SAC | single avoided crossing |
| SF | singlet fission |
| SH | surface hopping |
| SQC | symmetrical quasi-classical |
| TCF | time correlation function |
| TD-DMRG | time-dependent density matrix renormalization group |
| TWF | triangle window function. |
| $\mathbf{U}(\mathbf{R},\Delta t)$ | electronic propagator in the diabatic representation |
| $\tilde{\mathbf{U}}(\mathbf{R},\Delta t)$ | electronic propagator in the adiabatic representation |
| $w(\gamma)$ | normalized weight (quasi-probability distribution) function of constraint coordinate-momentum phase space |

# Supporting Information

## S1. Numerical Integrator for Effective Nonadiabatic Force of NaF

This section consists of two parts. The first part is the derivation of the effective EOMs corresponding to the integrator in refs [322, 324] (also listed in Section S2) in the infinitesimal time-step limit $\Delta t \to 0$ when no switching between electronic states occurs. The second part is the derivation of the analytical solution as well as numerical integrator for updating the nuclear momentum due to the contribution of the effective nonadiabatic nuclear force for a finite time step $\Delta t$.

**1) EOMs for the NaF integrator in the main text without switching between electronic states**

When no switching between electronic states occurs, the EOMs presented in eqs (25), (26) and (35) of the main text become

$$\begin{cases} \dot{\mathbf{P}} = -\nabla_{\mathbf{R}} E_{j_{occ}}(\mathbf{R}) - \sum_{n \neq m}^{F} \left(E_n(\mathbf{R}) - E_m(\mathbf{R})\right) \mathbf{d}_{mn}(\mathbf{R}) \tilde{\rho}_{nm} \\ \dot{\mathbf{R}} = \mathbf{M}^{-1}\mathbf{P} \\ \dot{\tilde{\mathbf{g}}} = -i\mathbf{V}^{(eff)}(\mathbf{R},\mathbf{P})\tilde{\mathbf{g}} \\ \dot{\tilde{\boldsymbol{\Gamma}}} = -i\mathbf{V}^{(eff)}(\mathbf{R},\mathbf{P})\tilde{\boldsymbol{\Gamma}} + i\tilde{\boldsymbol{\Gamma}}\mathbf{V}^{(eff)}(\mathbf{R},\mathbf{P}) \end{cases} \quad (S103)$$

where $\mathbf{V}^{(eff)}(\mathbf{R},\mathbf{P})$ and $\tilde{\boldsymbol{\rho}}$ are defined in eq (27) and eq (34) of the main text, respectively. Suppose that phase space variables $(\mathbf{R}_t, \mathbf{P}_t, \tilde{\boldsymbol{\rho}}_t)$ evolve to $(\mathbf{R}_{mid}, \mathbf{P}_{mid}, \tilde{\boldsymbol{\rho}}_{mid})$, as governed by eq (S103) with the *infinitesimal* time-step $dt$. Equation (S103) is equivalent to

$$\begin{cases} \mathbf{P}_{mid} = \mathbf{P}_t - \left[\nabla_{\mathbf{R}} E_{j_{occ}}(\mathbf{R}_t) + \sum_{n \neq m}^{F} \left(E_n(\mathbf{R}_t) - E_m(\mathbf{R}_t)\right) \mathbf{d}_{mn}(\mathbf{R}_t) \tilde{\rho}_{nm}\right] dt \\ \mathbf{R}_{mid} = \mathbf{R}_t + \mathbf{M}^{-1}\mathbf{P}_t dt, \\ \tilde{\mathbf{g}}_{mid} = \tilde{\mathbf{g}}_t - i\mathbf{V}^{(eff)}(\mathbf{R}_t,\mathbf{P}_t)\tilde{\mathbf{g}}_t dt, \\ \tilde{\boldsymbol{\Gamma}}_{mid} = \tilde{\boldsymbol{\Gamma}}_t - i\mathbf{V}^{(eff)}(\mathbf{R}_t,\mathbf{P}_t)\tilde{\boldsymbol{\Gamma}}_t dt + i\tilde{\boldsymbol{\Gamma}}_t \mathbf{V}^{(eff)}(\mathbf{R}_t,\mathbf{P}_t) dt, \end{cases} \quad (S104)$$

It is straightforward to obtain

$$\begin{aligned}
&H_{\text{NaF}}\left(\mathbf{R}_{\text{mid}}, \mathbf{P}_{\text{mid}}, \tilde{\boldsymbol{\rho}}_{\text{mid}}\right) \\
&= \mathbf{P}_{\text{mid}}^{\text{T}} \mathbf{M}^{-1} \mathbf{P}_{\text{mid}} / 2 + E_{j_{\text{occ}}}\left(\mathbf{R}_{\text{mid}}\right) \\
&= \mathbf{P}_t^{\text{T}} \mathbf{M}^{-1} \mathbf{P}_t / 2 - \mathbf{P}_t^{\text{T}} \mathbf{M}^{-1} \left( \nabla_{\mathbf{R}} E_{j_{\text{occ}}}\left(\mathbf{R}_t\right) + \sum_{n \neq m}^{F} \left(E_n\left(\mathbf{R}_t\right) - E_m\left(\mathbf{R}_t\right)\right) \mathbf{d}_{mn}\left(\mathbf{R}_t\right) \tilde{\rho}_{nm,t} \right) dt \\
&\quad + E_{j_{\text{occ}}}\left(\mathbf{R}_t + \mathbf{M}^{-1} \mathbf{P}_t dt\right) \\
&= \mathbf{P}_t^{\text{T}} \mathbf{M}^{-1} \mathbf{P}_t / 2 - \mathbf{P}_t^{\text{T}} \mathbf{M}^{-1} \left( \nabla_{\mathbf{R}} E_{j_{\text{occ}}}\left(\mathbf{R}_t\right) + \sum_{n \neq m}^{F} \left(E_n\left(\mathbf{R}_t\right) - E_m\left(\mathbf{R}_t\right)\right) \mathbf{d}_{mn}\left(\mathbf{R}_t\right) \tilde{\rho}_{nm,t} \right) dt \quad . \quad (\text{S105}) \\
&\quad + E_{j_{\text{occ}}}\left(\mathbf{R}_t\right) + \mathbf{P}_t^{\text{T}} \mathbf{M}^{-1} \nabla_{\mathbf{R}} E_{j_{\text{occ}}}\left(\mathbf{R}_t\right) dt \\
&= \mathbf{P}_t^{\text{T}} \mathbf{M}^{-1} \mathbf{P}_t / 2 + E_{j_{\text{occ}}}\left(\mathbf{R}_t\right) - \mathbf{P}_t^{\text{T}} \mathbf{M}^{-1} \sum_{n \neq m}^{F} \left(E_n\left(\mathbf{R}_t\right) - E_m\left(\mathbf{R}_t\right)\right) \mathbf{d}_{mn}\left(\mathbf{R}_t\right) \tilde{\rho}_{nm,t} dt \\
&= H_{\text{NaF}}\left(\mathbf{R}_t, \mathbf{P}_t, \tilde{\boldsymbol{\rho}}_t\right) - \mathbf{P}_t^{\text{T}} \mathbf{M}^{-1} \sum_{n \neq m}^{F} \left(E_n\left(\mathbf{R}_t\right) - E_m\left(\mathbf{R}_t\right)\right) \mathbf{d}_{mn}\left(\mathbf{R}_t\right) \tilde{\rho}_{nm,t} dt
\end{aligned}$$

In the energy-conservation procedure of the integrator of NaF in refs [322, 324], the nuclear kinematic momentum is rescaled along its own direction, i.e.,

$$\left(\mathbf{R}_{t+dt}, \mathbf{P}_{t+dt}, \tilde{\boldsymbol{\rho}}_{t+dt}\right) = \left(\mathbf{R}_{\text{mid}}, k_{\text{mid}} \mathbf{P}_{\text{mid}}, \tilde{\boldsymbol{\rho}}_{\text{mid}}\right) , \quad (\text{S106})$$

to ensure that

$$H_{\text{NaF}}\left(\mathbf{R}_{t+dt}, \mathbf{P}_{t+dt}, \tilde{\boldsymbol{\rho}}_{t+dt}\right) = H_{\text{NaF}}\left(\mathbf{R}_t, \mathbf{P}_t, \tilde{\boldsymbol{\rho}}_t\right) . \quad (\text{S107})$$

Because any terms that include $(dt)^n$ with $n \geq 2$ should disappear when we consider the first order of $dt$, we obtain from eq (S104)

$$\begin{aligned}
&\mathbf{P}_t^{\text{T}} \mathbf{M}^{-1} \sum_{n \neq m}^{F} \left(E_n\left(\mathbf{R}_t\right) - E_m\left(\mathbf{R}_t\right)\right) \mathbf{d}_{mn}\left(\mathbf{R}_t\right) \tilde{\rho}_{nm,t} dt \\
&= \mathbf{P}_{\text{mid}}^{\text{T}} \mathbf{M}^{-1} \sum_{n \neq m}^{F} \left(E_n\left(\mathbf{R}_{\text{mid}}\right) - E_m\left(\mathbf{R}_{\text{mid}}\right)\right) \mathbf{d}_{mn}\left(\mathbf{R}_{\text{mid}}\right) \tilde{\rho}_{nm,\text{mid}} dt
\end{aligned} \quad . \quad (\text{S108})$$

Substitution of eqs (S105), (S106) and (S108) into eq (S107) yields

$$\begin{aligned}
&k_{\text{mid}}^2 \mathbf{P}_{\text{mid}}^{\text{T}} \mathbf{M}^{-1} \mathbf{P}_{\text{mid}} / 2 \\
&= \mathbf{P}_{\text{mid}}^{\text{T}} \mathbf{M}^{-1} \mathbf{P}_{\text{mid}} / 2 + \mathbf{P}_{\text{mid}}^{\text{T}} \mathbf{M}^{-1} \sum_{n \neq m}^{F} \left(E_n\left(\mathbf{R}_{\text{mid}}\right) - E_m\left(\mathbf{R}_{\text{mid}}\right)\right) \mathbf{d}_{mn}\left(\mathbf{R}_{\text{mid}}\right) \tilde{\rho}_{nm,\text{mid}} dt
\end{aligned} \quad . \quad (\text{S109})$$

The solution to eq (S109) is

$$k_{\text{mid}} = \sqrt{1 + \frac{2\mathbf{P}_{\text{mid}}^{\text{T}}\mathbf{M}^{-1}\sum_{n\neq m}^{F}\left(E_{n}\left(\mathbf{R}_{\text{mid}}\right)-E_{m}\left(\mathbf{R}_{\text{mid}}\right)\right)\mathbf{d}_{mn}\left(\mathbf{R}_{\text{mid}}\right)\tilde{\rho}_{nm,\text{mid}}\mathrm{d}t}{\mathbf{P}_{\text{mid}}^{\text{T}}\mathbf{M}^{-1}\mathbf{P}_{\text{mid}}}}$$

$$= 1 + \frac{\mathbf{P}_{\text{mid}}^{\text{T}}\mathbf{M}^{-1}\sum_{n\neq m}^{F}\left(E_{n}\left(\mathbf{R}_{\text{mid}}\right)-E_{m}\left(\mathbf{R}_{\text{mid}}\right)\right)\mathbf{d}_{mn}\left(\mathbf{R}_{\text{mid}}\right)\tilde{\rho}_{nm,\text{mid}}\mathrm{d}t}{\mathbf{P}_{\text{mid}}^{\text{T}}\mathbf{M}^{-1}\mathbf{P}_{\text{mid}}}$$

(S110)

Thus, the energy-conservation procedure reads

$$\mathbf{P} + \frac{\mathbf{P}^{\text{T}}\mathbf{M}^{-1}\sum_{n\neq m}^{F}\left(E_{n}(\mathbf{R})-E_{m}(\mathbf{R})\right)\mathbf{d}_{mn}(\mathbf{R})\tilde{\rho}_{nm}\mathrm{d}t}{\mathbf{P}^{\text{T}}\mathbf{M}^{-1}\mathbf{P}}\mathbf{P} \mapsto \mathbf{P} ,$$

(S111)

which corresponds to the following EOM,

$$\dot{\mathbf{P}} = \frac{\mathbf{P}^{\text{T}}\mathbf{M}^{-1}\sum_{n\neq m}^{F}\left(E_{n}(\mathbf{R})-E_{m}(\mathbf{R})\right)\mathbf{d}_{mn}(\mathbf{R})\tilde{\rho}_{nm}}{\mathbf{P}^{\text{T}}\mathbf{M}^{-1}\mathbf{P}}\mathbf{P} .$$

(S112)

After we combine eqs (S103) and (S112), the effective EOMs corresponding to the NaF integrator in refs [322, 324] in the infinitesimal time-step limit then read

$$\begin{cases} \dot{\mathbf{P}} = -\nabla_{\mathbf{R}}E_{j_{\text{occ}}}(\mathbf{R}) - \sum_{n\neq m}^{F}\left(E_{n}(\mathbf{R})-E_{m}(\mathbf{R})\right)\mathbf{d}_{mn}(\mathbf{R})\tilde{\rho}_{nm} \\ \quad + \frac{\mathbf{P}^{\text{T}}\mathbf{M}^{-1}\sum_{n\neq m}^{F}\left(E_{n}(\mathbf{R})-E_{m}(\mathbf{R})\right)\mathbf{d}_{mn}(\mathbf{R})\tilde{\rho}_{nm}}{\mathbf{P}^{\text{T}}\mathbf{M}^{-1}\mathbf{P}}\mathbf{P}, \\ \dot{\mathbf{R}} = \mathbf{M}^{-1}\mathbf{P}, \\ \dot{\tilde{\mathbf{g}}} = -i\mathbf{V}^{(\text{eff})}(\mathbf{R},\mathbf{P})\tilde{\mathbf{g}}, \\ \dot{\tilde{\mathbf{\Gamma}}} = -i\mathbf{V}^{(\text{eff})}(\mathbf{R},\mathbf{P})\tilde{\mathbf{\Gamma}} + i\tilde{\mathbf{\Gamma}}\mathbf{V}^{(\text{eff})}(\mathbf{R},\mathbf{P}) \end{cases}$$

(S113)

In the RHS of the first equation of eq (S113), the last two terms generate the *effective* nonadiabatic force. The EOM of the nuclear momentum with the *effective* nonadiabatic force then reads

$$\dot{\mathbf{P}} = -\sum_{n \neq m}^{F} \left( E_n(\mathbf{R}) - E_m(\mathbf{R}) \right) \mathbf{d}_{mn}(\mathbf{R}) \tilde{\rho}_{nm} + \frac{\mathbf{P}^T \mathbf{M}^{-1} \sum_{n \neq m}^{F} \left( E_n(\mathbf{R}) - E_m(\mathbf{R}) \right) \mathbf{d}_{mn}(\mathbf{R}) \tilde{\rho}_{nm}}{\mathbf{P}^T \mathbf{M}^{-1} \mathbf{P}} \mathbf{P} \ . \quad (S114)$$

The effective nonadiabatic force conserves the total kinetic energy, as it is straightforward to show that eq (S114) leads to

$$\frac{\mathrm{d}}{\mathrm{d}t} \left( \frac{1}{2} \mathbf{P}^T \mathbf{M}^{-1} \mathbf{P} \right) = \mathbf{P}^T \mathbf{M}^{-1} \dot{\mathbf{P}} = 0 \ . \quad (S115)$$

We note that the rescaling direction of $\mathbf{P}$ is, however, not unique in the energy-conservation procedure. For example, eq (S36) in the Supporting Information[351] of ref [324] corresponds to

$$\begin{aligned}\dot{\mathbf{P}} &= -\sum_{n \neq m}^{F} \left( E_n(\mathbf{R}) - E_m(\mathbf{R}) \right) \mathbf{d}_{mn}(\mathbf{R}) \tilde{\rho}_{nm} \\ &+ \frac{(\mathbf{M}^{-1}\mathbf{P}) \cdot \sum_{n \neq m}^{F} \left( E_n(\mathbf{R}) - E_m(\mathbf{R}) \right) \mathbf{d}_{mn}(\mathbf{R}) \tilde{\rho}_{nm}}{\mathbf{P}^T \mathbf{M}^{-2} \mathbf{P}} \mathbf{M}^{-1}\mathbf{P} \end{aligned} , \quad (S116)$$

which also conserves the kinetic energy. We recommend eq (S114) due to its simplicity.

**2) Numerical integrator for the effective nonadiabatic force**

The analytical solution to eq (S114) produces the numerical integrator for updating the nuclear momentum from the contribution of the effective nonadiabatic force, which conserves the total kinetic energy. For convenience, we define

$$\begin{cases} \boldsymbol{\Pi} = \mathbf{M}^{-1/2} \mathbf{P} \\ \mathbf{B} = \mathbf{M}^{-1/2} \sum_{n \neq m} \tilde{\rho}_{nm} \left( E_n(\mathbf{R}) - E_m(\mathbf{R}) \right) \mathbf{d}_{mn}(\mathbf{R}) \end{cases} . \quad (S117)$$

Equation (S114) then becomes

$$\dot{\boldsymbol{\Pi}} = -\mathbf{B} + \frac{\boldsymbol{\Pi} \cdot \mathbf{B}}{\boldsymbol{\Pi} \cdot \boldsymbol{\Pi}} \boldsymbol{\Pi} \ . \quad (S118)$$

When only one nuclear DOF is involved, the EOM of eq (S118) is exactly $\dot{\mathbf{\Pi}}=\mathbf{0}$. Thus, the *effective* nonadiabatic force plays *no* role for the $N_{nuc}=1$ case. We then focus on the cases for $N_{nuc} \geq 2$ and obtain the solution to eq (S118) for a constant vector, $\mathbf{B}$, where $\mathbf{R}$ is not updated during the update of $\mathbf{P}$ or $\mathbf{\Pi}$.

We employ the vector decomposition,

$$\mathbf{\Pi} = \mathbf{\Pi}_{\|} + \mathbf{\Pi}_{\perp} = \alpha_{\|}\mathbf{e}_{\|} + \alpha_{\perp}\mathbf{e}_{\perp} ,  \tag{S119}$$

where $\mathbf{e}_{\|}$ is the unit vector in the direction of $\mathbf{B}$, and its coefficient is

$$\alpha_{\|} = \mathbf{\Pi} \cdot \mathbf{e}_{\|} . \tag{S120}$$

Denote the kinetic energy as $E_{\text{kin}} = \mathbf{\Pi} \cdot \mathbf{\Pi}/2$, which is conserved in eq (S118). In eq (S119), $\mathbf{e}_{\perp}$ is the unit vector in the direction of $\mathbf{\Pi} - \alpha_{\|}\mathbf{e}_{\|}$ (i.e., the unit vector perpendicular to the direction of $\mathbf{B}$), and its coefficient reads

$$\alpha_{\perp} = \sqrt{2E_{\text{kin}} - \alpha_{\|}^2} . \tag{S121}$$

Substituting eqs (S119)-(S121) into eq (S118), we obtain the following EOMs for the coefficients $(\alpha_{\|}, \alpha_{\perp})$,

$$\begin{cases} \dot{\alpha}_{\|} = -B + \dfrac{B\alpha_{\|}^2}{2E_{\text{kin}}}, \\ \dot{\alpha}_{\perp} = \dfrac{B\alpha_{\|}\alpha_{\perp}}{2E_{\text{kin}}} \end{cases} \tag{S122}$$

where $B$ is the scalar length of vector $\mathbf{B}$. The solution to eq (S122) reads

$$\begin{cases} \alpha_\parallel(\tau) = \sqrt{2E_{\text{kin}}} \tanh\left[-\frac{B\tau}{\sqrt{2E_{\text{kin}}}} + \text{arctanh}\frac{\alpha_\parallel(0)}{\sqrt{2E_{\text{kin}}}}\right] \\ \qquad = \sqrt{2E_{\text{kin}}} \frac{\left(\alpha_\parallel(0) - \sqrt{2E_{\text{kin}}}\right) + \left(\alpha_\parallel(0) + \sqrt{2E_{\text{kin}}}\right)\exp\left[-\frac{2B\tau}{\sqrt{2E_{\text{kin}}}}\right]}{\left(\sqrt{2E_{\text{kin}}} - \alpha_\parallel(0)\right) + \left(\alpha_\parallel(0) + \sqrt{2E_{\text{kin}}}\right)\exp\left[-\frac{2B\tau}{\sqrt{2E_{\text{kin}}}}\right]} \\ \alpha_\perp(\tau) = \sqrt{2E_{\text{kin}}} \,\text{sech}\left[\frac{B\tau}{\sqrt{2E_{\text{kin}}}} - \text{arctanh}\frac{\alpha_\parallel(0)}{\sqrt{2E_{\text{kin}}}}\right] \\ \qquad = \frac{2\sqrt{2E_{\text{kin}}}\,\alpha_\perp(0)\exp\left[-\frac{B\tau}{\sqrt{2E_{\text{kin}}}}\right]}{\left(\sqrt{2E_{\text{kin}}} - \alpha_\parallel(0)\right) + \left(\alpha_\parallel(0) + \sqrt{2E_{\text{kin}}}\right)\exp\left[-\frac{2B\tau}{\sqrt{2E_{\text{kin}}}}\right]} \end{cases} \quad (S123)$$

By substituting eq (S123) into eqs (S117) and (S119), we recommend the following numerical integrator for updating the nuclear momentum due to the contribution of the effective nonadiabatic force for a finite time-step $\Delta\tau$,

$$\begin{aligned}\mathbf{P} \leftarrow &\sqrt{2E_{\text{kin}}}\frac{\left(\alpha_\parallel - \sqrt{2E_{\text{kin}}}\right) + \left(\alpha_\parallel + \sqrt{2E_{\text{kin}}}\right)\exp\left[-\frac{2B\Delta\tau}{\sqrt{2E_{\text{kin}}}}\right]}{\left(\sqrt{2E_{\text{kin}}} - \alpha_\parallel\right) + \left(\alpha_\parallel + \sqrt{2E_{\text{kin}}}\right)\exp\left[-\frac{2B\Delta\tau}{\sqrt{2E_{\text{kin}}}}\right]}\mathbf{M}^{1/2}\mathbf{e}_\parallel \\ &+ \frac{2\sqrt{2E_{\text{kin}}}\exp\left[-\frac{B\Delta\tau}{\sqrt{2E_{\text{kin}}}}\right]}{\left(\sqrt{2E_{\text{kin}}} - \alpha_\parallel\right) + \left(\alpha_\parallel + \sqrt{2E_{\text{kin}}}\right)\exp\left[-\frac{2B\Delta\tau}{\sqrt{2E_{\text{kin}}}}\right]}\mathbf{M}^{1/2}\mathbf{\Pi}_\perp\end{aligned} \quad (S124)$$

Equation (S124) is robust except when $\frac{B\Delta\tau}{\sqrt{2E_{\text{kin}}}}$ is either too small or too large.

When $\frac{B\Delta\tau}{\sqrt{2E_{kin}}}$ is small, which is because $B$, the magnitude of $\mathbf{B}$ is small, it is difficult to numerically determine the unit vector $\mathbf{e}_\parallel$ for $\mathbf{B}$. When $\frac{B\Delta\tau}{\sqrt{2E_{kin}}} < 10^{-20}$, it is sufficient to truncate eq (S124) to the first order of $\frac{B\Delta\tau}{\sqrt{2E_{kin}}}$, which produces

$$\mathbf{P} \leftarrow \left(1 + \Delta\tau \frac{\mathbf{B}\cdot\mathbf{M}^{-1/2}\mathbf{P}}{2E_{kin}}\right)\mathbf{P} - \Delta\tau \mathbf{M}^{1/2}\mathbf{B} \quad . \tag{S125}$$

Equation (S125) is more convenient to use when $B$ is small.

When $\frac{B\Delta\tau}{\sqrt{2E_{kin}}}$ is large and vectors $\mathbf{M}^{-1/2}\mathbf{P}$ and $\mathbf{B}$ are nearly parallel in the same direction ($\alpha_\parallel / \sqrt{2E_{kin}} \to 1$), the denominator of eq (S124) approaches zero and causes numerical instability. When $\frac{B\Delta\tau}{\sqrt{2E_{kin}}} \to +\infty$, eq (S124) yields

$$\mathbf{P} \to \begin{cases} \sqrt{2E_{kin}}\mathbf{M}^{1/2}\mathbf{e}_\parallel & \text{for } \alpha_\parallel = \sqrt{2E_{kin}}, \\ -\sqrt{2E_{kin}}\mathbf{M}^{1/2}\mathbf{e}_\parallel & \text{for } \alpha_\parallel \neq \sqrt{2E_{kin}} \end{cases} \tag{S126}$$

after a finite time-step $\Delta\tau$. Thus, for $N_{nuc} \geq 2$, when $\frac{B\Delta\tau}{\sqrt{2E_{kin}}} > 100$ and $\left|1 - \frac{\alpha_\parallel}{\sqrt{2E_{kin}}}\right| < 10^{-15}$, we should employ the self-adaptive time-step strategy to decrease the size of the time-step until $\frac{B\Delta\tau}{\sqrt{2E_{kin}}} < 100$, then use eq (S124) as the integrator for updating the nuclear momentum due to the effective nonadiabatic force.

## S2. Further Details of the Numerical Integrator of NaF

As shown in refs [322, 324], several models defined in the diabatic representation (e.g., FMO and SF models) require a significantly short time-step when employing the propagator in eq (33) of the main text to propagate the electronic phase space variables in the adiabatic representation. This is because the electronic propagation defined in eqs (31)-(33) of the main text (originally in the adiabatic representation) is different from that defined in eqs (28)-(30) of the main text (originally in the diabatic representation) with *nonfrozen* nuclear DOFs. For these models, the electronic phase space variables should be propagated with eqs (28)-(30) of the main text in the diabatic representation.

When eqs (28)-(29) of the main text are used in Step 3 of the numerical integrator in Sub-Section 2.3 of the main text (instead of eqs (44)-(45) of the main text), it is straightforward to prove that eqs (28)-(29) of the main text for a half time-step $\Delta t/2$ correspond to

$$\begin{aligned}
\mathbf{g}_{t+\Delta t/2} &= e^{-i\mathbf{V}(\mathbf{R}_t)\Delta t/2}\mathbf{g}_t \\
\Rightarrow \mathbf{T}(\mathbf{R}_t)\tilde{\mathbf{g}}_{t+\Delta t/2}(\mathbf{R}_t) &= e^{-i\mathbf{V}(\mathbf{R}_t)\Delta t/2}\mathbf{T}(\mathbf{R}_t)\tilde{\mathbf{g}}_t(\mathbf{R}_t) \\
\Rightarrow \tilde{\mathbf{g}}_{t+\Delta t/2}(\mathbf{R}_t) &= e^{-i\mathbf{E}(\mathbf{R}_t)\Delta t/2}\tilde{\mathbf{g}}_t(\mathbf{R}_t)
\end{aligned} \quad (S127)$$

and

$$\begin{aligned}
\mathbf{\Gamma}_{t+\Delta t/2} &= e^{-i\mathbf{V}(\mathbf{R}_t)\Delta t/2}\mathbf{\Gamma}_t e^{i\mathbf{V}(\mathbf{R}_t)\Delta t/2} \\
\Rightarrow \mathbf{T}(\mathbf{R}_t)\tilde{\mathbf{\Gamma}}_{t+\Delta t/2}(\mathbf{R}_t)\mathbf{T}^\dagger(\mathbf{R}_t) &= e^{-i\mathbf{V}(\mathbf{R}_t)\Delta t/2}\mathbf{T}(\mathbf{R}_t)\tilde{\mathbf{\Gamma}}_t(\mathbf{R}_t)\mathbf{T}^\dagger(\mathbf{R}_t)e^{i\mathbf{V}(\mathbf{R}_t)\Delta t/2} \\
\Rightarrow \tilde{\mathbf{\Gamma}}_{t+\Delta t/2}(\mathbf{R}_t) &= e^{-i\mathbf{E}(\mathbf{R}_t)\Delta t/2}\tilde{\mathbf{\Gamma}}_t(\mathbf{R}_t)e^{i\mathbf{E}(\mathbf{R}_t)\Delta t/2}
\end{aligned} \quad (S128)$$

in the adiabatic representation, respectively. Similarly, when eqs (28)-(29) of the main text for a half time-step $\Delta t/2$ are used in Step 5 of the numerical integrator of the main text (instead of eqs (47)-(48) of the main text), they correspond to

$$\begin{aligned}\mathbf{g}_{t+\Delta t} &= e^{-i\mathbf{V}(\mathbf{R}_{t+\Delta t})\Delta t/2}\mathbf{g}_{t+\Delta t/2}\\ \Rightarrow \mathbf{T}(\mathbf{R}_{t+\Delta t})\tilde{\mathbf{g}}_{t+\Delta t/2}(\mathbf{R}_{t+\Delta t}) &= e^{-i\mathbf{V}(\mathbf{R}_{t+\Delta t})\Delta t/2}\mathbf{T}(\mathbf{R}_t)\tilde{\mathbf{g}}_{t+\Delta t/2}(\mathbf{R}_t)\\ \Rightarrow \tilde{\mathbf{g}}_{t+\Delta t}(\mathbf{R}_{t+\Delta t}) &= e^{-i\mathbf{E}(\mathbf{R}_{t+\Delta t})\Delta t/2}\mathbf{T}^\dagger(\mathbf{R}_{t+\Delta t})\mathbf{T}(\mathbf{R}_t)\tilde{\mathbf{g}}_{t+\Delta t/2}(\mathbf{R}_t)\end{aligned} \quad (S129)$$

and

$$\begin{aligned}\boldsymbol{\Gamma}_{t+\Delta t} &= e^{-i\mathbf{V}(\mathbf{R}_{t+\Delta t})\Delta t/2}\boldsymbol{\Gamma}_{t+\Delta t/2}e^{i\mathbf{V}(\mathbf{R}_{t+\Delta t})\Delta t/2}\\ \Rightarrow \mathbf{T}(\mathbf{R}_{t+\Delta t})\tilde{\boldsymbol{\Gamma}}_{t+\Delta t}(\mathbf{R}_{t+\Delta t})\mathbf{T}^\dagger(\mathbf{R}_{t+\Delta t}) &= e^{-i\mathbf{V}(\mathbf{R}_{t+\Delta t})\Delta t/2}\mathbf{T}(\mathbf{R}_t)\tilde{\boldsymbol{\Gamma}}_{t+\Delta t/2}(\mathbf{R}_t)\mathbf{T}^\dagger(\mathbf{R}_t)e^{i\mathbf{V}(\mathbf{R}_{t+\Delta t})\Delta t/2}\\ \Rightarrow \tilde{\boldsymbol{\Gamma}}_{t+\Delta t}(\mathbf{R}_{t+\Delta t}) &= e^{-i\mathbf{E}(\mathbf{R}_{t+\Delta t})\Delta t/2}\mathbf{T}^\dagger(\mathbf{R}_{t+\Delta t})\mathbf{T}(\mathbf{R}_t)\tilde{\boldsymbol{\Gamma}}_{t+\Delta t/2}(\mathbf{R}_t)\mathbf{T}^\dagger(\mathbf{R}_t)\mathbf{T}(\mathbf{R}_{t+\Delta t})e^{i\mathbf{E}(\mathbf{R}_{t+\Delta t})\Delta t/2}\end{aligned} \quad (S130)$$

in the adiabatic representation, respectively. Equations (S127)-(S128) and eqs (S129)-(S130) suggest directly using $e^{-i\mathbf{E}(\mathbf{R}_t)\Delta t/2}$ and $e^{-i\mathbf{E}(\mathbf{R}_{t+\Delta t})\Delta t/2}\mathbf{T}^\dagger(\mathbf{R}_{t+\Delta t})\mathbf{T}(\mathbf{R}_t)$ as the alternative propagators in the adiabatic representation for Step 3 and Step 5 of the numerical integrator in the main text, respectively. The resulting numerical integrator is listed as follows.

1. Update the nuclear kinematic momentum within a half time-step $\Delta t/2$ using the adiabatic force

$$\mathbf{P}_{t+\Delta t/2} \leftarrow \mathbf{P}_t - \nabla_{\mathbf{R}} E_{j_{\text{old}}}(\mathbf{R}_t)\frac{\Delta t}{2}. \quad (S131)$$

2. Update the nuclear kinematic momentum within a half time-step $\Delta t/2$ using the numerical integrator for the effective nonadiabatic force for the $N_{\text{nuc}} \geq 2$ case

$$\begin{aligned}\mathbf{P}_{t+\Delta t/2} \leftarrow\ &c_1(\mathbf{R}_t, \mathbf{P}_{t+\Delta t/2}, \tilde{\boldsymbol{\rho}}_t, \Delta t/2)\mathbf{M}^{1/2}\mathbf{e}_\parallel(\mathbf{R}_t, \tilde{\boldsymbol{\rho}}_t)\\ &+c_2(\mathbf{R}_t, \mathbf{P}_{t+\Delta t/2}, \tilde{\boldsymbol{\rho}}_t, \Delta t/2)\mathbf{M}^{1/2}\boldsymbol{\Pi}_\perp(\mathbf{R}_t, \mathbf{P}_{t+\Delta t/2}, \tilde{\boldsymbol{\rho}}_t),\end{aligned} \quad (S132)$$

where the definitions of $c_1(\mathbf{R}, \mathbf{P}, \tilde{\boldsymbol{\rho}}, \Delta t)$, $c_2(\mathbf{R}, \mathbf{P}, \tilde{\boldsymbol{\rho}}, \Delta t)$, $\mathbf{e}_\parallel(\mathbf{R}, \tilde{\boldsymbol{\rho}})$ and $\boldsymbol{\Pi}_\perp(\mathbf{R}, \mathbf{P}, \tilde{\boldsymbol{\rho}})$ can be found in Sub-Section 2.3 of the main text. For the $N_{\text{nuc}} = 1$ case, this step is skipped. Please refer to Section S1 for details of additional treatments to prevent numerical instability.

3. Update phase space variables of electronic DOFs within a half time-step $\Delta t/2$ according to

$$\tilde{\mathbf{g}}_{t+\Delta t/2} \leftarrow e^{-i\mathbf{E}(\mathbf{R}_t)\Delta t/2}\tilde{\mathbf{g}}_t \ . \tag{S133}$$

$$\tilde{\mathbf{\Gamma}}_{t+\Delta t/2} \leftarrow e^{-i\mathbf{E}(\mathbf{R}_t)\Delta t/2}\tilde{\mathbf{\Gamma}}_t e^{i\mathbf{E}(\mathbf{R}_t)\Delta t/2} \ . \tag{S134}$$

4. Update the nuclear coordinate within a full time-step $\Delta t$

$$\mathbf{R}_{t+\Delta t} \leftarrow \mathbf{R}_t + \mathbf{M}^{-1}\mathbf{P}_{t+\Delta t/2}\Delta t \ . \tag{S135}$$

5. Update phase space variables of electronic DOFs within the other half time-step $\Delta t/2$ according to

$$\tilde{\mathbf{g}}_{t+\Delta t} \leftarrow e^{-i\mathbf{E}(\mathbf{R}_{t+\Delta t})\Delta t/2}\mathbf{T}^{\dagger}(\mathbf{R}_{t+\Delta t})\mathbf{T}(\mathbf{R}_t)\tilde{\mathbf{g}}_{t+\Delta t/2} \ . \tag{S136}$$

$$\tilde{\mathbf{\Gamma}}_{t+\Delta t} \leftarrow e^{-i\mathbf{E}(\mathbf{R}_{t+\Delta t})\Delta t/2}\mathbf{T}^{\dagger}(\mathbf{R}_{t+\Delta t})\mathbf{T}(\mathbf{R}_t)\tilde{\mathbf{\Gamma}}_{t+\Delta t/2}\mathbf{T}^{\dagger}(\mathbf{R}_t)\mathbf{T}(\mathbf{R}_{t+\Delta t})e^{i\mathbf{E}(\mathbf{R}_{t+\Delta t})\Delta t/2} \ . \tag{S137}$$

Calculate the effective electronic density matrix $\tilde{\boldsymbol{\rho}}$ according to eq (34) of the main text.

6. Determine a new occupied state $j_{\text{new}}$ based on $\tilde{\boldsymbol{\rho}}$ and rescale $\mathbf{P}$ if $j_{\text{new}} \neq j_{\text{old}}$

$$\mathbf{P}_{t+\Delta t/2} \leftarrow \mathbf{P}_{t+\Delta t/2}\sqrt{\left(H_{\text{NaF}}\left(\mathbf{R}_{t+\Delta t},\mathbf{P}_{t+\Delta t/2},\tilde{\boldsymbol{\rho}}_{t+\Delta t}\right) - E_{j_{\text{new}}}\left(\mathbf{R}_{t+\Delta t}\right)\right)/\left(\mathbf{P}_{t+\Delta t/2}^{\text{T}}\mathbf{M}^{-1}\mathbf{P}_{t+\Delta t/2}/2\right)} \ . \tag{S138}$$

If $H_{\text{NaF}}\left(\mathbf{R}_{t+\Delta t},\mathbf{P}_{t+\Delta t/2},\tilde{\boldsymbol{\rho}}_{t+\Delta t}\right) < E_{j_{\text{new}}}\left(\mathbf{R}_{t+\Delta t}\right)$, the switching of the adiabatic nuclear force component is frustrated. In such a case we keep $j_{\text{new}} = j_{\text{old}}$ and the rescaling step (for the nuclear kinematic momentum) eq (S138) is skipped.

7. Similar to Step 2, update the nuclear kinematic momentum within the other half time-step $\Delta t/2$ using the numerical integrator for the effective nonadiabatic force for the $N_{\text{nuc}} \geq 2$ case

$$\begin{aligned}\mathbf{P}_{t+\Delta t} \leftarrow &\, c_1\left(\mathbf{R}_{t+\Delta t},\mathbf{P}_{t+\Delta t/2},\tilde{\boldsymbol{\rho}}_{t+\Delta t},\Delta t/2\right)\mathbf{M}^{1/2}\mathbf{e}_{\parallel}\left(\mathbf{R}_{t+\Delta t},\tilde{\boldsymbol{\rho}}_{t+\Delta t}\right) \\ &+ c_2\left(\mathbf{R}_{t+\Delta t},\mathbf{P}_{t+\Delta t/2},\tilde{\boldsymbol{\rho}}_{t+\Delta t},\Delta t/2\right)\mathbf{M}^{1/2}\mathbf{\Pi}_{\perp}\left(\mathbf{R}_{t+\Delta t},\mathbf{P}_{t+\Delta t/2},\tilde{\boldsymbol{\rho}}_{t+\Delta t}\right) \end{aligned} \ . \tag{S139}$$

Please refer to Section S1 for details of additional treatments to prevent numerical instability.

8. Update the nuclear kinematic momentum within the other half time-step $\Delta t/2$ using the adiabatic force

$$\mathbf{P}_{t+\Delta t} \leftarrow \mathbf{P}_{t+\Delta t} - \nabla_{\mathbf{R}} E_{j_{\text{new}}}(\mathbf{R}_{t+\Delta t}) \frac{\Delta t}{2} . \tag{S140}$$

The above numerical integrator is equivalent to propagating electronic DOFs according to eqs (28)-(30) of the main text in the diabatic representation while nuclear DOFs are propagated in the adiabatic representation. In fact, the nuclear force of NaF can also be represented in the diabatic representation, as detailed in Section S3. The elements of the matrix $\mathbf{O}(\mathbf{R}_{t+\Delta t}, \mathbf{R}_t) = \mathbf{T}^\dagger(\mathbf{R}_{t+\Delta t}) \mathbf{T}(\mathbf{R}_t)$, which read $O_{nm}(\mathbf{R}_{t+\Delta t}, \mathbf{R}_t) = \langle \phi_n(\mathbf{R}_{t+\Delta t}) | \phi_m(\mathbf{R}_t) \rangle$, represent the overlap of the adiabatic electronic wave function between two adjacent time-steps. (This overlap matrix has been defined in the imaginary time when path integral is applied to nonadiabatic systems in our previous work[376].) In principle, one can also employ eqs (S129)-(S130) for simulations with *ab initio* calculations if the overlap matrix $\mathbf{O}(\mathbf{R}_{t+\Delta t}, \mathbf{R}_t)$ can be obtained, regardless of whether the diabatic representation is rigorously defined or not. The overlap matrix $\mathbf{O}(\mathbf{R}_{t+\Delta t}, \mathbf{R}_t)$ is also employed to ensure a smooth change in the adiabatic basis. Specifically, the order and sign of $|\phi_n(\mathbf{R}_{t+\Delta t})\rangle$ are carefully tracked by guaranteeing that $O_{nm}(\mathbf{R}_{t+\Delta t}, \mathbf{R}_t) - \delta_{nm}$ remains sufficiently small.

## S3. Nuclear Force of NaF in the Diabatic Representation

In this section, we represent the nuclear force of NaF in the diabatic representation, and then provide the approach to evolve the NaF trajectory using the nuclear force in the diabatic representation. We first rewrite the nuclear force of NaF in eqs (35)-(37) of the main text for the $I$-th nuclear DOF

$$\dot{P}_I = -\frac{\partial E_{j_{occ}}(\mathbf{R})}{\partial R_I} - \sum_{n \neq m}^{F} \left( E_n(\mathbf{R}) - E_m(\mathbf{R}) \right) d_{mn}^{(I)}(\mathbf{R}) \tilde{\rho}_{nm} \tag{S141}$$

as

$$\dot{P}_I = \mathrm{Tr}_e \left[ \tilde{\mathbf{F}}_I(\mathbf{R}) \left( \tilde{\mathbf{Q}}^{(a)} + \tilde{\mathbf{Q}}^{(na)} \right) \right]. \tag{S142}$$

Here, $\tilde{\mathbf{F}}_I(\mathbf{R})$ denotes the force matrix in the adiabatic representation for the $I$-th nuclear DOF, whose elements read

$$\left( \tilde{\mathbf{F}}_I(\mathbf{R}) \right)_{nm} = -\frac{\partial E_n(\mathbf{R})}{\partial R_I} \delta_{nm} - \left( E_n(\mathbf{R}) - E_m(\mathbf{R}) \right) d_{mn}^{(I)}(\mathbf{R}), \tag{S143}$$

and the elements of the Hermitian matrices $\tilde{\mathbf{Q}}^{(a)}$ and $\tilde{\mathbf{Q}}^{(na)}$ are

$$\tilde{Q}_{nm}^{(a)} = \delta_{nj_{occ}} \delta_{nm}, \quad \tilde{Q}_{nm}^{(na)} = \tilde{\rho}_{nm}(1 - \delta_{nm}), \tag{S144}$$

where the effective density matrix $\tilde{\boldsymbol{\rho}}$ is defined in eq (34) of the main text. It is straightforward to show that $\mathrm{Tr}_e \left[ \tilde{\mathbf{F}}_I(\mathbf{R}) \tilde{\mathbf{Q}}^{(a)} \right]$ and $\mathrm{Tr}_e \left[ \tilde{\mathbf{F}}_I(\mathbf{R}) \tilde{\mathbf{Q}}^{(na)} \right]$ are the adiabatic force and nonadiabatic force, respectively. Note that the force matrix $\tilde{\mathbf{F}}_I(\mathbf{R})$ is covariant under the adiabatic-to-diabatic transformation, i.e., $\tilde{\mathbf{F}}_I(\mathbf{R}) = \mathbf{T}^\dagger(\mathbf{R}) \mathbf{F}_I(\mathbf{R}) \mathbf{T}(\mathbf{R})$ with $\left( \mathbf{F}_I(\mathbf{R}) \right)_{nm} = -\frac{\partial V_{nm}(\mathbf{R})}{\partial R_I}$ denotes the corresponding element of the force matrix $\mathbf{F}_I(\mathbf{R})$ in the diabatic representation. Therefore, eq (S142) can be rewritten as

$$\begin{aligned} \dot{P}_I &= \mathrm{Tr}_e\left[\tilde{\mathbf{F}}_I(\mathbf{R})\left(\tilde{\mathbf{Q}}^{(a)} + \tilde{\mathbf{Q}}^{(na)}\right)\right] \\ &= \mathrm{Tr}_e\left[\mathbf{T}^\dagger(\mathbf{R})\mathbf{F}_I(\mathbf{R})\mathbf{T}(\mathbf{R})\left(\tilde{\mathbf{Q}}^{(a)} + \tilde{\mathbf{Q}}^{(na)}\right)\right] \\ &= \mathrm{Tr}_e\left[\mathbf{F}_I(\mathbf{R})\left(\mathbf{Q}^{(a)} + \mathbf{Q}^{(na)}\right)\right] \end{aligned} \qquad (S145)$$

where

$$\mathbf{Q}^{(a)} = \mathbf{T}(\mathbf{R})\tilde{\mathbf{Q}}^{(a)}\mathbf{T}^\dagger(\mathbf{R}) \qquad (S146)$$

and

$$\mathbf{Q}^{(na)} = \mathbf{T}(\mathbf{R})\tilde{\mathbf{Q}}^{(na)}\mathbf{T}^\dagger(\mathbf{R}). \qquad (S147)$$

Therefore, we only need to obtain the matrices $\mathbf{Q}^{(a)}$ and $\mathbf{Q}^{(na)}$ to calculate the nuclear force in each time-step. This can be achieved through the diabatic-to-adiabatic transformation. In the following, we provide the algorithm of NaF based on the nuclear force represented in the diabatic representation. By referring to the numerical integrator for effective nonadiabatic force in Section S1 and the numerical integrator scheme "P-e-R-e-P" in the main text, the algorithm reads:

1. Sample the initial values of nuclear and electronic phase space variables. Diagonalize the diabatic potential matrix: $\mathbf{V}(\mathbf{R}_0) = \mathbf{T}(\mathbf{R}_0)\mathbf{E}(\mathbf{R}_0)\mathbf{T}^\dagger(\mathbf{R}_0)$. Calculate the effective density matrix $\boldsymbol{\rho}_0 = \frac{1+\mathrm{Tr}_e[\boldsymbol{\Gamma}_0]}{\left(\mathbf{x}_0^\mathrm{T}\mathbf{x}_0 + \mathbf{p}_0^\mathrm{T}\mathbf{p}_0\right)}(\mathbf{x}_0 + i\mathbf{p}_0)(\mathbf{x}_0 - i\mathbf{p}_0)^\mathrm{T} - \boldsymbol{\Gamma}_0$ in the diabatic representation and that of adiabatic representation $\tilde{\boldsymbol{\rho}}_0 = \mathbf{T}^\dagger(\mathbf{R}_0)\boldsymbol{\rho}_0\mathbf{T}(\mathbf{R}_0)$. Choose the index of the occupied state as $j_{occ} = \arg\max_n \tilde{\rho}_{nn}$. Calculate the matrices $\tilde{\mathbf{Q}}_0^{(a)}$ and $\tilde{\mathbf{Q}}_0^{(na)}$ according to eq (S144), and obtain the matrices $\mathbf{Q}_0^{(a)} = \mathbf{T}(\mathbf{R}_0)\tilde{\mathbf{Q}}_0^{(a)}\mathbf{T}^\dagger(\mathbf{R}_0)$ and $\mathbf{Q}_0^{(na)} = \mathbf{T}(\mathbf{R}_0)\tilde{\mathbf{Q}}_0^{(na)}\mathbf{T}^\dagger(\mathbf{R}_0)$. Calculate the electronic propagator $U(\mathbf{R}_0; \Delta t/2) = e^{-i\mathbf{V}(\mathbf{R}_0)\Delta t/2} = \mathbf{T}(\mathbf{R}_0) e^{-i\mathbf{E}(\mathbf{R}_0)\Delta t/2} \mathbf{T}^\dagger(\mathbf{R}_0)$ in the diabatic representation. Initialize the time to $t = 0$.

2. Update the nuclear canonical momentum in the diabatic representation within a half time-step $\Delta t/2$ using the adiabatic force

$$\mathbf{P}_{t+\Delta t/2} \leftarrow \mathbf{P}_t + \mathrm{Tr}_e \left[ \mathbf{F}(\mathbf{R}_t) \mathbf{Q}_t^{(a)} \right] \frac{\Delta t}{2} . \tag{S148}$$

3. Update the nuclear kinematic momentum within a half time-step $\Delta t/2$ using the numerical integrator for the effective nonadiabatic force for the $N_{\mathrm{nuc}} \geq 2$ case

$$\begin{aligned}\mathbf{P}_{t+\Delta t/2} \leftarrow\ & c_1\left(\mathbf{R}_t, \mathbf{P}_{t+\Delta t/2}, \mathbf{Q}_t^{(\mathrm{na})}, \Delta t/2\right) \mathbf{M}^{1/2} \mathbf{e}_{\parallel}\left(\mathbf{R}_t, \mathbf{Q}_t^{(\mathrm{na})}\right) \\ & + c_2\left(\mathbf{R}_t, \mathbf{P}_{t+\Delta t/2}, \mathbf{Q}_t^{(\mathrm{na})}, \Delta t/2\right) \mathbf{M}^{1/2} \mathbf{\Pi}_{\perp}\left(\mathbf{R}_t, \mathbf{P}_{t+\Delta t/2}, \mathbf{Q}_t^{(\mathrm{na})}\right),\end{aligned} \tag{S149}$$

where

$$c_1\left(\mathbf{R}, \mathbf{P}, \mathbf{Q}^{(\mathrm{na})}, \Delta t\right) = \sqrt{2E_{\mathrm{kin}}} \frac{\left(\alpha_{\parallel} - \sqrt{2E_{\mathrm{kin}}}\right) + \left(\alpha_{\parallel} + \sqrt{2E_{\mathrm{kin}}}\right)\exp\left[-\frac{2B\Delta t}{\sqrt{2E_{\mathrm{kin}}}}\right]}{\left(\sqrt{2E_{\mathrm{kin}}} - \alpha_{\parallel}\right) + \left(\alpha_{\parallel} + \sqrt{2E_{\mathrm{kin}}}\right)\exp\left[-\frac{2B\Delta t}{\sqrt{2E_{\mathrm{kin}}}}\right]}, \tag{S150}$$

$$c_2\left(\mathbf{R}, \mathbf{P}, \mathbf{Q}^{(\mathrm{na})}, \Delta t\right) = \frac{2\sqrt{2E_{\mathrm{kin}}}\exp\left[-\frac{B\Delta t}{\sqrt{2E_{\mathrm{kin}}}}\right]}{\left(\sqrt{2E_{\mathrm{kin}}} - \alpha_{\parallel}\right) + \left(\alpha_{\parallel} + \sqrt{2E_{\mathrm{kin}}}\right)\exp\left[-\frac{2B\Delta t}{\sqrt{2E_{\mathrm{kin}}}}\right]} . \tag{S151}$$

Here, $E_{\mathrm{kin}} = \mathbf{P}^{\mathrm{T}} \mathbf{M}^{-1} \mathbf{P}/2$ is the total kinetic energy, $\mathbf{e}_{\parallel}(\mathbf{R}, \mathbf{Q}^{(\mathrm{na})})$ denotes the unit vector for the direction of the vector $\mathbf{B} = \mathbf{M}^{-1/2} \sum_{n,m} \nabla_{\mathbf{R}} V_{nm}(\mathbf{R}) Q_{mn}^{(\mathrm{na})}$, $B$ represents the scalar length of the vector $\mathbf{B}$, $\mathbf{\Pi}_{\parallel} = \alpha_{\parallel} \mathbf{e}_{\parallel} = \left(\mathbf{M}^{-1/2}\mathbf{P} \cdot \mathbf{e}_{\parallel}\right)\mathbf{e}_{\parallel}$ and $\mathbf{\Pi}_{\perp}(\mathbf{R}, \mathbf{P}, \tilde{\mathbf{\rho}}) = \mathbf{M}^{-1/2}\mathbf{P} - \alpha_{\parallel}\mathbf{e}_{\parallel}$ are the components of $\mathbf{\Pi} \equiv \mathbf{M}^{-1/2}\mathbf{P}$ parallel and perpendicular to $\mathbf{B}$, respectively. For the $N_{\text{nuc}} = 1$ case, this step is skipped. When $\dfrac{B_t \Delta t / 2}{\sqrt{2E_{\text{kin}}(t)}}$ is very small or very large, please refer to Section S1 for details of additional treatments to prevent numerical instability.

4. Update the phase space variables of electronic DOFs in the diabatic representation within a half time-step $\Delta t / 2$ according to

$$\mathbf{g}_{t+\Delta t/2} \leftarrow \mathbf{U}(\mathbf{R}_t; \Delta t / 2)\mathbf{g}_t \ . \tag{S152}$$

$$\mathbf{\Gamma}_{t+\Delta t/2} \leftarrow \mathbf{U}(\mathbf{R}_t; \Delta t / 2)\mathbf{\Gamma}_t \mathbf{U}^{\dagger}(\mathbf{R}_t; \Delta t / 2) \ . \tag{S153}$$

5. Update the nuclear coordinate within a full time-step $\Delta t$

$$\mathbf{R}_{t+\Delta t} \leftarrow \mathbf{R}_t + \mathbf{M}^{-1}\mathbf{P}_{t+\Delta t/2}\Delta t \ . \tag{S154}$$

6. Diagonalize the diabatic potential matrix: $\mathbf{V}(\mathbf{R}_{t+\Delta t}) = \mathbf{T}(\mathbf{R}_{t+\Delta t})\mathbf{E}(\mathbf{R}_{t+\Delta t})\mathbf{T}^{\dagger}(\mathbf{R}_{t+\Delta t})$. Calculate the electronic propagator in the diabatic representation

$$\mathbf{U}(\mathbf{R}_{t+\Delta t}; \Delta t / 2) = e^{-i\mathbf{V}(\mathbf{R}_{t+\Delta t})\Delta t/2} = \mathbf{T}(\mathbf{R}_{t+\Delta t})e^{-i\mathbf{E}(\mathbf{R}_{t+\Delta t})\Delta t/2}\mathbf{T}^{\dagger}(\mathbf{R}_{t+\Delta t}) \ . \tag{S155}$$

Update the phase space variables of electronic DOFs in the diabatic representation within the other half time-step $\Delta t / 2$ according to

$$\mathbf{g}_{t+\Delta t} \leftarrow \mathbf{U}(\mathbf{R}_{t+\Delta t}; \Delta t / 2)\mathbf{g}_{t+\Delta t/2} \ . \tag{S156}$$

$$\mathbf{\Gamma}_{t+\Delta t} \leftarrow \mathbf{U}(\mathbf{R}_{t+\Delta t}; \Delta t / 2)\mathbf{\Gamma}_{t+\Delta t/2}\mathbf{U}^{\dagger}(\mathbf{R}_{t+\Delta t}; \Delta t / 2) \ . \tag{S157}$$

7. Calculate the effective density matrix $\boldsymbol{\rho}_{t+\Delta t} = \dfrac{1 + \text{Tr}_e[\mathbf{\Gamma}_{t+\Delta t}]}{\left(\mathbf{x}_{t+\Delta t}^{\text{T}}\mathbf{x}_{t+\Delta t} + \mathbf{p}_{t+\Delta t}^{\text{T}}\mathbf{p}_{t+\Delta t}\right)}(\mathbf{x}_{t+\Delta t} + i\mathbf{p}_{t+\Delta t})(\mathbf{x}_{t+\Delta t} - i\mathbf{p}_{t+\Delta t})^{\text{T}} - \mathbf{\Gamma}_{t+\Delta t}$ in the diabatic representation and that in the adiabatic representation $\tilde{\boldsymbol{\rho}}_{t+\Delta t} = \mathbf{T}^{\dagger}(\mathbf{R}_{t+\Delta t})\boldsymbol{\rho}_{t+\Delta t}\mathbf{T}(\mathbf{R}_{t+\Delta t})$. Determine a new occupied state $j_{new}$ based on the maximum diagonal elements of $\tilde{\boldsymbol{\rho}}_{t+\Delta t}$ and rescale the nuclear canonical momentum in the diabatic representation if $j_{new} \neq j_{old}$

$$\mathbf{P}_{t+\Delta t/2} \leftarrow \mathbf{P}_{t+\Delta t/2}\sqrt{\frac{\mathbf{P}_{t+\Delta t/2}^{\mathrm{T}}\mathbf{M}^{-1}\mathbf{P}_{t+\Delta t/2}/2 + E_{j_{old}}(\mathbf{R}_{t+\Delta t}) - E_{j_{new}}(\mathbf{R}_{t+\Delta t})}{\mathbf{P}_{t+\Delta t/2}^{\mathrm{T}}\mathbf{M}^{-1}\mathbf{P}_{t+\Delta t/2}/2}} . \tag{S158}$$

If $\mathbf{P}_{t+\Delta t/2}^{\mathrm{T}}\mathbf{M}^{-1}\mathbf{P}_{t+\Delta t/2}/2 + E_{j_{old}}(\mathbf{R}_{t+\Delta t}) < E_{j_{new}}(\mathbf{R}_{t+\Delta t})$, the switching of $j_{occ}$ is frustrated. In such a case we keep $j_{new} = j_{old}$ and the rescaling step eq (S158) is skipped. Calculate the matrices $\tilde{\mathbf{Q}}_{t+\Delta t}^{(a)}$ and $\tilde{\mathbf{Q}}_{t+\Delta t}^{(na)}$ according to eq (S144), and obtain the matrices $\mathbf{Q}_{t+\Delta t}^{(a)} = \mathbf{T}(\mathbf{R}_{t+\Delta t})\tilde{\mathbf{Q}}_{t+\Delta t}^{(a)}\mathbf{T}^{\dagger}(\mathbf{R}_{t+\Delta t})$ and $\mathbf{Q}_{t+\Delta t}^{(na)} = \mathbf{T}(\mathbf{R}_{t+\Delta t})\tilde{\mathbf{Q}}_{t+\Delta t}^{(na)}\mathbf{T}^{\dagger}(\mathbf{R}_{t+\Delta t})$.

8. Similar to Step 3, update the nuclear kinematic momentum within the other half time-step $\Delta t/2$ using the numerical integrator for the effective nonadiabatic force for the $N_{nuc} \geq 2$ case

$$\begin{aligned}\mathbf{P}_{t+\Delta t} \leftarrow\ &c_1\left(\mathbf{R}_{t+\Delta t},\mathbf{P}_{t+\Delta t/2},\mathbf{Q}_{t+\Delta t}^{(na)},\Delta t/2\right)\mathbf{M}^{1/2}\mathbf{e}_{\|}\left(\mathbf{R}_{t+\Delta t},\mathbf{Q}_{t+\Delta t}^{(na)}\right)\\&+c_2\left(\mathbf{R}_{t+\Delta t},\mathbf{P}_{t+\Delta t/2},\mathbf{Q}_{t+\Delta t}^{(na)},\Delta t/2\right)\mathbf{M}^{1/2}\mathbf{\Pi}_{\perp}\left(\mathbf{R}_{t+\Delta t},\mathbf{P}_{t+\Delta t/2},\mathbf{Q}_{t+\Delta t}^{(na)}\right)\end{aligned}. \tag{S159}$$

When $\dfrac{B_{t+\Delta t}\Delta t/2}{\sqrt{2E_{kin}(t+\Delta t/2)}}$ is very small or very large, please refer to Section S1 for details of additional treatments to prevent numerical instability.

9. Update the nuclear canonical momentum in the diabatic representation within the other half time-step $\Delta t/2$ using the adiabatic force

$$\mathbf{P}_{t+\Delta t} \leftarrow \mathbf{P}_{t+\Delta t} + \mathrm{Tr}_e\left[\mathbf{F}(\mathbf{R}_{t+\Delta t})\mathbf{Q}_{t+\Delta t}^{(a)}\right]\frac{\Delta t}{2} , \tag{S160}$$

10. Update the time variable, $t \leftarrow t + \Delta t$. Repeat Steps 2-9 until the evolution of the trajectory

ends.

The algorithm described above is capable of evolving NaF trajectories equivalently *without* the need for nonadiabatic coupling vectors and derivatives of potential energy surfaces in the adiabatic representation. In other words, it does not require calculating the element $\langle \phi_m(\mathbf{R}) | (-\nabla_\mathbf{R} \hat{H}_{el}(\mathbf{R})) | \phi_n(\mathbf{R}) \rangle$ of the force matrix in the adiabatic representation via the diabatic-to-adiabatic transformation. For system-bath models with constant diabatic coupling, one can disregard the nuclear force contributions from the off-diagonal terms, thereby further enhance the efficiency of NaF. The results of NaF-TW in different representations for the 7-state FMO model are illustrated in Figure S1.

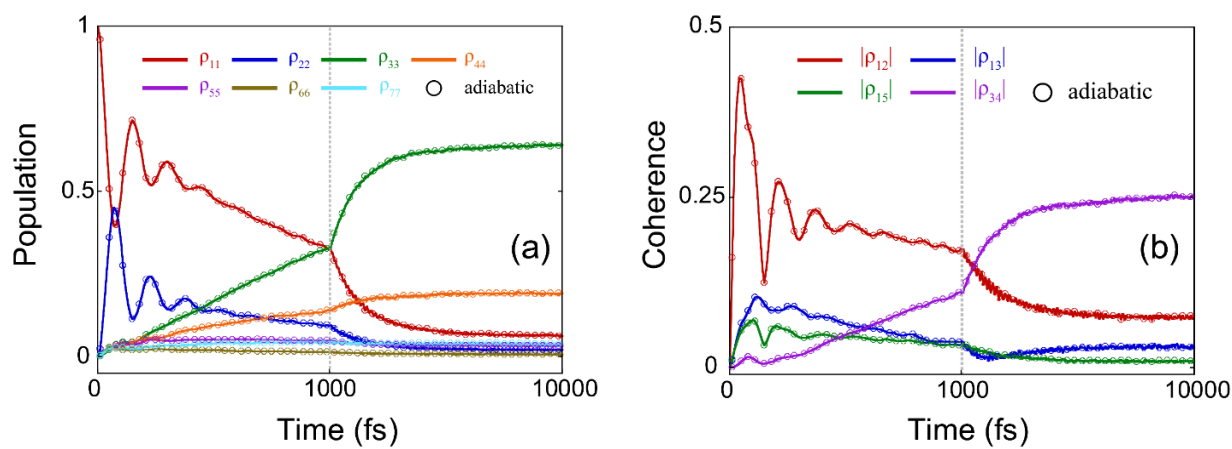

**Figure S1.** Panels (a) and (b) are similar to panel (d) of Figure 11 and Figure 12 of the main text, respectively, but the solid lines represent the results of NaF-TW with nuclear force represented in the diabatic representation, while the hollow circles with corresponding colors denote the results of NaF-TW with nuclear force represented in the adiabatic representation.

## S4. Derivation of the Covariant-Covariant TCF with Action-Angle Variables

This section only serves for *education purposes* for illustrating the strategy of using action-angle variables to develop the CPS formulation, which was first presented in eqs (A3)-(A20) of Appendix A of ref [318] for general $F$-state systems ($F \geq 2$), where the $\gamma = 0$ case was used as only an example for demonstration because the $\gamma = 0$ case avoids the "inverted potential energy surface problem" as mentioned in Section IV of ref [318].

Following Appendix A of ref [318], we first represent phase space variables $\{\mathbf{x}, \mathbf{p}\}$ of CPS $\mathcal{S}(\mathbf{x}, \mathbf{p}, \gamma)$ with parameter $\gamma$ in the action-angle representation[201, 318] as

$$x^{(n)} = \sqrt{2e^{(n)}} \cos\theta^{(n)}, \quad p^{(n)} = \sqrt{2e^{(n)}} \sin\theta^{(n)}, \quad n = 1, \cdots, F \tag{S161}$$

where action variables $e^{(n)} \in [0, 1 + F\gamma]$ and angle variables $\theta^{(n)} \in [0, 2\pi)$. It is straightforward to determine that $\mathbf{dxdp} = \mathbf{ded\theta}$. The phase space constraint of CPS with parameter $\gamma$ in the action-angle representation reads

$$\mathcal{S}(\mathbf{e}, \boldsymbol{\theta}, \gamma) = \delta\left(\sum_{n=1}^{F} e^{(n)} - (1 + F\gamma)\right) . \tag{S162}$$

Define $S_F(\lambda) = \int_0^\lambda \mathbf{de}\, \delta\left(\sum_{n=1}^{F} e^{(n)} - \lambda\right)$. Following Appendix A of ref [318], it is straightforward to obtain the relation between $S_F(\lambda)$ and $S_F(1)$ as

$$\begin{aligned}
S_F(\lambda) &= \int_0^\lambda \mathbf{de}\, \delta\left(\sum_{n=1}^{F} e^{(n)} - \lambda\right) = \left(\prod_{n=1}^{F} \int_0^\lambda de^{(n)}\right) \delta\left(\sum_{n=1}^{F} e^{(n)} - \lambda\right) \\
&\stackrel{e^{(n)} = \lambda \bar{e}^{(n)}}{=} \lambda^F \left(\prod_{n=1}^{F} \int_0^1 d\bar{e}^{(n)}\right) \delta\left(\lambda\left(\sum_{n=1}^{F} \bar{e}^{(n)} - 1\right)\right) \\
&= \lambda^{F-1} \left(\prod_{n=1}^{F} \int_0^1 d\bar{e}^{(n)}\right) \delta\left(\sum_{n=1}^{F} \bar{e}^{(n)} - 1\right) = \lambda^{F-1} S_F(1)
\end{aligned} \tag{S163}$$

Utilizing eq (S163), we also obtain the relation between $S_F(1)$ and $S_{F-1}(1)$ as

$$\begin{aligned}
S_F(1) &= \int_0^1 \mathrm{d}\mathbf{e}\,\delta\!\left(\sum_{n=1}^F e^{(n)} - 1\right) = \left(\prod_{n=1}^F \int_0^1 \mathrm{d}e^{(n)}\right) \delta\!\left(\sum_{n\neq j}^F e^{(n)} - \left(1 - e^{(j)}\right)\right) \\
&= \int_0^1 \mathrm{d}e^{(j)} \left(\prod_{n\neq j}^F \int_0^{1-e^{(j)}} \mathrm{d}e^{(n)}\right) \delta\!\left(\sum_{n\neq j}^F e^{(n)} - \left(1 - e^{(j)}\right)\right) \\
&= \int_0^1 \mathrm{d}e^{(j)} S_{F-1}\!\left(1 - e^{(j)}\right) = \int_0^1 \mathrm{d}e^{(j)} \left(1 - e^{(j)}\right)^{F-2} S_{F-1}(1) = \frac{S_{F-1}(1)}{F-1}
\end{aligned} \quad . \tag{S164}$$

It is straightforward to employ eqs (S163) and (S164) to express the expectation of a given function $f(e^{(j)})$ on CPS $\mathcal{S}(\mathbf{e},\boldsymbol{\theta},\gamma)$,

$$\begin{aligned}
&\left\langle f(e^{(j)}) \right\rangle_{1+F\gamma} \\
&= \int_{\mathcal{S}(\mathbf{e},\boldsymbol{\theta},\gamma)} \mathrm{d}\mathbf{e}\,\mathrm{d}\boldsymbol{\theta}\, f(e^{(j)}) \\
&= \frac{\int_0^{2\pi} \mathrm{d}\boldsymbol{\theta} \int_0^{1+F\gamma} \mathrm{d}\mathbf{e}\, f(e^{(j)}) \mathcal{S}(\mathbf{e},\boldsymbol{\theta},\gamma)}{\int_0^{2\pi} \mathrm{d}\boldsymbol{\theta} \int_0^{1+F\gamma} \mathrm{d}\mathbf{e}\, \mathcal{S}(\mathbf{e},\boldsymbol{\theta},\gamma)} \\
&= \frac{\int_0^{1+F\gamma} \mathrm{d}\mathbf{e}\, f(e^{(j)})\, \delta\!\left(\sum_{n=1}^F e^{(n)} - (1+F\gamma)\right)}{\int_0^{1+F\gamma} \mathrm{d}\mathbf{e}\, \delta\!\left(\sum_{n=1}^F e^{(n)} - (1+F\gamma)\right)} \\
&= \frac{1}{S_F(1+F\gamma)} \int_0^{1+F\gamma} \mathrm{d}\mathbf{e}\, f(e^{(j)})\, \delta\!\left(\sum_{n=1}^F e^{(n)} - (1+F\gamma)\right) \\
&= \frac{1}{S_F(1+F\gamma)} \int_0^{1+F\gamma} \mathrm{d}e^{(j)} f(e^{(j)}) \left(\prod_{n\neq j}^F \int_0^{1+F\gamma - e^{(j)}} \mathrm{d}e^{(j)}\right) \delta\!\left(\sum_{n\neq j}^F e^{(n)} - (1+F\gamma - e^{(j)})\right) \\
&= \frac{1}{S_F(1+F\gamma)} \int_0^{1+F\gamma} \mathrm{d}e^{(j)} f(e^{(j)})\, S_{F-1}(1+F\gamma - e^{(j)}) \\
&= \frac{S_{F-1}(1)}{S_F(1+F\gamma)} \int_0^{1+F\gamma} \mathrm{d}e^{(j)} f(e^{(j)}) \left(1+F\gamma - e^{(j)}\right)^{F-2} \\
&= \frac{F-1}{(1+F\gamma)^{F-1}} \int_0^{1+F\gamma} \mathrm{d}e^{(j)} f(e^{(j)}) \left(1+F\gamma - e^{(j)}\right)^{F-2}
\end{aligned} \quad . \tag{S165}$$

Equation (S165) yields

$$\left\langle e^{(j)} \right\rangle_{1+F\gamma} = \frac{F-1}{(1+F\gamma)^{F-1}} \int_0^{1+F\gamma} \mathrm{d}e^{(j)} \left(1+F\gamma - e^{(j)}\right)^{F-2} e^{(j)} = \frac{1+F\gamma}{F} \tag{S166}$$

and

$$\left\langle \left(e^{(j)}\right)^2 \right\rangle_{1+F\gamma} = \frac{F-1}{\left(1+F\gamma\right)^{F-1}} \int_0^{1+F\gamma} \mathrm{d}e^{(j)} \left(1+F\gamma - e^{(j)}\right)^{F-2} \left(e^{(j)}\right)^2 = \frac{2\left(1+F\gamma\right)^2}{F(F+1)} \quad . \tag{S167}$$

Because

$$\begin{aligned}
\left\langle \left(1+F\gamma\right) e^{(j)} \right\rangle_{1+F\gamma} &= \left\langle e^{(j)} \sum_{n=1}^{F} e^{(n)} \right\rangle_{1+F\gamma} = \left\langle \left(e^{(j)}\right)^2 \right\rangle_{1+F\gamma} + \sum_{n\neq j}^{F} \left\langle e^{(j)} e^{(n)} \right\rangle_{1+F\gamma} \\
&= \left\langle \left(e^{(j)}\right)^2 \right\rangle_{1+F\gamma} + (F-1) \left\langle e^{(j)} e^{(n\neq j)} \right\rangle_{1+F\gamma}
\end{aligned} \tag{S168}$$

we obtain that $\left\langle e^{(j)} e^{(n\neq j)} \right\rangle_{1+F\gamma} = (1+F\gamma)^2 / F(F+1)$, or equivalently

$$\left\langle e^{(j)} e^{(n)} \right\rangle_{1+F\gamma} = \frac{\left(1+\delta_{jn}\right)\left(1+F\gamma\right)^2}{F(F+1)} \quad . \tag{S169}$$

It is trivial to use eqs (S166) and (S169) to obtain the following phase space integrals:

$$\begin{aligned}
&\int_{\mathcal{S}(\mathbf{x},\mathbf{p},\gamma)} \mathrm{d}\mathbf{x}\mathrm{d}\mathbf{p} \frac{\left(x^{(n)} + \mathrm{i}p^{(n)}\right)\left(x^{(m)} - \mathrm{i}p^{(m)}\right)}{2} \\
&= \frac{\int_0^{2\pi} \mathrm{d}\boldsymbol{\theta} \int_0^{1+F\gamma} \mathrm{d}\mathbf{e}\, \mathcal{S}(\mathbf{e},\boldsymbol{\theta},\gamma) \sqrt{e^{(n)} e^{(m)}} e^{\mathrm{i}(\theta^{(n)} - \theta^{(m)})}}{\int_0^{2\pi} \mathrm{d}\boldsymbol{\theta} \int_0^{1+F\gamma} \mathrm{d}\mathbf{e}\, \mathcal{S}(\mathbf{e},\boldsymbol{\theta},\gamma)} \\
&= \frac{1}{(2\pi)^F S_F(1+F\gamma)} \int_0^{2\pi} \mathrm{d}\boldsymbol{\theta} \int_0^{1+F\gamma} \mathrm{d}\mathbf{e}\, \mathcal{S}(\mathbf{e},\boldsymbol{\theta},\gamma) \sqrt{e^{(n)} e^{(m)}} e^{\mathrm{i}(\theta^{(n)} - \theta^{(m)})} \\
&= \frac{1}{S_F(1+F\gamma)} \int_0^{1+F\gamma} \mathrm{d}\mathbf{e}\, \delta\left(\sum_{j=1}^F e^{(j)} - (1+F\gamma)\right) \sqrt{e^{(n)} e^{(m)}} \times \frac{1}{(2\pi)^F} \int_0^{2\pi} \mathrm{d}\boldsymbol{\theta}\, e^{\mathrm{i}(\theta^{(n)} - \theta^{(m)})} \\
&= \left\langle e^{(n)} \right\rangle_{1+F\gamma} \delta_{nm} = \frac{(1+F\gamma)}{F} \delta_{nm}
\end{aligned} \tag{S170}$$

and

$$\begin{aligned}
&\int_{\mathcal{S}(\mathbf{x},\mathbf{p},\gamma)} d\mathbf{x} d\mathbf{p} \frac{\left(x^{(n)}+\mathrm{i}p^{(n)}\right)\left(x^{(m)}-\mathrm{i}p^{(m)}\right)}{2} \times \frac{\left(x^{(k)}+\mathrm{i}p^{(k)}\right)\left(x^{(l)}-\mathrm{i}p^{(l)}\right)}{2} \\
&= \frac{\int_0^{2\pi} d\boldsymbol{\theta} \int_0^{1+F\gamma} d\mathbf{e}\, \mathcal{S}(\mathbf{e},\boldsymbol{\theta},\gamma) \sqrt{e^{(n)}e^{(m)}e^{(k)}e^{(l)}}\, e^{\mathrm{i}(\theta^{(n)}-\theta^{(m)}+\theta^{(k)}-\theta^{(l)})}}{\int_0^{2\pi} d\boldsymbol{\theta} \int_0^{1+F\gamma} d\mathbf{e}\, \mathcal{S}(\mathbf{e},\boldsymbol{\theta},\gamma)} \\
&= \frac{1}{S_F(1+F\gamma)} \int_0^{1+F\gamma} d\mathbf{e}\, \delta\!\left(\sum_{j=1}^{F} e^{(j)} - (1+F\gamma)\right) \sqrt{e^{(n)}e^{(m)}e^{(k)}e^{(l)}} \\
&\quad \times \frac{1}{(2\pi)^F} \int_0^{2\pi} d\boldsymbol{\theta}\, e^{\mathrm{i}(\theta^{(n)}-\theta^{(m)}+\theta^{(k)}-\theta^{(l)})} \\
&= \left\langle e^{(n)} e^{(m\ne n)} \right\rangle_{1+F\gamma} \delta_{mk}\delta_{nl}(1-\delta_{mn}) + \left\langle e^{(m)} e^{(k)} \right\rangle_{1+F\gamma} \delta_{mn}\delta_{kl} \\
&= \frac{(1+F\gamma)^2}{F(F+1)} \delta_{mk}\delta_{nl}(1-\delta_{mn}) + \frac{(1+\delta_{mk})(1+F\gamma)^2}{F(F+1)} \delta_{mn}\delta_{kl} \\
&= \frac{(1+F\gamma)^2}{F(F+1)} \left(\delta_{mk}\delta_{nl} + \delta_{mn}\delta_{kl}\right)
\end{aligned} \qquad (\text{S171})$$

For the given covariant mapping kernel of eq (56) of the main text

$$\hat{K}_{\mathrm{ele}}(\mathbf{x},\mathbf{p}) = \sum_{n,m=1}^{F} \left[\frac{1}{2}\left(x^{(n)}+\mathrm{i}p^{(n)}\right)\left(x^{(m)}-\mathrm{i}p^{(m)}\right) - \gamma\delta_{nm}\right] |n\rangle\langle m| , \qquad (\text{S172})$$

suppose the corresponding inverse mapping kernel is also covariant,

$$\hat{K}_{\mathrm{ele}}^{-1}(\mathbf{x},\mathbf{p}) = \sum_{k,l=1}^{F} \left[A \frac{\left(x^{(k)}+\mathrm{i}p^{(k)}\right)\left(x^{(l)}-\mathrm{i}p^{(l)}\right)}{2} - B\delta_{kl}\right] |k\rangle\langle l| , \qquad (\text{S173})$$

where $A$ and $B$ are undetermined coefficients. For operators $\hat{O}_1 = |m\rangle\langle n|$ and $\hat{O}_2 = |l\rangle\langle k|$, we have $\mathrm{Tr}\!\left[\hat{O}_1 \hat{O}_2\right] = \delta_{mk}\delta_{nl}$. After utilizing eqs (S170) and (S171), the phase space expression of $\mathrm{Tr}\!\left[\hat{O}_1 \hat{O}_2\right]$ reads

$$\mathrm{Tr}\left[\hat{O}_1\hat{O}_2\right]$$
$$= \int_{\mathcal{S}(\mathbf{x},\mathbf{p},\gamma)} F\mathrm{d}\mathbf{x}\mathrm{d}\mathbf{p} \left[\frac{\left(x^{(n)}+\mathrm{i}p^{(n)}\right)\left(x^{(m)}-\mathrm{i}p^{(m)}\right)}{2} - \gamma\delta_{nm}\right]\left[\frac{A\left(x^{(k)}+\mathrm{i}p^{(k)}\right)\left(x^{(l)}-\mathrm{i}p^{(l)}\right)}{2} - B\delta_{kl}\right].\text{(S174)}$$
$$= \frac{A(1+F\gamma)^2}{(F+1)}\left(\delta_{mk}\delta_{nl}+\delta_{nm}\delta_{kl}\right) - (1+F\gamma)B\delta_{nm}\delta_{kl} - (1+F\gamma)A\gamma\delta_{nm}\delta_{kl} + \gamma FB\delta_{nm}\delta_{kl}$$
$$= \frac{A(1+F\gamma)^2}{(F+1)}\delta_{mk}\delta_{nl} + \left[\frac{A(1+F\gamma)^2}{(F+1)} - B - (1+F\gamma)A\gamma\right]\delta_{nm}\delta_{kl}$$

Because the RHS of eq (S174) should be equal to $\delta_{mk}\delta_{nl}$, it yields

$$A = \frac{1+F}{(1+F\gamma)^2}, \quad B = \frac{1-\gamma}{1+F\gamma}. \tag{S175}$$

Substitution of eq (S175) into eq (S173) leads to the expression of the (covariant) inverse mapping kernel, which is eq (57) of the main text.

The strategy of using action-angle variables to establish the CPS formulation, which we first presented in Appendix A of ref [318], was later employed in ref [323] with the *Abel integral equation* to derive a novel class of CPS representations for two-state systems, which yields exact (electronic) population dynamics in the frozen nuclei limit. In any case of this class, each trajectory on CPS makes non-negative contribution to the electronic population dynamics. The TWF approach of Cotton and Miller[195], which had been used for population dynamics in the SQC/MM method, was proved as a special case of this class in ref [323] by us. The strategy of using action-angle variables presented in Appendix A of ref [318] was also implemented to construct the electronic population-coherence, coherence-population, coherence-coherence correlation functions with TWFs in ref [324].

References [12, 319] used the same strategy but with coordinate-momentum variables instead, which was introduced only to help understand Appendix A of ref [318]. For instance, the authors of ref [204] (published in 2020) even failed to understand that Appendix A of ref [318] (published in 2019) simply leads to an exact phase space formulation of quantum mechanics for general $F$-state systems ($F \geq 2$).

**S5. Simulation Details of One-Dimensional Holstein Model**

Here we report a benchmark test for a one-dimensional Holstein model[435, 436] of an organic semiconductor investigated by one of us in 2020[412]. The Hamiltonian of the one-dimensional Holstein model reads

$$\hat{H} = \hat{H}_e + \hat{H}_p + \hat{H}_c \ . \tag{S176}$$

The first term of the RHS of eq (S176) denotes the tight-binding electronic Hamiltonian

$$\hat{H}_e = \sum_{n=1}^{F} \left( \hat{c}_{n+1}^\dagger \hat{c}_n + \hat{c}_n^\dagger \hat{c}_{n+1} \right) V \ , \tag{S177}$$

where $V$ is the electronic coupling/transfer integral, and $\hat{c}_n^\dagger$ ($\hat{c}_n$) represents the creation (annihilation) operator of electron DOFs of $n$-th site. The periodic boundary condition is applied so that $\hat{c}_{F+1} \equiv \hat{c}_1$. The pure phonon term of the RHS of eq (S176) reads

$$\hat{H}_p = \sum_{n=1}^{F} \sum_{j=1}^{N_p} \omega_j \left( \hat{b}_{jn}^\dagger \hat{b}_{jn} + \frac{1}{2} \right) \ , \tag{S178}$$

where $\hat{b}_{jn}^\dagger$ ($\hat{b}_{jn}$) denotes the raising (lowering) operator of the $j$-th phonon on the $n$-th site with the corresponding vibrational frequency $\omega_n$, and $N_p = N_{\text{nuc}} / F$ is the number of phonon modes on each site. The electron-phonon coupling term of eq (S176) is

$$\hat{H}_c = \sum_{n=1}^{F} \sum_{j=1}^{N_p} g_j \omega_j (\hat{b}_{jn}^\dagger + \hat{b}_{jn}) \hat{c}_n^\dagger \hat{c}_n \ , \tag{S179}$$

where $g_j$ is the dimensionless coupling coefficient between the *j*-th mode and the electronic DOFs. When the case with only one electron is considered, the total Hamiltonian (of eq (S176)) is isomorphic to a multi-state Hamiltonian, as illustrated by the isomorphism relation in ref [317] in 2017. Consequently, the creation and annihilation operators of electronic DOFs can be represented as

$$\hat{c}_n^\dagger = |0_1, \cdots, 1_n, \cdots, 0_F\rangle\langle 0_1, \cdots, 0_n, \cdots, 0_F| \triangleq |n\rangle\langle\tilde{0}| \\ \hat{c}_n = |0_1, \cdots, 0_n, \cdots, 0_F\rangle\langle 0_1, \cdots, 1_n, \cdots, 0_F| \triangleq |\tilde{0}\rangle\langle n| \quad (S180)$$

with $|\tilde{0}\rangle$ denoting the vacuum state. Equation (S180) yields

$$\hat{c}_n^\dagger \hat{c}_m = |n\rangle\langle m| \quad . \quad (S181)$$

After substitution of eq (S181) for electronic DOFs into eqs (S177) and (S179) and implementation of the relation between the raising/lowering operators and the canonical coordinate/momentum operators for phonon modes

$$\hat{b}_{nj} = \sqrt{\frac{\omega_j}{2}}\left(\hat{R}_{nj} + \frac{i}{\omega_j}\hat{P}_{nj}\right) \\ \hat{b}_{nj}^\dagger = \sqrt{\frac{\omega_j}{2}}\left(\hat{R}_{nj} - \frac{i}{\omega_j}\hat{P}_{nj}\right) \quad (S182)$$

into eqs (S178) and (S179), it is trivial to show that the (one-dimensional) Holstein model of eq (S176) is isomorphic to the *F*-state Hamiltonian,

$$\hat{H} = \sum_{n=1}^{F}\sum_{j=1}^{N_p}\left(\frac{1}{2}\hat{P}_{nj}^2 + \frac{\omega_j^2}{2}\hat{R}_{nj}^2\right) \\ + V\sum_{n=1}^{F}(|n+1\rangle\langle n| + |n\rangle\langle n+1|) + \sum_{n=1}^{F}\sum_{j=1}^{N_p}g_j\omega_j\sqrt{2\omega_j}\hat{R}_{nj}|n\rangle\langle n| \quad . \quad (S183)$$

As suggested in ref [317], it can be simulated within the framework of nonadiabatic dynamics. NaF methods then offer a practical tool for studying the Holstein model of eq (S176) or equivalently eq (S183).

We calculate the physical quantity,

$$\mu = \frac{\beta}{e_0} \int_0^{+\infty} \mathrm{d}t\, C(t) \ , \tag{S184}$$

with the current correlation function

$$C(t) = \mathrm{Re}\left(\mathrm{Tr}\left[\hat{\rho}_0 \hat{j}(0)\hat{j}(t)\right]\right) \ . \tag{S185}$$

Here,

$$\begin{aligned}\hat{j} &= \frac{e_0 V d}{\mathrm{i}} \sum_{n=1}^{F} \left(\hat{c}_{n+1}^\dagger \hat{c}_n - \hat{c}_n^\dagger \hat{c}_{n+1}\right) \\ &= \frac{e_0 V d}{\mathrm{i}} \sum_{n=1}^{F} \left(|n+1\rangle\langle n| - |n\rangle\langle n+1|\right)\end{aligned} \ , \tag{S186}$$

with $e_0$ representing the elementary charge and $d$ denoting the intermolecular distance, and $\hat{j}(t)$ is the Heisenberg operator

$$\hat{j}(t) = e^{i\hat{H}t/\hbar} \hat{j} e^{-i\hat{H}t/\hbar} \ . \tag{S187}$$

In eq (S185), $\hat{\rho}_0$ is the initial density[437]

$$\hat{\rho}_0 = e^{-\beta \hat{H}_{\mathrm{eff}}} \otimes e^{-\beta \hat{H}_p} / (Z_e Z_p) \ , \tag{S188}$$

where $Z_e = \mathrm{Tr}_e\left[e^{-\beta \hat{H}_{\mathrm{eff}}}\right]$ and $Z_p = \mathrm{Tr}_n\left[e^{-\beta \hat{H}_p}\right]$ are the partition functions for electronic and nuclear DOFs, respectively, and the effective Hamiltonian for electronic DOFs reads[437]

$$\begin{aligned}\hat{H}_{\text{eff}} &= e^{-\beta\lambda/3}V\sum_{n=1}^{F}\left(\hat{c}_{n+1}^{\dagger}\hat{c}_{n}+\hat{c}_{n}^{\dagger}\hat{c}_{n+1}\right)\\&=e^{-\beta\lambda/3}V\sum_{n=1}^{F}\left(|n+1\rangle\langle n|+|n\rangle\langle n+1|\right)\end{aligned} \qquad (S189)$$

with $\lambda=\sum_{j=1}^{N_p}g_j^2\omega_j$ denoting the total reorganization energy of each site. The physical quantity, $\mu$, is related to the carrier mobility of organic semiconductors[438]. All the parameters of the model investigated in this section are taken from refs [412, 439], which were derived from *ab initio* data of rubrene. As described in ref [412], the electronic coupling/transfer integral is set to $V = 0.083$ eV, the number of sites is $F = 21$, and each site includes $N_p = 9$ phonon modes. We examine the temperature dependence of $\mu$.

All simulations employ the same Hamiltonian of eq (S176) as well as the same initial condition of eq (S188). The numerically exact results are obtained by TD-DMRG for comparison. When NaF methods are used, the initial occupied state is uniformly randomly selected from all sites. According to eq (74) of the main text, eq (S185) can be expressed in the generalized coordinate-momentum phase space formulation,

$$\begin{aligned}C(t) &\mapsto \text{Re} \sum_{n,m,k,l=1}^{F} \frac{1}{\overline{C}_{nm,kl}(t)} \int d\mu(\mathbf{R},\mathbf{P}) \int d\gamma w(\gamma) \int_{\mathcal{S}(\mathbf{x},\mathbf{p},\Gamma;\gamma)} d\mu(\mathbf{x},\mathbf{p},\Gamma) \\&\times \rho_p(\mathbf{R},\mathbf{P})\left(\hat{j}e^{-\beta\hat{H}_{\text{eff}}}/Z_e\right)_{nm}(\hat{j})_{kl}\overline{Q}_{nm,kl}(\mathbf{x},\mathbf{p},\Gamma;\gamma;t)\end{aligned} \qquad (S190)$$

where

$$\rho_p(\mathbf{R},\mathbf{P}) \propto \prod_{n=1}^{F}\prod_{j=1}^{N_p}\exp\left(-\frac{\beta}{2Q(\omega_j)}\left(P_{nj}^2+\omega_j^2 R_{nj}^2\right)\right) \qquad (S191)$$

denotes the Wigner distribution of phonons with the quantum correction factor[21] $Q(\omega) = \dfrac{\beta\hbar\omega/2}{\tanh(\beta\hbar\omega/2)}$, and $\left(\hat{j}e^{-\beta\hat{H}_{\text{eff}}}/Z_e\right)_{nm}$ and $\left(\hat{j}\right)_{kl}$ represent the corresponding matrix elements of the operators $\hat{j}e^{-\beta\hat{H}_{\text{eff}}}/Z_e$ and $\hat{j}$, respectively. The matrix elements of $\hat{j}$ can trivially be obtained from eq (S186). The matrix elements of operator $\hat{j}e^{-\beta\hat{H}_{\text{eff}}}/Z_e$ can also be readily achieved by the eigen-decomposition. Assume that $\mathbf{H}_{\text{eff}}$ is the matrix for operator $\hat{H}_{\text{eff}}$ and $\mathbf{j}$ is the matrix for operator $\hat{j}$. Matrix $\mathbf{H}_{\text{eff}}$ can be expressed as $\mathbf{H}_{\text{eff}} = \mathbf{T}_e \mathbf{E}_e \mathbf{T}_e^{\text{T}}$, where $\mathbf{E}_e$ is a diagonal (eigenvalue) matrix and $\mathbf{T}_e$ is an orthogonal matrix. The matrix for operator $\hat{j}e^{-\beta\hat{H}_{\text{eff}}}/Z_e$ is then $\mathbf{j}e^{-\beta\mathbf{H}_{\text{eff}}}/Z_e = \mathbf{j}\mathbf{T}_e e^{-\beta\mathbf{E}_e}\mathbf{T}_e^{\text{T}}/\text{Tr}_e\left[e^{-\beta\mathbf{E}_e}\right]$.

When NaF methods are used, the results for the correlation function $C(t)$ of eq (S185) decay to zero before 6000 au. A Gaussian damping term $\exp(-t^2/2\sigma^2)$ with $\sigma = 6000$ au is introduced when integrating $C(t)$ over the time region, $[0, 6000]$ au, to obtain $\mu$ for NaF methods. Similarly, when TD-DMRG is employed for the computation of $\mu$, we also use a Gaussian damping term $\exp(-t^2/2\sigma^2)$ for integrating $C(t)$ over the time region, $[0, 20670]$ au, because the long-time tail of $C(t)$ calculated by TD-DMRG is not meaningful due to finite size effects. The result for $\mu$ is the average of the value of $\mu$ obtained with $\sigma = 6000$ au and that with $\sigma = 3000$ au, and its error bar is estimated as half of the absolute difference between the two values. Figure 16 of the main text demonstrates the results of $\mu$ of NaF methods compared with those of TD-DMRG in a wide temperature range. Panels (a)-(b) of Figure S2 illustrates the comparison between NaF-TW and TD-DMRG for $C(t)/d^2$ at 200 K and that at 400 K. We also demonstrate the NaF-TW results, for which the initial condition of electronic DOFs is $e^{-\beta \hat{H}_{\text{eff}}}/Z_e$ quantum mechanically but the initial condition of nuclear DOFs is sampled from the corresponding classical Boltzmann distribution of $e^{-\beta \hat{H}_p}/Z_p$ instead, i.e.,

$$\rho_p^{(\text{Classical})}(\mathbf{R},\mathbf{P}) \propto \prod_{n=1}^{F} \prod_{j=1}^{N_p} \exp\left(-\frac{\beta}{2}\left(P_{nj}^2 + \omega_j^2 R_{nj}^2\right)\right) . \quad (S192)$$

That is, the initial condition is the mixed quantum-classical limit of the initial density operator, eq (S188). Figure S2 demonstrates the comparison among results obtained by TD-DMRG, those by NaF-TW with the full quantum initial density of eq (S188), those by NaF-TW with the mixed quantum-classical limit of eq (S188). Panels (a) and (b) of Figure S2 demonstrate that the correlation function $C(t)$ of eq (S185) predicted in the mixed quantum-classical limit decays much slowly before 2000 au, while NaF-TW agrees reasonably well with TD-DMRG. When we calculate the value of $\mu$ with NaF-TW in the mixed quantum-classical limit, we integrate $C(t)$ in the time region, $[0,12000]$ au, which is sufficiently longer than the length of the correlation time. A Gaussian damping term $\exp(-t^2/2\sigma^2)$ with $\sigma = 6000$ au is also used in the integration.

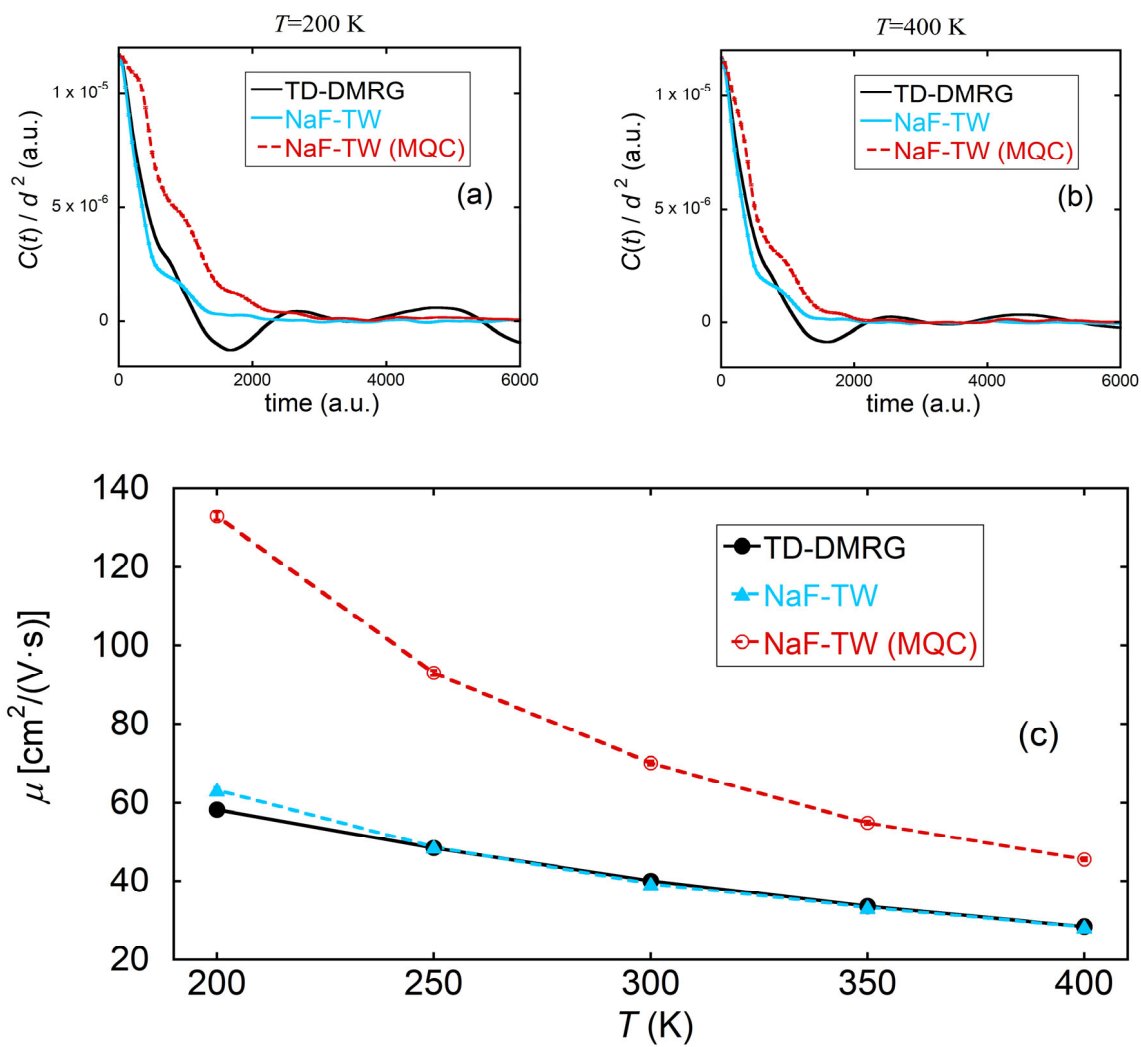

**Figure S2**. Panels (a)-(b): $C(t)/d^2$ of the one-dimensional Holstein model investigated in ref [412] at 200 K and 400 K, respectively. Cyan solid lines: NaF-TW; Red dashed lines: NaF-TW in the mixed quantum-classical limit; Black solid lines: TD-DMRG. Panel (c): Benchmark results of $\mu$ as functions of temperature. Cyan triangles with cyan dashed lines: NaF-TW; Red hollow circles with red dashed lines: NaF-TW in the mixed quantum-classical limit; Black points with black solid lines: TD-DMRG.

## S6. Details of TD-DMRG Simulations

In the TD-DMRG simulations of the 7-state FMO model, we first apply the thermofield dynamics algorithm to transform the original Hamiltonian into an effective one[151]:

$$\tilde{H} = \sum_i \varepsilon_i a_i^\dagger a_i + \sum_{ij} V_{ij} a_i^\dagger a_j + \sum_{ik} \omega_{ik} \left( b_{ik}^\dagger b_{ik} - \tilde{b}_{ik}^\dagger \tilde{b}_{ik} \right) \\ + \sum_{ik} g_{ik} \omega_{ik} a_i^\dagger a_i \left[ \cosh(\theta_{ik})\left(b_{ik}^\dagger + b_{ik}\right) + \sinh(\theta_{ik})\left(\tilde{b}_{ik}^\dagger + \tilde{b}_{ik}\right) \right] \quad \text{(S193)}$$

where $\theta_k = \text{arctanh}\left(e^{-\beta\omega_k/2}\right)$. The unitary transformation makes the dynamics of the electronic part of the original Hamiltonian at finite temperature equivalent to that of the effective Hamiltonian (eq (S193)) at zero temperature. We then convert the effective Hamiltonian from the "star" representation to the "chain" representation using the Lanczos iteration algorithm, which recasts eq (S193) into

$$\tilde{H}_2 = \sum_i \varepsilon_i a_i^\dagger a_i + \sum_{ij} V_{ij} a_i^\dagger a_j + \sum_i \beta_{i0} a_i^\dagger a_i (b_{i0}^\dagger + b_{i0}) \\ + \sum_{ik} \beta_{ik} (b_{ik+1}^\dagger b_{ik} + b_{ik}^\dagger b_{ik+1}) + \sum_{ik} \alpha_{ik} b_{ik}^\dagger b_{ik} \quad \text{(S194)}$$

The unitary transformation modifies the interaction pattern to be more numerically favorable for DMRG, without altering the dynamics of the electronic part. To map all DOFs onto a linear chain, we place the electronic site at the leftmost position, followed by the vibrational modes $v_{ik}$ arranged as $v_{0,1}, v_{0,2}, \cdots, v_{0,7}, v_{1,1}, v_{1,2}, \cdots, v_{1,7}, \cdots$. The number of energy levels of each vibrational mode is truncated to 5. The bond dimension is set to 300 at 77 K and 100 at 0 K. The real-time evolution is carried out by using the projector-splitting algorithm based on the time-dependent variational principle, with a time-step of 1 fs. All TD-DMRG simulations are performed using the *Renormalizer* package.

In the TD-DMRG simulation of the one-dimensional Holstein model, we use the same setup as for the 7-site FMO model, except that the bond dimension is set to 350.

## S7. Additional Numerical Results for the 3-State Photodissociation Models and for the FMO Model at Zero Temperature

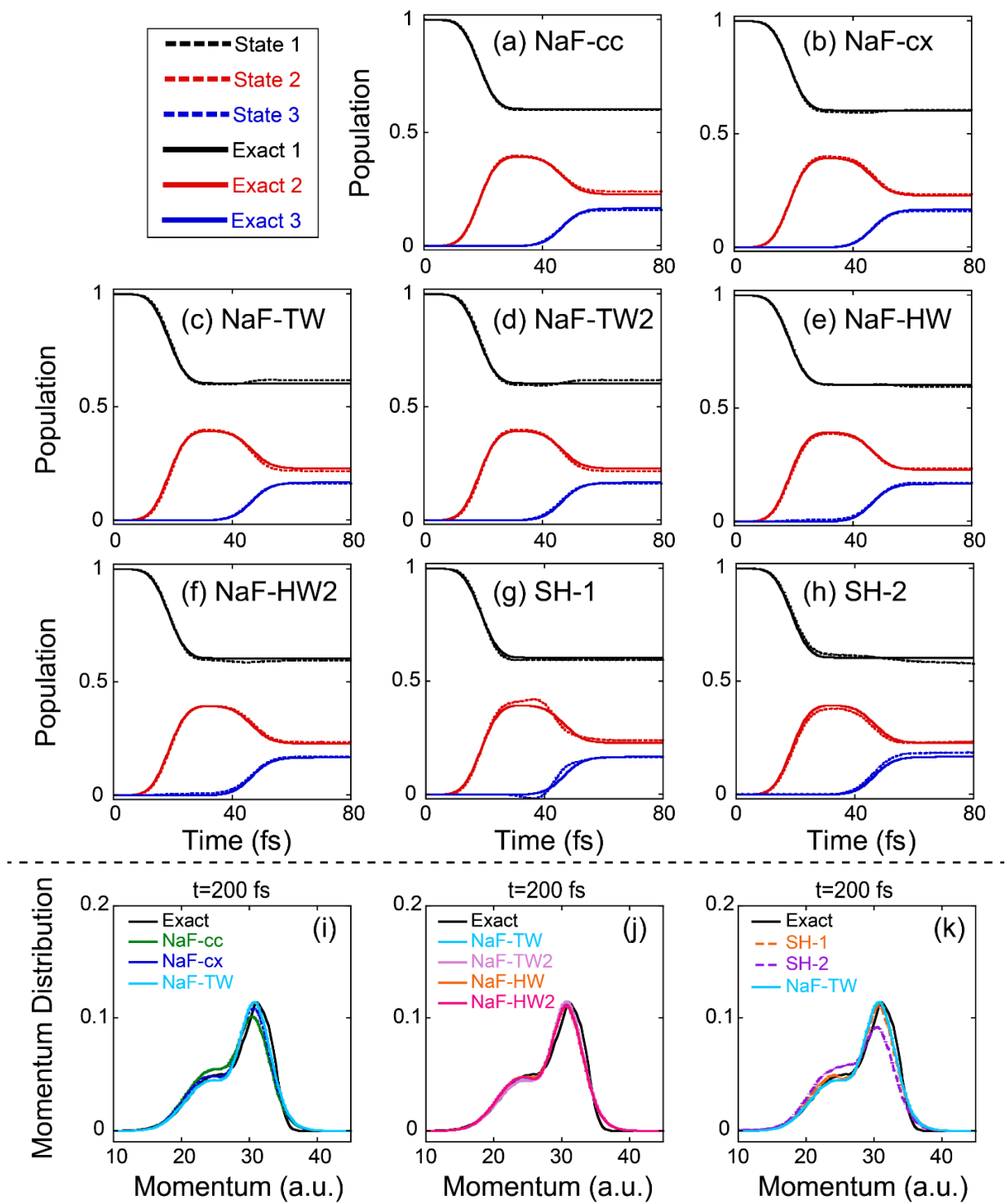

**Figure S3.** Results of Model 1 of the 3-state photodissociation models. In panels (a)-(h), the black, red and blue dashed lines represent the population dynamics of states 1-3, respectively. Panel (a): NaF-cc; Panel (b): NaF-cx; Panel (c): NaF-TW; Panel (d): NaF-TW2; Panel (e): NaF-HW; Panel (f): NaF-HW2; Panel (g): SH-1; Panel (h): SH-2. Note that SH-3 is not applicable for this 3-state model. The numerically exact results produced by DVR[363] are demonstrated by solid lines with corresponding colors in panels (a)-(h). Panels (i)-(k) illustrate the nuclear momentum distribution at 200 fs. The green, blue, and cyan solid lines in panel (i) represent the results of NaF-cc, NaF-cx and NaF-TW, respectively. The cyan, pink, orange and magenta solid lines in panel (j) denote the results of NaF-TW, NaF-TW2, NaF-HW and NaF-HW2, respectively. The orange dashed lines, purple dashed lines and cyan solid lines in panel (k) denote the results of SH-1, SH-2 and NaF-TW, respectively. The black solid lines in panels (i)-(k) denote the numerically exact results produced by DVR[363].

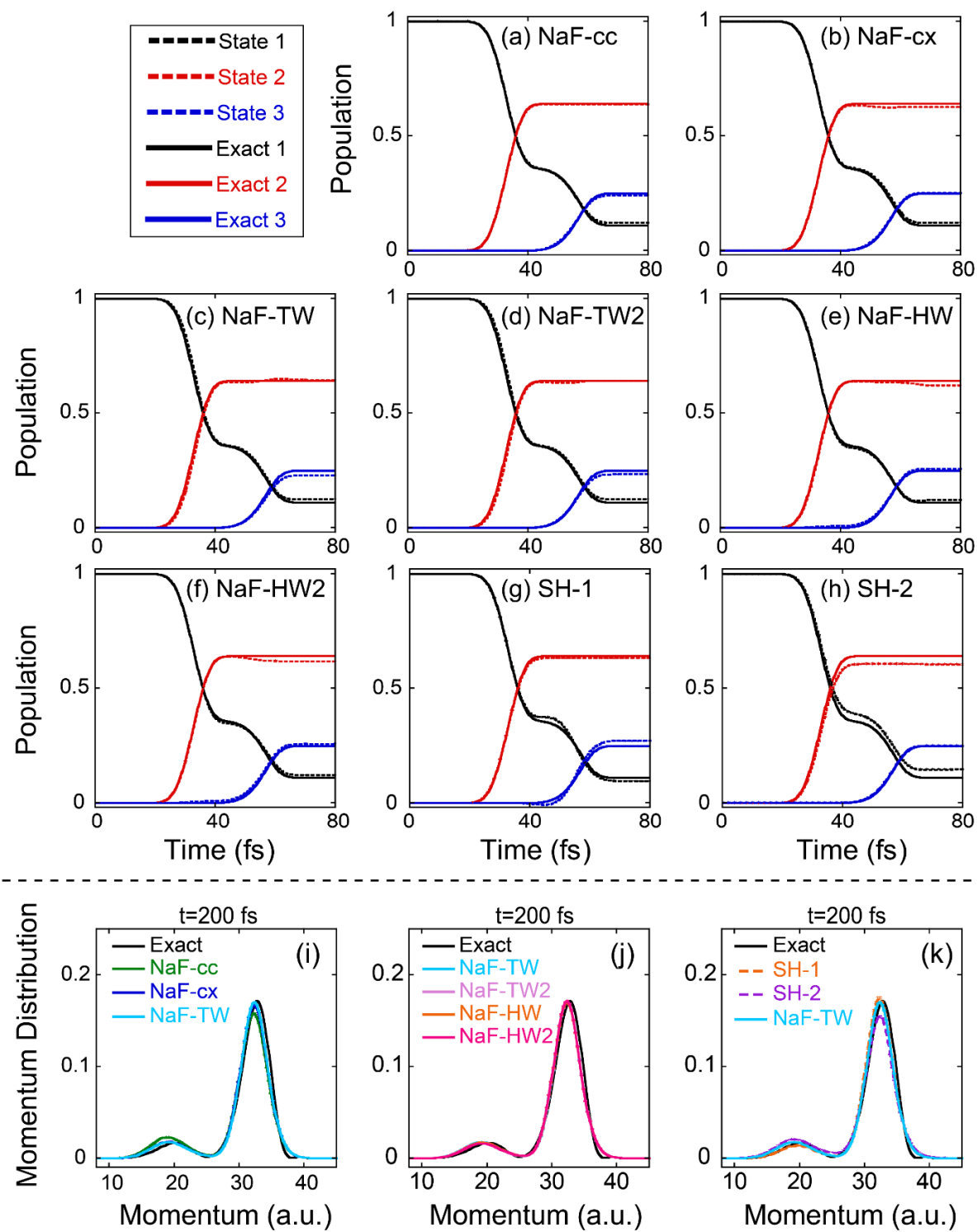

**Figure S4.** Results of Model 3 of the 3-state photodissociation models. In panels (a)-(h), the black, red and blue dashed lines represent the population dynamics of states 1-3, respectively. Panel (a):

NaF-cc; Panel (b): NaF-cx; Panel (c): NaF-TW; Panel (d): NaF-TW2; Panel (e): NaF-HW; Panel (f): NaF-HW2; Panel (g): SH-1; Panel (h): SH-2. Note that SH-3 is not applicable for this 3-state model. The numerically exact results produced by DVR[363] are demonstrated by solid lines with corresponding colors in panels (a)-(h). Panels (i)-(k) illustrate the nuclear momentum distribution at 200 fs. The green, blue, and cyan solid lines in panel (i) represent the results of NaF-cc, NaF-cx and NaF-TW, respectively. The cyan, pink, orange and magenta solid lines in panel (j) denote the results of NaF-TW, NaF-TW2, NaF-HW and NaF-HW2, respectively. The orange dashed lines, purple dashed lines and cyan solid lines in panel (k) denote the results of SH-1, SH-2 and NaF-TW, respectively. The black solid lines in panels (i)-(k) denote the numerically exact results produced by DVR[363].

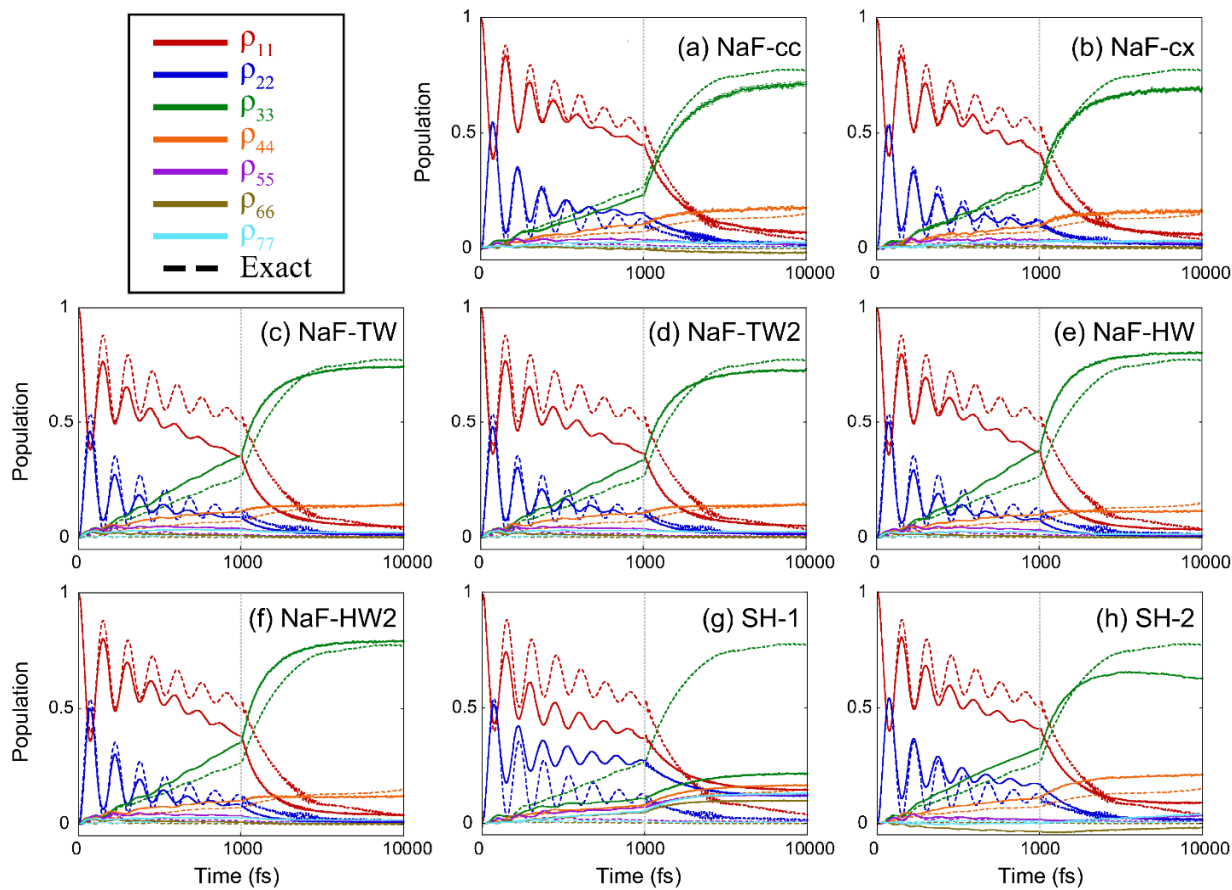

**Figure S5.** Results of population dynamics of the 7-state FMO model at zero temperature. 100 nuclear (bath) modes in the discretization scheme are employed for each state in the simulations. In each panel, the red, blue, green, orange, purple, brown and cyan solid lines represent the population of states 1-7, respectively. Panel (a): NaF-cc; Panel (b): NaF-cx; Panel (c): NaF-TW; Panel (d): NaF-TW2; Panel (e): NaF-HW; Panel (f): NaF-HW2; Panel (g): SH-1; Panel (h): SH-2. Note that SH-3 is not applicable for this 7-state model. The numerically exact results produced by TD-DMRG[151-160] for the same effective Hamiltonian in the discretization scheme are demonstrated by dashed lines with corresponding colors in each panel.

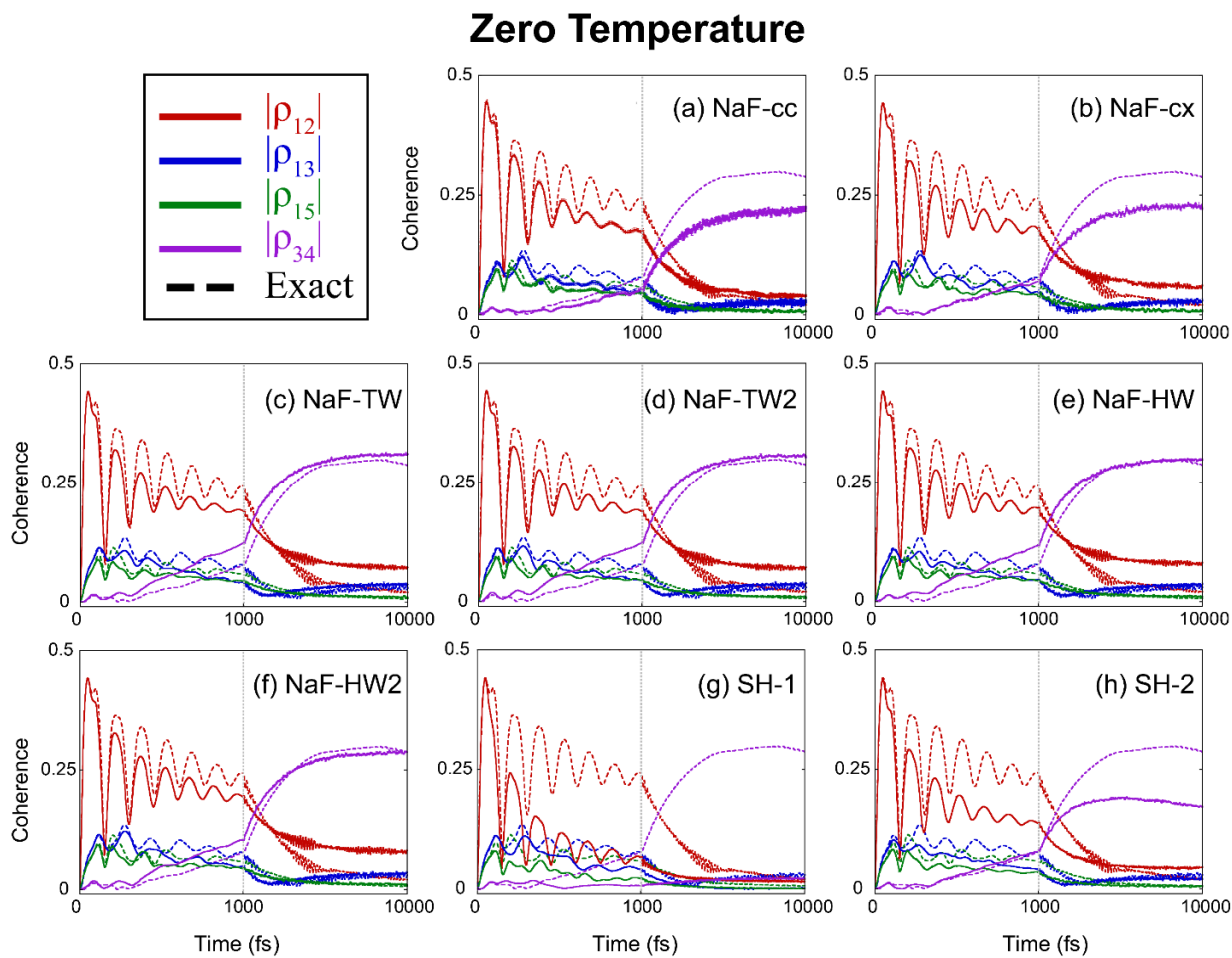

**Figure S6.** Results of coherence dynamics of the 7-state FMO model at zero temperature. 100 nuclear (bath) modes in the discretization scheme are employed for each state in the simulations. In each panel, the red, blue, green, and purple solid lines represent the moduli of the off-diagonal terms $\rho_{12}$, $\rho_{13}$, $\rho_{15}$ and $\rho_{34}$, respectively. Panel (a): NaF-cc; Panel (b): NaF-cx; Panel (c): NaF-TW; Panel (d): NaF-TW2; Panel (e): NaF-HW; Panel (f): NaF-HW2; Panel (g): SH-1; Panel (h): SH-2. Note that SH-3 is not applicable for this 7-state model. The numerically exact results produced by TD-DMRG[151-160] for the same effective Hamiltonian in the discretization scheme are demonstrated by dashed lines with corresponding colors in each panel.

## S8. Transition Path Flight Time Simulation

In this section, we use NaF-TW to study the transition path flight times of the SAC model, a benchmark test proposed in our previous work[347]. All computational details are described in ref [347]. Below we present a brief summary.

The parameters of the SAC model are listed in Sub-Section 3.2 of the main text. The barrier center of the adiabatic ground state is located at 0 au. The initial nuclear wave function is described by eq (82) of the main text with $R_0 = -77.8617$ au. The width parameter $\alpha \in \{0.006, 0.03\}$ corresponds to the narrow and wide initial wavepacket cases, respectively. We focus on the most challenging resonance region in ref [347], where the center of the nuclear momentum is defined by $P_0 = \sqrt{2ME_{\text{kin}}}$ with the initial mean kinetic energy $E_{\text{kin}} \in [0.015, 0.02]$ au. Two screens for detecting the scattering time are placed at $Y_i = \pm 145.723$ au. The trajectories start from the adiabatic ground state and end when they reach either screen. The method for counting the scattering time of NaF-TW is the same as that used for SH-1 in ref [347]. Specifically, trajectory ensembles in different scattering channels are distinguished by the index of the occupied state $j_{occ}$ and the nuclear coordinate. The average arrival time of the corresponding trajectory ensemble is calculated. We demonstrate the difference between the mean flight time and the free-particle flight time $t_{\text{fp}}$. In both transmission and reflection channels of the adiabatic ground state, the free-particle flight time is defined as $t_{\text{fp}} = M(|Y_i| + |R_0|)/P_0$. Figure S7 presents the difference between the mean flight time obtained from NaF-TW and the free-particle flight time as a function of the initial mean kinetic energy, in comparison to that obtained by SH-1 and the exact result calculated by DVR.

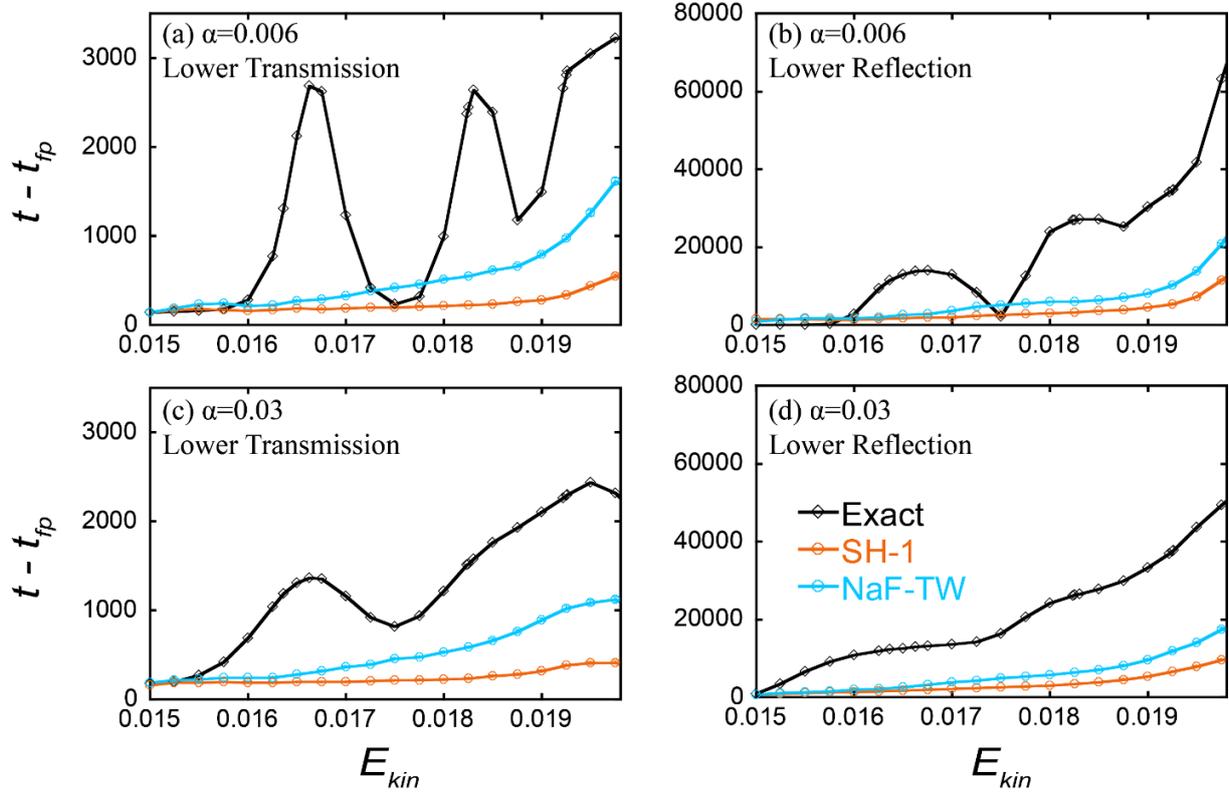

**Figure S7**. Results of the mean flight time difference of the SAC model as a function of the initial mean kinetic energy are illustrated. Panels (a) and (b) illustrate the transmission and reflection results on the adiabatic ground state with width parameter $\alpha = 0.006$. Panels (c) and (d) are the same as panels (a) and (b) but with width parameter $\alpha = 0.03$. Black hollow diamonds with black solid lines: numerically exact results by DVR[347]; Orange circles with orange solid lines: SH-1 results; Cyan circles with cyan solid lines: NaF-TW results. All physical qualities are defined in atomic units.

## S9. Relation between the EOMs of NaF and the Exact EOMs in the Generalized Coordinate-Momentum Phase Space Formulation of Quantum Mechanics

In this section, we keep $\hbar$ in all equations for clarity. As discussed in Appendix 5 of ref [12], the phase space expression of the quantum Liouville theorem (or the von Neumann equation)

$$\frac{\partial}{\partial t}\hat{\rho} = \frac{1}{i\hbar}\left[\hat{H},\hat{\rho}\right] \tag{S195}$$

leads to the exact general Wigner-Moyal equation on quantum coordinate-momentum phase space

$$\frac{\partial}{\partial t}\rho(\mathbf{X}) = \{\{H(\mathbf{X}),\rho(\mathbf{X})\}\} = \frac{1}{i\hbar}\left(H(\mathbf{X})\star\rho(\mathbf{X})-\rho(\mathbf{X})\star H(\mathbf{X})\right) = \mathcal{L}\rho(\mathbf{X}) \ , \tag{S196}$$

where $\mathbf{X}=(\mathbf{R},\mathbf{P},\mathbf{x},\mathbf{p},\boldsymbol{\Gamma})$ is the set of variables of the generalized coordinate-momentum phase space (in the diabatic representation), the symbol $\star$ denotes the (general) Moyal product[8], and $\{\{A,B\}\} = (A\star B - B\star A)/i\hbar$ represents the corresponding Moyal bracket. In eq (S196), $\mathcal{L}$ is the quantum Liouville operator, and $\mathcal{L}^*$ is its adjoint operator.

Equation (S196) is a generalized version of the Wigner-Moyal equation for continuous-variable systems, e.g., used in refs [14-17, 36, 38-40] for developing practical quantum dynamics methods on Wigner phase space. When we use the mapping kernel $\hat{K}_{\text{nuc}}(\mathbf{R},\mathbf{P})\otimes\hat{K}_{\text{ele}}(\mathbf{x},\mathbf{p},\boldsymbol{\Gamma})$ defined by eqs (18)-(19) of the main text, the mapping Hamiltonian $H(\mathbf{X})$ in the diabatic representation is eq (17) of the main text, and the corresponding (generalized) Moyal product is $\star = \star_{\text{nuc}}\otimes\star_{\text{ele}}$ in the diabatic representation, where

$$\star_{\text{nuc}} = \exp\left(\frac{i\hbar\left(\overleftarrow{\nabla}_{\mathbf{R}}\cdot\overrightarrow{\nabla}_{\mathbf{P}}-\overleftarrow{\nabla}_{\mathbf{P}}\cdot\overrightarrow{\nabla}_{\mathbf{R}}\right)}{2}\right) \tag{S197}$$

denotes the Moyal product of nuclear DOFs on Wigner phase space[8, 13, 15, 40, 440], and

$$\star_{\text{ele}} = \hat{\Xi}_{\mathbf{x},\mathbf{p}} - \sum_{k=1}^{F} s_k \hat{\Xi}_{\xi_k,\pi_k} \tag{S198}$$

denotes the (generalized) Moyal product of electronic DOFs on CPS[12, 441]. Here we express commutator matrix $\mathbf{\Gamma}$ by using the corresponding commutator variables $\{\xi_k, \pi_k\}$[320]:

$$\mathbf{\Gamma} = \sum_{k=1}^{F} \frac{s_k}{2} (\xi_k + i\pi_k)(\xi_k - i\pi_k)^{\text{T}} \tag{S199}$$

with $s_k$ denoting the sign of the $k$-th eigenvalue of $\mathbf{\Gamma}$, and operators $\hat{\Xi}_{\mathbf{x},\mathbf{p}}$ and $\hat{\Xi}_{\xi_k,\pi_k}$ read[12]

$$\begin{cases} \hat{\Xi}_{\mathbf{x},\mathbf{p}} = \frac{1}{2}(\overleftarrow{\nabla}_{\mathbf{x}} - i\overleftarrow{\nabla}_{\mathbf{p}}) \cdot (\overrightarrow{\nabla}_{\mathbf{x}} + i\overrightarrow{\nabla}_{\mathbf{p}}), \\ \hat{\Xi}_{\xi_k,\pi_k} = \frac{1}{2}(\overleftarrow{\nabla}_{\xi_k} - i\overleftarrow{\nabla}_{\pi_k}) \cdot (\overrightarrow{\nabla}_{\xi_k} + i\overrightarrow{\nabla}_{\pi_k}) \end{cases}. \tag{S200}$$

Throughout this section, eqs (S198) and (S200) are defined on the Euclidean space spanned by the variables $\{\mathbf{R},\mathbf{P},\mathbf{x},\mathbf{p},\{\xi_k,\pi_k\}\}$. Equation (S196) involves *no* approximation. As mentioned in ref [12], the exact EOMs of eq (S196) do not lead to practical approaches for generating (independent) trajectories even with the quantum-classical approximation.

1) **EOMs of CMMcv**

When only the terms up to the first-order of $\hbar$ of the Moyal product of nuclear DOFs are retained in eq (S197), eq (S196) is approximated by

$$\begin{aligned}\frac{\partial}{\partial t}\rho(\mathbf{X}) \approx &\frac{1}{2}\big(\nabla_{\mathbf{R}}H(\mathbf{X}) \star_{\text{ele}} \nabla_{\mathbf{P}}\rho(\mathbf{X}) + \nabla_{\mathbf{P}}\rho(\mathbf{X}) \star_{\text{ele}} \nabla_{\mathbf{R}}H(\mathbf{X})\big) \\ &-\nabla_{\mathbf{P}}H(\mathbf{X}) \cdot \nabla_{\mathbf{R}}\rho(\mathbf{X}) \\ &+\frac{1}{\hbar}\big(\nabla_{\mathbf{x}}H(\mathbf{X}) \cdot \nabla_{\mathbf{p}}\rho(\mathbf{X}) - \nabla_{\mathbf{p}}H(\mathbf{X}) \cdot \nabla_{\mathbf{x}}\rho(\mathbf{X})\big) \\ &-\frac{1}{\hbar}\sum_{k=1}^{F}s_k\big(\nabla_{\xi_k}H(\mathbf{X}) \cdot \nabla_{\pi_k}\rho(\mathbf{X}) - \nabla_{\pi_k}H(\mathbf{X}) \cdot \nabla_{\xi_k}\rho(\mathbf{X})\big)\end{aligned}, \tag{S201}$$

which does not generate practical trajectory evolution for nuclear DOFs. The strategy of the linearized semiclassical initial value representation (LSC-IVR, also known as the classical Wigner model) for nuclear DOFs in principle approximates the Moyal bracket to the corresponding classical Poisson bracket[11, 12, 14, 22, 318-320], leading to the following approximation of eq (S201)

$$\begin{aligned}\frac{\partial}{\partial t}\rho(\mathbf{X}) &\approx \{H(\mathbf{X}),\rho(\mathbf{X})\}_{\text{Poisson}} \\ &= \nabla_{\mathbf{R}}H(\mathbf{X})\cdot\nabla_{\mathbf{P}}\rho(\mathbf{X})-\nabla_{\mathbf{P}}H(\mathbf{X})\cdot\nabla_{\mathbf{R}}\rho(\mathbf{X}) \\ &+ \frac{1}{\hbar}\left(\nabla_{\mathbf{x}}H(\mathbf{X})\cdot\nabla_{\mathbf{p}}\rho(\mathbf{X})-\nabla_{\mathbf{p}}H(\mathbf{X})\cdot\nabla_{\mathbf{x}}\rho(\mathbf{X})\right) \\ &- \frac{1}{\hbar}\sum_{k=1}^{F}s_k\left(\nabla_{\xi_k}H(\mathbf{X})\cdot\nabla_{\pi_k}\rho(\mathbf{X})-\nabla_{\pi_k}H(\mathbf{X})\cdot\nabla_{\xi_k}\rho(\mathbf{X})\right)\end{aligned} \quad (S202)$$

Equation (S202) produces the EOMs in the diabatic representation[320, 442]

$$\begin{aligned}\dot{\mathbf{R}} &= \nabla_{\mathbf{P}}H(\mathbf{X})=\mathbf{M}^{-1}\mathbf{P} \\ \dot{\mathbf{P}} &= -\nabla_{\mathbf{R}}H(\mathbf{X})=\mathbf{F}^{(\text{MF})}(\mathbf{X}) \\ \dot{\mathbf{x}} &= \frac{1}{\hbar}\nabla_{\mathbf{p}}H(\mathbf{X})=\frac{1}{\hbar}\mathbf{V}(\mathbf{R})\mathbf{p} \\ \dot{\mathbf{p}} &= -\frac{1}{\hbar}\nabla_{\mathbf{x}}H(\mathbf{X})=-\frac{1}{\hbar}\mathbf{V}(\mathbf{R})\mathbf{x} \quad , \\ \dot{\xi}_k &= -\frac{s_k}{\hbar}\nabla_{\pi_k}H(\mathbf{X})=\frac{1}{\hbar}\mathbf{V}(\mathbf{R})\pi_k \\ \dot{\pi}_k &= \frac{s_k}{\hbar}\nabla_{\xi_k}H(\mathbf{X})=-\frac{1}{\hbar}\mathbf{V}(\mathbf{R})\xi_k\end{aligned} \quad (S203)$$

where we assume that $\mathbf{V}(\mathbf{R})$ is real and symmetric for simplicity. (It is trivial to generalize it to the complex Hermitian case.) Equation (S203) yields Ehrenfest-like dynamics employed by CMMcv in our previous work[320] with the nuclear force

$$\begin{aligned}\mathbf{F}^{(\text{MF})}(\mathbf{X}) &= -\text{Tr}\left[\nabla_{\mathbf{R}}\mathbf{V}(\mathbf{R})\mathbf{K}_{\text{ele}}(\mathbf{x},\mathbf{p},\mathbf{\Gamma})\right] \\ &= -\sum_{n,m=1}^{F}\nabla_{\mathbf{R}}V_{mn}(\mathbf{R})\left(\frac{(x^{(n)}+\mathrm{i}p^{(n)})(x^{(m)}-\mathrm{i}p^{(m)})}{2}-\sum_{k=1}^{F}\frac{s_k(\xi_k^{(n)}+\mathrm{i}\pi_k^{(n)})(\xi_k^{(m)}-\mathrm{i}\pi_k^{(m)})}{2}\right)\end{aligned} \quad (S204)$$

in the diabatic representation. The covariant EOMs of eq (S203) in the adiabatic representation reads[12, 320]

$$\begin{aligned}
\dot{\mathbf{R}} &= \mathbf{M}^{-1}\mathbf{P} \\
\dot{\mathbf{P}} &= \tilde{\mathbf{F}}^{(\mathrm{MF})}(\tilde{\mathbf{X}}) \\
\dot{\tilde{\mathbf{x}}} &= \frac{1}{\hbar}\nabla_{\tilde{\mathbf{p}}}V^{(\mathrm{eff})}(\tilde{\mathbf{X}}) = \frac{1}{\hbar}\mathrm{Im}\left[\mathbf{V}^{(\mathrm{eff})}(\mathbf{R},\mathbf{P})(\tilde{\mathbf{x}}+i\tilde{\mathbf{p}})\right] \\
\dot{\tilde{\mathbf{p}}} &= -\frac{1}{\hbar}\nabla_{\tilde{\mathbf{x}}}V^{(\mathrm{eff})}(\tilde{\mathbf{X}}) = -\frac{1}{\hbar}\mathrm{Re}\left[\mathbf{V}^{(\mathrm{eff})}(\mathbf{R},\mathbf{P})(\tilde{\mathbf{x}}+i\tilde{\mathbf{p}})\right] \quad , \\
\dot{\tilde{\boldsymbol{\xi}}}_k &= -\frac{s_k}{\hbar}\nabla_{\tilde{\boldsymbol{\pi}}_k}V^{(\mathrm{eff})}(\tilde{\mathbf{X}}) = \frac{1}{\hbar}\mathrm{Im}\left[\mathbf{V}^{(\mathrm{eff})}(\mathbf{R},\mathbf{P})(\tilde{\boldsymbol{\xi}}_k+i\tilde{\boldsymbol{\pi}}_k)\right] \\
\dot{\tilde{\boldsymbol{\pi}}}_k &= \frac{s_k}{\hbar}\nabla_{\tilde{\boldsymbol{\xi}}_k}V^{(\mathrm{eff})}(\tilde{\mathbf{X}}) = -\frac{1}{\hbar}\mathrm{Re}\left[\mathbf{V}^{(\mathrm{eff})}(\mathbf{R},\mathbf{P})(\tilde{\boldsymbol{\xi}}_k+i\tilde{\boldsymbol{\pi}}_k)\right]
\end{aligned} \quad (\mathrm{S205})$$

where $\{\tilde{\boldsymbol{\xi}}_k, \tilde{\boldsymbol{\pi}}_k\}$ are the covariant commutator variables in the adiabatic representation that satisfy $\tilde{\boldsymbol{\xi}}_k + i\tilde{\boldsymbol{\pi}}_k = \mathbf{T}^{\dagger}(\mathbf{R})(\boldsymbol{\xi}_k + i\boldsymbol{\pi}_k)$, and $\tilde{\mathbf{X}} = (\mathbf{R},\mathbf{P},\tilde{\mathbf{x}},\tilde{\mathbf{p}},\{\tilde{\boldsymbol{\xi}}_k,\tilde{\boldsymbol{\pi}}_k\})$. The effective potential matrix, $\mathbf{V}^{(\mathrm{eff})}(\mathbf{R},\mathbf{P})$, is defined by eq (27) of the main text, and in eq (S205) we have $V^{(\mathrm{eff})}(\tilde{\mathbf{X}}) = \mathrm{Tr}\left[\mathbf{V}^{(\mathrm{eff})}(\mathbf{R},\mathbf{P})\mathbf{K}_{\mathrm{ele}}(\tilde{\mathbf{x}},\tilde{\mathbf{p}},\tilde{\boldsymbol{\Gamma}})\right]$. The nuclear force in the adiabatic representation reads

$$\begin{aligned}
\tilde{\mathbf{F}}^{(\mathrm{MF})}(\tilde{\mathbf{X}}) = &-\sum_{n=1}^{F}\nabla_{\mathbf{R}}E_n(\mathbf{R})\left(\frac{(\tilde{x}^{(n)})^2+(\tilde{p}^{(n)})^2}{2} - \sum_{k=1}^{F}\frac{s_k\left((\tilde{\xi}_k^{(n)})^2+(\tilde{\pi}_k^{(n)})^2\right)}{2}\right) \\
&-\sum_{n\neq m}^{F}(E_n(\mathbf{R})-E_m(\mathbf{R}))\mathbf{d}_{mn}(\mathbf{R}) \\
&\times\left(\frac{(\tilde{x}^{(n)}+i\tilde{p}^{(n)})(\tilde{x}^{(m)}-i\tilde{p}^{(m)})}{2} - \sum_{k=1}^{F}\frac{s_k(\tilde{\xi}_k^{(n)}+i\tilde{\pi}_k^{(n)})(\tilde{\xi}_k^{(m)}-i\tilde{\pi}_k^{(m)})}{2}\right)
\end{aligned} \quad . \quad (\mathrm{S206})$$

The EOMs in eq (S205) lead to the counterpart of eq (S202) in the adiabatic representation

$$\begin{aligned}
\frac{\partial}{\partial t}\rho(\tilde{\mathbf{X}}) \approx &\left(-\tilde{\mathbf{F}}^{(\mathrm{MF})}(\tilde{\mathbf{X}})\cdot\nabla_{\mathbf{P}}\rho(\tilde{\mathbf{X}}) - \mathbf{M}^{-1}\mathbf{P}\cdot\nabla_{\mathbf{R}}\rho(\tilde{\mathbf{X}})\right) \\
&+\frac{1}{\hbar}\left(\nabla_{\tilde{\mathbf{x}}}V^{(\mathrm{eff})}(\tilde{\mathbf{X}})\cdot\nabla_{\tilde{\mathbf{p}}}\rho(\tilde{\mathbf{X}}) - \nabla_{\tilde{\mathbf{p}}}V^{(\mathrm{eff})}(\tilde{\mathbf{X}})\cdot\nabla_{\tilde{\mathbf{x}}}\rho(\tilde{\mathbf{X}})\right) \\
&-\frac{1}{\hbar}\sum_{k=1}^{F}s_k\left(\nabla_{\tilde{\boldsymbol{\xi}}_k}V^{(\mathrm{eff})}(\tilde{\mathbf{X}})\cdot\nabla_{\tilde{\boldsymbol{\pi}}_k}\rho(\tilde{\mathbf{X}}) - \nabla_{\tilde{\boldsymbol{\pi}}_k}V^{(\mathrm{eff})}(\tilde{\mathbf{X}})\cdot\nabla_{\tilde{\boldsymbol{\xi}}_k}\rho(\tilde{\mathbf{X}})\right)
\end{aligned} \quad . \quad (\mathrm{S207})$$

We note that either of eq (S203) and eq (S205) preserves the mapping Hamiltonian with commutator matrix [eq (17) or eq (20)].

### 2) EOMs of NaF

Instead of the introduction of the approximation of the Poisson bracket, a more profound perspective for developing approximate quantum dynamics methods with independent trajectories on quantum phase space is presented in ref [40]. The strategy of ref [40] can straightforwardly be implemented on generalized coordinate-momentum phase space. The essential idea is to solve the partial differential equation of eq (S196) by using a set of ordinary differential equations as well as an exact series expansion of the phase space propagator.

In the adiabatic representation, the EOMs of NaF are eqs (25), (26), (35)-(37) of the main text, which leads to the adjoint Liouville operator (for physical properties),

$$\tilde{\mathcal{L}}^*_{\text{NaF}} = \left( \mathbf{F}^{(\text{NaF})}_{\text{eff}}(\tilde{\mathbf{X}}) \cdot \nabla_{\mathbf{P}} + \mathbf{M}^{-1}\mathbf{P} \cdot \nabla_{\mathbf{R}} \right)$$
$$+ \frac{1}{\hbar}\left( -\nabla_{\tilde{\mathbf{x}}} V^{(\text{eff})}(\tilde{\mathbf{X}}) \cdot \nabla_{\tilde{\mathbf{p}}} + \nabla_{\tilde{\mathbf{p}}} V^{(\text{eff})}(\tilde{\mathbf{X}}) \cdot \nabla_{\tilde{\mathbf{x}}} \right) \qquad (S208)$$
$$- \frac{1}{\hbar}\sum_{k=1}^{F} s_k \left( -\nabla_{\tilde{\xi}_k} V^{(\text{eff})}(\tilde{\mathbf{X}}) \cdot \nabla_{\tilde{\pi}_k} + \nabla_{\tilde{\pi}_k} V^{(\text{eff})}(\tilde{\mathbf{X}}) \cdot \nabla_{\tilde{\xi}_k} \right)$$

and the corresponding evolution of the phase space density reads

$$\frac{\partial}{\partial t}\rho(\tilde{\mathbf{X}}) \approx \tilde{\mathcal{L}}_{\text{NaF}}\rho(\tilde{\mathbf{X}})$$
$$= \left( -\nabla_{\mathbf{P}} \cdot \left( \tilde{\mathbf{F}}^{(\text{NaF})}_{\text{eff}}(\tilde{\mathbf{X}})\rho(\tilde{\mathbf{X}}) \right) - \mathbf{M}^{-1}\mathbf{P} \cdot \nabla_{\mathbf{R}}\rho(\tilde{\mathbf{X}}) \right)$$
$$+ \frac{1}{\hbar}\left( \nabla_{\tilde{\mathbf{x}}} V^{(\text{eff})}(\tilde{\mathbf{X}}) \cdot \nabla_{\tilde{\mathbf{p}}}\rho(\tilde{\mathbf{X}}) - \nabla_{\tilde{\mathbf{p}}} V^{(\text{eff})}(\tilde{\mathbf{X}}) \cdot \nabla_{\tilde{\mathbf{x}}}\rho(\tilde{\mathbf{X}}) \right) \qquad (S209)$$
$$- \frac{1}{\hbar}\sum_{k=1}^{F} s_k \left( \nabla_{\tilde{\xi}_k} V^{(\text{eff})}(\tilde{\mathbf{X}}) \cdot \nabla_{\tilde{\pi}_k}\rho(\tilde{\mathbf{X}}) - \nabla_{\tilde{\pi}_k} V^{(\text{eff})}(\tilde{\mathbf{X}}) \cdot \nabla_{\tilde{\xi}_k}\rho(\tilde{\mathbf{X}}) \right)$$

As described in Section S1, the mapping energy conservation procedure of NaF implies an *effective* nonadiabatic force. The effective nuclear force of NaF in the adiabatic representation is

$$\tilde{\mathbf{F}}_{\text{eff}}^{(\text{NaF})}(\tilde{\mathbf{X}}) = -\nabla_{\mathbf{R}} E_{j_{occ}}(\mathbf{R}) - \sum_{n \neq m}^{F} (E_n(\mathbf{R}) - E_m(\mathbf{R})) \mathbf{d}_{mn}(\mathbf{R}) \tilde{\rho}_{nm}$$
$$+ \frac{\mathbf{P}^T \mathbf{M}^{-1} \sum_{n \neq m}^{F} (E_n(\mathbf{R}) - E_m(\mathbf{R})) \mathbf{d}_{mn}(\mathbf{R}) \tilde{\rho}_{nm}}{\mathbf{P}^T \mathbf{M}^{-1} \mathbf{P}} \mathbf{P} \quad , \tag{S210}$$

with $\tilde{\boldsymbol{\rho}}$ defined in eq (34) of the main text as the effective electronic density matrix in the adiabatic representation.

In the diabatic representation, the EOMs of NaF become eqs (23), (24), (35) of the main text, and eq (S145), in addition to the energy conservation. Following Section S3, we yield the evolution of the phase space density

$$\frac{\partial}{\partial t} \rho(\mathbf{X}) \approx \mathcal{L}_{\text{NaF}} \rho(\mathbf{X}) \quad , \tag{S211}$$

where the NaF Liouville operator in the diabatic representation is

$$\begin{aligned}\mathcal{L}_{\text{NaF}} &= \left( -\nabla_{\mathbf{P}} \cdot \left( \mathbf{F}_{\text{eff}}^{(\text{NaF})}(\mathbf{X}) \cdot \right) - \nabla_{\mathbf{R}} \cdot \left( \mathbf{M}^{-1} \mathbf{P} \cdot \right) \right) \\ &+ \frac{1}{\hbar} \left( \nabla_{\mathbf{p}} \cdot \left( \nabla_{\mathbf{x}} H(\mathbf{X}) \cdot \right) - \nabla_{\mathbf{x}} \cdot \left( \nabla_{\mathbf{p}} H(\mathbf{X}) \cdot \right) \right) \\ &- \frac{1}{\hbar} \sum_{k=1}^{F} s_k \left( \nabla_{\boldsymbol{\pi}_k} \cdot \left( \nabla_{\boldsymbol{\xi}_k} H(\mathbf{X}) \cdot \right) - \nabla_{\boldsymbol{\xi}_k} \cdot \left( \nabla_{\boldsymbol{\pi}_k} H(\mathbf{X}) \cdot \right) \right)\end{aligned} \quad . \tag{S212}$$

The effective nuclear force in the diabatic representation reads

$$\mathbf{F}_{\text{eff}}^{(\text{NaF})}(\mathbf{X}) = -\sum_{n,m=1}^{F} \nabla_{\mathbf{R}} V_{mn}(\mathbf{R}) \left( Q_{nm}^{(a)} + Q_{nm}^{(na)} \right) + \frac{\mathbf{P}^T \mathbf{M}^{-1} \left( \sum_{n,m=1}^{F} \nabla_{\mathbf{R}} V_{mn}(\mathbf{R}) Q_{nm}^{(na)} \right)}{\mathbf{P}^T \mathbf{M}^{-1} \mathbf{P}} \mathbf{P} \quad , \tag{S213}$$

with $\mathbf{Q}^{(a)}$ and $\mathbf{Q}^{(na)}$ defined by eqs (S146) and (S147), respectively.

The operator adjoint to $\mathcal{L}_{\text{NaF}}$ is

$$\mathcal{L}^*_{\text{NaF}} = \left( \mathbf{F}^{(\text{NaF})}_{\text{eff}}(\mathbf{X}) \cdot \nabla_{\mathbf{P}} + \mathbf{M}^{-1}\mathbf{P} \cdot \nabla_{\mathbf{R}} \right)$$
$$+ \frac{1}{\hbar}\left( -\nabla_{\mathbf{x}} H(\mathbf{X}) \cdot \nabla_{\mathbf{p}} + \nabla_{\mathbf{p}} H(\mathbf{X}) \cdot \nabla_{\mathbf{x}} \right) \quad . \quad \text{(S214)}$$
$$- \frac{1}{\hbar}\sum_{k=1}^{F} s_k \left( -\nabla_{\xi_k} H(\mathbf{X}) \cdot \nabla_{\pi_k} + \nabla_{\pi_k} H(\mathbf{X}) \cdot \nabla_{\xi_k} \right)$$

In NaF, we use the zeroth order propagator $U_0(t) = \exp(\mathcal{L}^*_{\text{NaF}} t)$ instead of the exact phase space propagator $U(t) = \exp(\mathcal{L}^* t)$ in the evaluation of eq (73) or eq (74) of the main text for general time-dependent properties. We define a correction operator

$$\mathcal{C}(t) = \left( \mathcal{L}^* - \mathcal{L}^*_{\text{NaF}} \right) U_0(t) \quad \text{(S215)}$$

and the recursion relation

$$U_{j+1}(t) = \int_0^t d\tau\, U_j(t-\tau) \mathcal{C}(\tau) \quad . \quad \text{(S216)}$$

It is straightforward to show the exact propagator can be represented by an exact series expansion, i.e.,

$$U(t) = \sum_{j=0}^{\infty} U_j(t) \quad . \quad \text{(S217)}$$

In principle, we can construct the exact series expansion of the phase space propagator of eq (S217) by using the NaF approach. It is, however, often numerically demanding to accomplish the task for studying general nonadiabatic systems. The estimation of the first correction $U_1(t) = \int_0^t d\tau\, U_0(t-\tau) \mathcal{C}(\tau)$ sheds light on the performance of NaF even when exact results are not available. In the main text as well as refs [322, 324], we have shown in various benchmark tests that NaF predicts reasonably reliable results for both electronic and nuclear motion in a wide region, which includes where relevant (electronic) states always keep coupled in a broad range or all the time and where the bifurcation characteristic of nuclear motion is essentially important. Provided that computational effort is affordable, lower-order correction terms of eq (S217) can be constructed and should often be enough to improve over NaF in a rigorous and systematic framework. The strategy can in principle be applied to other nonadiabatic dynamics approaches on quantum phase space. Because NaF is superior to many other approaches for the zero-order term, the convergence of the exact series expansion is expected to be much more efficient.

When only electronic DOFs are involved, i.e., in the frozen-nuclei limit, all terms involving the derivatives of nuclear DOFs disappear in the evolution of the phase space density. Both eqs (S202) and (S211) become identical to eq (S201) as well as eq (S196). Pure electronic dynamics on quantum phase space in either NaF or CMMcv is simply exact.

**S10. Electronic TCF of the Weighted Mapping Model with the Born-Oppenheimer Limit**

The weighted CPS formulation[12] includes the (electronic) TCF of our earlier work with one phase space parameter[443] in 2021, which is generalized to the (electronic) TCF of weighted mapping model[12] with the Born-Oppenheimer limit (wMM-BO) as described in refs [441, 444].

The TCF of wMM-BO employs the initial sampling over $F$ sets of angle variables $\{\theta_1,\cdots,\theta_F\}$, where each $\theta_\nu \in [0, 2\pi)$. The mapping kernel of wMM-BO is

$$\hat{K}_{\text{ele}}(\mathbf{x},\mathbf{p},\boldsymbol{\Gamma}) = \mathcal{N}_{\text{wMM-BO}}^{-1} \sum_{\nu=1}^{F} \hat{K}^\nu(\mathbf{x},\mathbf{p},\boldsymbol{\Gamma}) \mathcal{R}_{\mathcal{S}_\nu}(\mathbf{x},\mathbf{p},\boldsymbol{\Gamma}) \ , \tag{S218}$$

where

$$\left[\hat{K}^\nu(\mathbf{x},\mathbf{p},\boldsymbol{\Gamma})\right]_{ij} = \begin{cases} 1, & \text{if } i=j=\nu, \\ 0, & \text{if } i=j\neq\nu, \\ \gamma e^{\mathrm{i}(\theta_j-\theta_i)}, & \text{if } i\neq\nu \text{ and } j\neq\nu \text{ and } i\neq j, \\ \alpha e^{\mathrm{i}(\theta_j-\theta_i)}, & \text{otherwise,} \end{cases} \tag{S219}$$

and $\mathcal{R}_{S_v}(\mathbf{X})$ denotes the restriction from the total phase space to the subset $[0, 2\pi)^F$ with variables $\{\theta_1, \cdots, \theta_F\}$, and $\mathcal{N}_{\text{wMM-BO}}^{-1}$ is the normalization factor. The (electronic) phase space is defined as the envelope of all trajectories that start from the initial sampling points[326], which is usually much larger than the initial sampling subset.

Following refs [325, 326], the structure of the phase space and the inverse mapping kernel can be directly obtained from the structure of the eigenvalue set of $\hat{K}^v(\mathbf{x}, \mathbf{p}, \mathbf{\Gamma})$. The eigenvalues of $\hat{K}^v(\mathbf{x}, \mathbf{p}, \mathbf{\Gamma})$ are

$$\begin{cases} \lambda_1 = \frac{1}{2}\left(1 + (F-2)\gamma + \sqrt{((F-2)\gamma - 1)^2 + (4F-4)\alpha^2}\right) \\ \lambda_2 = \frac{1}{2}\left(1 + (F-2)\gamma - \sqrt{((F-2)\gamma - 1)^2 + (4F-4)\alpha^2}\right) \\ \lambda_3 = \cdots = \lambda_F = -\gamma \end{cases} \quad \text{(S220)}$$

Only when $\alpha^2 = \gamma(\gamma + 1)$ and thus $\lambda_2 = -\gamma$, eq (S220) yields an $(F-1)$-fold degeneracy in the eigenvalue set. In such a special case, the corresponding phase space structure becomes $\text{U}(F)/\text{U}(F-1)$[326] and the inverse mapping kernel $\hat{K}_{\text{ele}}^{-1}(\mathbf{x}, \mathbf{p}, \mathbf{\Gamma})$ of wMM-BO is equal to that of the covariant-covariant TCF. In all other cases, eq (S220) includes two distinct eigenvalues that are different from $-\gamma$, and the phase space structure is $\text{U}(F)/\text{U}(F-2)$[326]. When $\alpha^2 > \gamma(\gamma + 1)$, one obtains $\lambda_2 < -\gamma$, and *vice versa*. The inverse mapping kernel of wMM-BO in this case then reads

$$\hat{K}_{\text{ele}}^{-1}(\mathbf{x}, \mathbf{p}, \mathbf{\Gamma}) = \sum_{m,n=1}^{F} \left[ \frac{1}{2}\left(x^{(n)} + ip^{(n)}\right)\left(x^{(m)} - ip^{(m)}\right) \right. \\ \left. + \frac{s}{2}\left(\overline{x}^{(n)} + i\overline{p}^{(n)}\right)\left(\overline{x}^{(m)} - i\overline{p}^{(m)}\right) - \gamma\delta_{nm} \right] |n\rangle\langle m| \quad \text{(S221)}$$

with $s = \text{sgn}(\lambda_2 + \gamma) = \text{sgn}(\gamma(\gamma+1) - \alpha^2)$.

It is trivial to show that the (electronic) TCF of the generalized discrete truncated Wigner approximation (GDTWA) method[329] is a special case of wMM-BO when $\gamma = 0$ and $\alpha = 1/\sqrt{2}$ and that the (electronic) TCF of the focused spin-LSC method[445] corresponds to another special case of wMM-BO when $\alpha = \sqrt{\gamma(\gamma+1)}$. Such a result was already included in the TCF of our earlier work with one phase space parameter[443] in 2021. The TCF of wMM-BO is more general.

When Ehrenfest-like dynamics is employed in wMM-BO, the GDTWA and focused spin-LSC methods are only two special cases of this class. In addition to the failure for capturing the bifurcation behavior of nuclear motion in the asymptotic region where the nonadiabatic coupling vanishes, such a class of Ehrenfest-like dynamics (with the TCF of wMM-BO) also suffers much from the negative population problem, which occurs, e.g., in the simulations of the 2-state LVCM with 24 modes for pyrazine[359] and the 7-site FMO model. Some more examples are listed in Section S6 of the Supporting Information[340] of ref [322].

The negative population problem embedded in the TCF of wMM-BO, however, cannot be solved by the NaF approach. Figure S12 of the Supporting Information[340] of ref [322] shows that the NaF approach can alleviate it, but the negative population problem is still considerable in the simulation of the 7-site FMO model. When NaF is used, the TCF of wMM-BO is then not as generally useful as those phase space expressions listed in the main text of the present paper.